\documentclass[a4paper,11pt]{article}
\usepackage{heppub}
\pdfoutput=1
\graphicspath{{plots/}}
\usepackage{xspace}
\usepackage{mathtools}

\newcommand{\abs}[1]{\lvert#1\rvert}
\renewcommand{\Re}{\operatorname{Re}} 
\newcommand{\ord}[1]{\mathcal{O}(#1)}

\newcommand{\df}{\mathrm{d}} 
\newcommand\defeq{\equiv} 

\newcommand{\GeV}{\,\mathrm{GeV}}
\newcommand{\nn}{\nonumber}

\newcommand{\as}{\alpha_s}
\newcommand{\MSbar}{\ensuremath{\overline{\text{MS}}}\xspace}

\newcommand{\bra}[1]{\ensuremath{\left\langle#1\right|}}
\newcommand{\ket}[1]{\ensuremath{\left|#1\right\rangle}}

\newcommand\mb{\widehat{m}_b} 
\newcommand\mbPole{m_b^{\rm pole}} 
\newcommand\deltam{\delta_m}
\newcommand\ShapeFunction{\mathcal{F}} 
\newcommand\Hard{\widehat{h}_s} 
\newcommand\HardPole{h_s} 
\newcommand\Jet{\widehat{J}} 
\newcommand\JetPole{J} 
\newcommand\HadronicSoft{S} 
\newcommand\PartonicSoft{\widehat{C}_0} 
\newcommand\PartonicSoftPole{C_0} 
\newcommand\HardEvolution{\widehat{U}_H} 
\newcommand\HardEvolutionPole{U_H} 
\newcommand\JetEvolution{\widehat{U}_J} 
\newcommand\SoftEvolution{U_S} 
\newcommand\CSevenInclusive{C_7^{\rm incl}}
\newcommand\CSevenInclusiveSquared{\bigl|C_7^{\rm incl}\bigr|^2}
\newcommand\PartonicSpectrum{\widehat{P}}
\newcommand\lambdaOne{\widehat{\lambda}_1} 
\newcommand\rhoOne{\widehat{\rho}_1} 
\newcommand{\OneS}{{1\ensuremath{S}}\xspace}

\newcommand\DeltaScalevar{\Delta_{\rm profile}}
\newcommand\DeltaResummation{\Delta_{\rm resum}}
\newcommand\DeltaNonsingular{\Delta_{\rm ns}}
\newcommand\DeltaMatching{\Delta_{\rm match}}
\newcommand\DeltaNonsingularNP{\Delta_{c^{\rm ns}}}
\newcommand\DeltaHardNP{\Delta_{h_3}}

\newcommand{\NLLp}{NLL$'$\xspace}
\newcommand{\NNLLp}{NNLL$'$\xspace}
\newcommand{\NNNLLp}{N$^3$LL$'$\xspace}
\newcommand{\NLLpMatched}{NLL$'+$NLO\xspace}
\newcommand{\NNLLpMatched}{NNLL$'+$NNLO\xspace}
\newcommand{\NNNLLpMatched}{N$^3$LL$'+$N$^3$LO$(c_k)$\xspace}
\newcommand{\NNNLO}{N$^3$LO\xspace}

\newcommand{\TwoLoop}{2-loop\xspace}
\newcommand{\ThreeLoop}{3-loop\xspace}

\newcommand{\sing}{\mathrm{s}}
\newcommand{\nons}{\mathrm{ns}}
\newcommand{\full}{\mathrm{full}}
\newcommand{\cusp}{\mathrm{cusp}}
\newcommand{\expandedHard}{\mathrm{exp}}
\newcommand{\unexpandedHard}{\mathrm{unexp}}

\newcommand{\scetlib}{{\tt SCETlib}}

\newcommand{\btosgamma}{$B\to X_s\gamma$\xspace}
\newcommand{\btou}{$B\to X_u\ell\nu$\xspace}

\allowdisplaybreaks[2]

\newcommand{\WidthOneSubfig}{0.6\textwidth}
\newcommand{\WidthTwoSubfigs}{0.48\textwidth}

\title{\boldmath The photon energy spectrum in \btosgamma at \NNNLLp}

\author{Bahman Dehnadi,}
\emailAdd{bahman.dehnadi@desy.de}

\author{Ivan Novikov,}
\emailAdd{ivan.novikov@desy.de}

\author{and Frank J.~Tackmann}
\emailAdd{frank.tackmann@desy.de}

\affiliation{Deutsches Elektronen-Synchrotron DESY, Notkestr. 85, 22607 Hamburg, Germany}

\abstract{%
We present predictions for the photon energy spectrum in inclusive \btosgamma decays mediated by
the electromagnetic penguin operator $O_7$ to \NNNLLp. We use soft-collinear
effective theory (SCET) to resum the singular contributions in the peak region
at large photon energy. In the tail region the resummed predictions
are matched to fixed order at \NNNLO, where we include the known fixed-order
contributions for $O_7$ up to $\ord{\alpha_s^2}$. We develop a method to suitably
parametrize the still unknown $\ord{\alpha_s^3}$ nonsingular corrections in
terms of theory nuisance parameters, whose variations provide an estimate of the
associated theory uncertainty. In this context, we also study
different ways to treat higher-order cross terms in the matching. Another
important aspect of our analysis is the short-distance scheme used for the
$b$-quark mass $m_b$. We find that in the present context, the \OneS mass
scheme, which was previously used up to \TwoLoop order, fails to work at
\ThreeLoop order, because the mass scheme enters at a soft
scale much smaller than $m_b$ here, for which the \OneS scheme was not devised. Using
instead the MSR mass scheme with $R\sim 1\GeV$, we obtain stable
results with good perturbative convergence up to \NNNLLp.
}

\date{2022-11-14}

\preprint{\vbox{%
\hbox{DESY 22-181}}
}

\begin{document}

\maketitle

\section{Introduction}
\label{sec:introdution}

The flavor-changing neutral-current $b\to s\gamma$ transition plays a key role in exploring the flavor sector of the Standard Model (SM)~\cite{Bertolini:1986th,Grinstein:1987vj,Misiak:2006zs,Misiak:2015xwa} and in searches for possible physics beyond the SM~\cite{Grinstein:1987pu,Hou:1987kf,Misiak:2017bgg}.
A prominent example is the inclusive \btosgamma decay, whose normalization is
sensitive to beyond-SM contributions.

Experimental measurements of \btosgamma are most sensitive in the peak region at large
photon energy $E_\gamma$, where as a result also most information on the
normalization of the \btosgamma rate comes from.
Furthermore, the shape of the $E_\gamma$ spectrum is directly sensitive to
the $b$-quark distribution function, known as shape function, which describes the relevant nonperturbative dynamics of the $b$ quark within the $B$ meson~\cite{Neubert:1993um,Bigi:1993ex,Neubert:1993ch}.
Recently, the first global fit exploiting all available experimental information on the \btosgamma spectrum~\cite{BaBar:2007yhb,Belle:2009nth,BaBar:2012idb,BaBar:2012eja}
was carried out by the SIMBA collaboration~\cite{Bernlochner:2020jlt}, simultaneously extracting the normalization of the \btosgamma rate, encoded in the effective inclusive Wilson coefficient $\CSevenInclusive$, the $b$-quark mass $m_b$, as well as the shape function.

The shape function is a universal object that also enters the description of inclusive \btou~\cite{Bigi:1993ex, Neubert:1993ch, Bauer:2003pi} and $B\to X_s\ell^+\ell^-$~\cite{Lee:2005pk, Lee:2005pwa} decays, where one restricts the phase space to small hadronic invariant masses to suppress the otherwise overwhelmingly large background from $b\to c\ell\nu$ transitions. In particular, it enters in the
extraction of the Cabibbo-Kobayashi-Maskawa (CKM) matrix element $|V_{ub}|$
from \btou, which is important for overconstraining the flavor sector of the SM as it is one of the few tree-level quantities. Inclusive determinations of
$|V_{ub}|$ show some tensions with its determination from exclusive decays
as well as the indirect determination from the CKM unitarity~\cite{Zyla:2020zbs}.

In the analysis of \refcite{Bernlochner:2020jlt}, the theoretical predictions
for the \btosgamma spectrum are obtained at \NNLLpMatched. The final fit results
exhibit a similar size of theoretical and experimental uncertainties. On the
experimental side, upcoming and future Belle II
measurements~\cite{Belle-II:2022hys, Belle-II:2022cgf}  of \btosgamma and
\btou will further reduce the experimental uncertainties.
To benefit from the improved experimental precision, the theoretical
predictions have to be improved likewise.

In this work, we take an important step in this direction by extending the
resummed predictions to the next order, \NNNLLp, taking advantage of the recent
computation of the \ThreeLoop jet function and heavy-to-light soft
function~\cite{Bruser:2018rad,Bruser:2019yjk}. At this order, the \ThreeLoop
hard function, corresponding to the $b\to s\gamma$ form factor, is also needed
but currently not known. Therefore, in our numerical results, we treat its
unknown nonlogarithmic constant term at $\mathcal{O}(\as^3)$ as a theory
nuisance parameter~\cite{TNPtalkSCET, TNPs} which we vary as part of our
perturbative uncertainties.

To obtain a complete
description of the spectrum away from the endpoint, we have to match to the full
fixed-order result. At \NNNLLp, this requires one to perform the
matching to \NNNLO. Since the full fixed-order results at this order are not
known, we devise a method to parametrize the missing ingredients in terms of a
set of theory nuisance parameters $c_k$~\cite{TNPtalkSCET, TNPs} in such a way
that the matching can be performed in a
consistent manner and the perturbative uncertainties due to the missing
ingredients can be estimated. We denote the so-constructed matched result as
\NNNLLpMatched. It provides a description of the
\btosgamma spectrum which benefits from the improved precision at \NNNLLp in the
peak region while consistently including the known fixed-order results up to
$\ord{\as^2}$~\cite{Blokland:2005uk,Melnikov:2005bx,Asatrian:2006sm}.

Throughout the paper, we focus on the contributions from the electromagnetic
penguin operator, $O_7$, in the electroweak Hamiltonian, which induces the
to-be-resummed singular contributions that dominate at large photon energies. At
sufficiently high order, operators other than $O_7$ also produce singular
contributions, but these are always $O_7$-like and are automatically included
via the definition of $\CSevenInclusive$. The purely nonsingular contributions
from non-$O_7$ operators can simply be added to the order they are known as in
\refcite{Bernlochner:2020jlt}, so we do not discuss them here further.

Extending the resummation order to \NNNLLp turns out to be more subtle than one
might naively expect. We find that the \OneS mass scheme~\cite{Hoang:1998ng,
Hoang:1998hm, Hoang:1999zc}, which was used at \NNLLp in
\refscite{Ligeti:2008ac,Bernlochner:2020jlt}, fails to provide a stable
prediction at \NNNLLp. The cause for this failure lies in the intrinsic
scale $R^\OneS(\mu_S)$ of the \OneS scheme, which at the soft
scale $\mu_S$ becomes large and incompatible with the power counting of HQET.
On the other hand, by using the MSR scheme~\cite{Hoang:2008yj} we are able to
obtain stable predictions that exhibit
good perturbative convergence. Furthermore, at \NNNLLp it turns out to become
necessary to switch to a short-distance mass scheme also in the jet and hard
functions, which makes the RGE of the hard function more involved, where
$m_b$ plays the role of setting the hard kinematic scale.

The outline of this paper is as follows. In \sec{theory-description} we present
our theory framework and discuss all the ingredients for computing the
\btosgamma photon energy spectrum at \NNNLLpMatched. We discuss the
implementation of the short-distance mass scheme in more detail in
\sec{short-distance-schemes}. Our methodology for estimating the perturbative
uncertainties of our predictions is given in \sec{uncertainty}. In
\sec{results}, we present our numerical results and discuss
alternative choices for the $b$-quark mass scheme and the treatment of
higher-order perturbative terms. We summarize our findings in \sec{conclusion}.

\section{Theory framework}
\label{sec:theory-description}

\subsection{Overview}
\label{sec:overview}

We follow the setup of \refcite{Bernlochner:2020jlt} and write
the \btosgamma photon energy spectrum as
\begin{equation}\label{eq:spectrum}
  \frac{\df\Gamma}{\df E_\gamma}=
  2\Gamma_0\biggl(\frac{2E_\gamma}{\mb}\biggr)^3
  \int\!\df k\, \PartonicSpectrum(k)\, \ShapeFunction(m_B-2E_\gamma-k)
  +\mathcal{O}\biggl(\frac{\Lambda_{\rm QCD}}{\mb}\biggr)
  \,,
\end{equation}
where $\mb$ denotes the $b$-quark mass in a short-distance scheme and
\begin{equation}\label{eq:Gamma0}
  \Gamma_0=\frac{G_F^2\,\mb^5}{8\pi^3}\frac{\alpha_{\rm em}}{4\pi}|V_{tb}V_{ts}^*|^2
  \,.
\end{equation}
The function $\PartonicSpectrum(k)$ is perturbatively calculable and corresponds to the partonic $b\to s\gamma$ spectrum with $k \sim \mb - 2E_\gamma$. We
write it as~\cite{Bernlochner:2020jlt}
\begin{equation}\label{eq:perturbative-contributions}
  \PartonicSpectrum(k)=\CSevenInclusiveSquared\Bigl[W_{77}^{\rm s}(k)+W_{77}^{\rm ns}(k)\Bigr]
  +2\Re(\CSevenInclusive)\sum_{i\neq7}\mathcal{C}_iW_{7i}^{\rm ns}(k)
  +\sum_{i,j\neq7}\mathcal{C}_i\mathcal{C}_jW_{ij}^{\rm ns}(k)
\,.\end{equation}
The coefficient $\CSevenInclusive$ contains by definition all virtual contributions from
operators in the electroweak Hamiltonian that give rise to singular contributions~\cite{Lee:2006gs, Bernlochner:2020jlt}.
It is dominated by the Wilson coefficient $C_7$ of the electromagnetic operator
$O_7$. In this paper, we focus on the contributions proportional to
$\CSevenInclusiveSquared$, which are discussed in more detail in the following.

The function $W_{77}^{\rm s}(k)$ in \eq{perturbative-contributions} accounts for
the contributions to the partonic spectrum $\sim \delta(k)$ and $\ln^n(k/\mb)/k$
that are singular and dominant in the peak of the spectrum where $k\ll\mb$. They
can be resummed to all orders based on their well-known
factorization~\cite{Korchemsky:1994jb,Bauer:2001yt}.
We will resum them to \NNNLLp, as discussed in more detail in \sec{singular}.
Note that the overall factor $(2E_\gamma/\mb)^3$ in \eq{spectrum} has a purely kinematic origin. It arises from the photon phase space integration and derivative operators in the photon field strength tensor of $O_7$.
As in \refcite{Bernlochner:2020jlt}, we factor it out of the singular contributions and keep it unexpanded in the endpoint region.

The function $W_{77}^{\rm ns}(k)$ in \eq{perturbative-contributions} contains
the remaining nonsingular contributions to the partonic spectrum. They start at
$\mathcal{O}(\as)$ and are accompanied by powers of $k/\mb$ relative to the
singular contributions in $W_{77}^{\sing}$. Thus, they are power-suppressed in the peak region where
$k \ll \mb$. On the other hand, in the tail region where $k\sim\mb$, they are of
similar size as the singular contributions. We will include their known results
at fixed order to $\mathcal{O}(\as^2)$, while the currently unknown corrections
at $\mathcal{O}(\as^3)$ are estimated by introducing appropriate theory nuisance
parameters, as discussed in \sec{nonsingular}.

The remaining non-77 contributions $W_{i7}^\nons$ and $W_{ij}^\nons$ in
\eq{perturbative-contributions} are purely nonsingular and thus only become
relevant in the tail region. Since they do not have singular counterparts,
they can be straightforwardly added to the order
they are known, as was done in \refcite{Bernlochner:2020jlt}.
We neglect them in the following, since our focus here is on the resummation
and fixed-order matching of the dominant 77 contributions.
Similarly, we neglect the remaining $\mathcal{O}(\Lambda_{\rm QCD}/\mb)$ terms in \eq{spectrum}, which contain subdominant resolved and unresolved contributions.

To ensure that \eq{spectrum} reproduces the full fixed-order result in the tail
with a smooth transition between the peak and tail regimes, we use profile
scales to gradually switch off the resummation away from the peak region, as
discussed in \sec{profile}.

The nonperturbative function $\ShapeFunction(k)$ in \eq{spectrum} contains the leading shape function as well as the combination of subleading shape functions that appear at tree level in \btosgamma. It is discussed in \sec{shape-function}.
The partonic spectrum $\PartonicSpectrum$ and the hadronic shape function $\ShapeFunction$ are completely factorized in \eq{spectrum}.
This factorization enables a coherent description of the spectrum in both peak and tail region. In the tail region only the first few moments of $\ShapeFunction$ are relevant, while in the peak region its full form is required.

\subsection{Singular contributions}
\label{sec:singular}

The singular contributions $W_{77}^\sing(k)$ are the leading contributions to
the spectrum in the limit $k\ll m_b$.
Their well-known factorization theorem~\cite{Korchemsky:1994jb,Bauer:2001yt}
allows us to systematically resum the large logarithmic
distributions to all orders in perturbation theory.
Here we make use of the SCET-based factorization theorem following
\refcite{Ligeti:2008ac}
\begin{align}\label{eq:w77s}
W_{77}^{\rm s}(k)
&= \Hard(\mb,\mu_H)\, \HardEvolution(\mb,\mu_H,\mu_J)\times
\nonumber\\&\quad\times
\int\!\df\omega\,\df\omega'\,\mb\,
  \Jet(\mb(k-\omega),\mu_J)\,
  \SoftEvolution(\omega-\omega',\mu_S,\mu_J)\,
  \PartonicSoft(\omega',\mu_S)
\,,\end{align}
where $\Hard$, $\Jet$, and $\PartonicSoft$ are the hard, jet, and partonic soft functions, respectively.
The hard and soft evolution kernels, $\HardEvolution$ and $\SoftEvolution$, evolve the hard and soft functions from their
characteristic hard and soft scales, $\mu_H$ and $\mu_S$, to the jet scale,
$\mu_J$, thereby summing logarithms
of the form $\ln(\mu_H/\mu_J)$ and $\ln(\mu_J/\mu_S)$.
Since we choose to evolve everything to the jet scale, the jet evolution
kernel $\JetEvolution(p^2,\mu_J,\mu_J) = \delta(p^2)$ drops out.
The hats indicate that an object is defined in a renormalon-free
short-distance scheme, as discussed in more detail in \sec{short-distance-schemes}.
Explicit results for the perturbative ingredients are collected in \app{resummation-ingredients}.

In \eq{w77s},
$\Hard(\mu_H)$, $\Jet(\mu_J)$, $\PartonicSoft(\mu_S)$ are the boundary conditions
for the evolution, which are evaluated at fixed order. When doing so, by
default we always strictly reexpand their product to the given order in $\as$,
i.e., we count $\as(\mu_H)\sim\as(\mu_J)\sim\as(\mu_S)$
and drop all higher-order cross terms in the product of their fixed-order
series. By doing so, the strict fixed-order expansion of $W_{77}^\sing$ in terms of a
common $\alpha_s(\mu)$ is reproduced simply by taking all scales to be equal $\mu_H = \mu_J = \mu_S = \mu$.
This is different to \refscite{Ligeti:2008ac,Bernlochner:2020jlt}, where only
the product $\Jet\otimes\PartonicSoft$ is reexpanded, while the hard function is
kept unexpanded as an overall multiplicative factor, which results in keeping
certain higher-order cross terms in the fixed-order limit.
The effect of these differences is studied in \sec{discussion-perturbative-treatment}.

To resum the singular corrections using \eq{w77s} to \NNNLLp order, we have to include the fixed-order boundary conditions $\Hard(\mu_H)$, $\Jet(\mu_J)$, $\PartonicSoft(\mu_S)$ to $\mathcal{O}(\as^3)$ and use the \ThreeLoop noncusp and $4$-loop cusp anomalous dimensions as well as the $4$-loop beta function in the evolution factors $\HardEvolution(\mu_H, \mu_J)$ and $\SoftEvolution(\mu_J, \mu_S)$.
The jet and soft functions have been computed up to three loops in \refscite{Bauer:2003pi, Becher:2006qw, Bruser:2018rad} and \refscite{Bauer:2003pi, Becher:2005pd, Bruser:2019yjk}, respectively.

Regarding the hard function, its \ThreeLoop anomalous dimension is also known via the consistency relation with the jet and soft anomalous dimensions and is given in \refcite{Bruser:2019yjk}.
The full hard function is currently only known up to NNLO~\cite{Bauer:2000yr, Ligeti:2008ac}.
We account for all the logarithmic terms at N$^3$LO, which are determined using the RGE in terms of the known anomalous dimensions and lower-order constant terms. The result
is given in \app{hard}.
Thus, the only missing ingredient to obtain $W_{77}^{\rm s}$ at full $\rm N^3LL'$ is the finite, nonlogarithmic \ThreeLoop constant of the hard function, $h_3$, which is defined by expanding the (pole-scheme) hard function at $\mu = m_b$ as
\begin{equation}
h_s(m_b, \mu_H = m_b)
= 1 + \frac{\as(m_b)}{\pi}\, h_1
   + \frac{\as^2(m_b)}{\pi^2}\, h_2
   + \frac{\as^3(m_b)}{\pi^3}\, h_3
   + \ord{\as^4}
\,.\end{equation}
We treat this unknown constant as a theory nuisance parameter,
\begin{equation}
h_3 = 0 \pm 80
\,,\end{equation}
where the range of variation is estimated using the Pad\'e approximation
\begin{equation}
h_3\sim\frac{h_2^2}{\abs{h_1}}\sim\frac{19.3^2}{4.55}\sim 80
\,.\end{equation}
We will see in \sec{results} that it only has a minor impact on the perturbative
precision of our results.

For future reference, we write the fixed-order expansion of the singular
contribution up to ${\rm N^3LO}$ as
\begin{align}\label{eq:singular-contribution}
W_{77}^{\rm s}(\mb x)
  &=\frac{C_F}{\mb}
  \biggl\{
    w^{\sing(0)}_{77}(x)
    +\frac{\as(\mu)}{\pi}\bigl[%
      w^{\sing(1)}_{77}(x)
      +\Delta w^{\sing(1)}_{77}(\mu, x)
    \bigr]
\nn\\ & \quad\qquad
    +\frac{\as^2(\mu)}{\pi^2}\biggl[%
      w^{\sing(2)}_{77}(x)
      +\frac{\beta_0}{2}w^{\sing(1)}_{77}(x)\ln\frac{\mu}{\mb}
      +\Delta w^{\sing(2)}_{77}(\mu, x)
    \biggr]
\nn\\ & \quad\qquad
    +\frac{\as^3(\mu)}{\pi^3}\biggl[%
      w^{\sing(3)}_{77}(x)
      +\biggl(%
        \beta_0w^{\sing(2)}_{77}(x)
        +\frac{\beta_1}{8}w^{\sing(1)}_{77}(x)
      \biggr)\ln\frac{\mu}{\mb}
\nn\\ & \qquad\qquad\qquad\quad
      +\frac{\beta_0^2}{4}w^{\sing(1)}_{77}(x)\ln^2\frac{\mu}{\mb}
      +\Delta w^{\sing(3)}_{77}(\mu, x)
    \biggr] + \ord{\as^4}
  \biggr\}
\,,\end{align}
where we have made the fixed-order $\mu$ dependence and its order-by-order
cancellation explicit.
Numerically, we have
\begin{align}
w^{\sing(0)}_{77}(x)&=0.75\,\delta(x)
\,,\nn\\
w^{\sing(1)}_{77}(x)&=
  -4.54\,\delta(x)
  -1.75\,\mathcal{L}_{0}(x)
  -1.00\,\mathcal{L}_{1}(x)
\,,\nn\\
w^{\sing(2)}_{77}(x)&=
  (-30.5+3.01\,n_f)\,\delta(x)
  +(5.94+0.316\,n_f)\,\mathcal{L}_{0}(x)
  +(12.4-0.181\,n_f)\,\mathcal{L}_{1}(x)
  \nn\\&\quad
  +(7.63-0.250\,n_f)\,\mathcal{L}_{2}(x)
  +0.667\,\mathcal{L}_{3}(x)
\,,\nn\\
w^{\sing(3)}_{77}(x)&=
  (88.2+0.75h_3-0.0269\,n_f-0.0309\,n_f^2)\,\delta(x)
  \\&\quad
  +(138-9.16\,n_f-0.00330\,n_f^2)\,\mathcal{L}_{0}(x)
  \nn\\&\quad
  +(68.7-8.75\,n_f+0.121\,n_f^2)\,\mathcal{L}_{1}(x)
  +(-25.1-0.815\,n_f+0.00694\,n_f^2)\,\mathcal{L}_{2}(x)
  \nn\\&\quad
  +(-43.0+3.16\,n_f-0.0648\,n_f^2) \mathcal{L}_{3}(x)
  +(-6.53+0.278\,n_f) \mathcal{L}_{4}(x)
  -0.222\, \mathcal{L}_{5}(x)
  \nn
\,,\end{align}
where $\mathcal{L}_n(x) \equiv [\ln^n(x)/x]_+$ are the usual plus distributions
defined in \eq{plus-distribution}.
Note that at this order the unknown \ThreeLoop constant $h_3$ appears only in the $\delta(x)$ coefficient, so $w^{\sing(3)}_{77}(x)$ is completely known for $x>0$.

The correction terms $\Delta w_{77}^{\sing(n)}(\mu, x)$ in
\eq{singular-contribution} arise from switching to the
short-distance $b$-quark mass $\mb$, and their $\mu$ dependence separately
cancels among them order by order. They are discussed in more detail in
\sec{subleading-delta-mb-corrections},
and their explicit expressions are provided in \eq{singular-delta-mb-corrections}.

\subsection{Nonsingular contributions}
\label{sec:nonsingular}

The nonsingular contribution $W_{77}^\nons(k)$ is included at fixed order. It is obtained by subtracting the fixed-order singular terms from the full fixed-order result for $\df\Gamma/\df E_\gamma$, which is known up to $\ord{\alpha_s^2}$~\cite{Blokland:2005uk,Melnikov:2005bx,Asatrian:2006sm}.
We write its perturbative expansion up to $\rm N^3LO$ as
\begin{align}\label{eq:nonsingular-contribution}
  W_{77}^{\rm ns}(\mb x)&=\frac{C_F}{\mb(1-x)^3}
  \biggl\{ \frac{\as(\mu_{\rm ns})}{\pi}\, w^{{\rm ns}(1)}_{77}(x)
\nn\\ &\quad
    +\frac{\as^2(\mu_{\rm ns})}{\pi^2}\biggl[%
      w^{{\rm ns}(2)}_{77}(x)
      +\frac{\beta_0}{2}\,w^{{\rm ns}(1)}_{77}(x)\,\ln\frac{\mu_{\rm ns}}{\mb}
      +\Delta w^{{\rm ns}(2)}_{77}(\mu_{\rm ns}, x)
    \biggr]
\nn\\ &\quad
    +\frac{\as^3(\mu_{\rm ns})}{\pi^3}\biggl[%
      w^{{\rm ns}(3)}_{77}(x)
      +\biggl(%
        \beta_0w^{{\rm ns}(2)}_{77}(x)
        +\frac{\beta_1}{8}w^{{\rm ns}(1)}_{77}(x)
      \biggr)\ln\frac{\mu_{\rm ns}}{\mb}
\nn\\ & \qquad\qquad\qquad
      +\frac{\beta_0^2}{4}\,w^{{\rm ns}(1)}_{77}(x)\ln^2\frac{\mu_{\rm ns}}{\mb}
      +\Delta w^{{\rm ns}(3)}_{77}(\mu_{\rm ns}, x)
    \biggr]
  \biggr\}
\,.\end{align}
The overall $1/(1-x)^3$ factor is included by convention.
Explicit expressions for $w^{{\rm ns}(1)}_{77}(x)$ and $w^{{\rm ns}(2)}_{77}(x)$ are given in eq.~(S21) in \refcite{Bernlochner:2020jlt}.
The $\ord{\as^3}$ function $w^{{\rm ns}(3)}_{77}(x)$ is currently unknown.
The $\Delta w^{\nons(n)}_{77}(\mu_\nons, x)$ terms arise from switching to the short-distance mass $\mb$
and are given in \eq{nonsingular-delta-mb-corrections}.

In the peak region for $k\ll \mb$, the nonsingular are power-suppressed by $k/\mb$ relative to the singular. Hence, the two can be considered as independent perturbative series, which are treated separately from each other. In particular, it is consistent to include the nonsingular only at fixed order, while the singular are being resummed. By contrast, for $k\sim\mb$, the separation into singular and nonsingular becomes ill-defined and only the full result given by their sum, $W_{77}^\full = W_{77}^\sing + W_{77}^\nons$, is meaningful. This is reflected by the fact that there are typically large numerical cancellations between the singular and nonsingular contributions for $k\to \mb$,
as we will see explicitly in \sec{profile}. Consequently,
$W_{77}^\sing$ and $W_{77}^\nons$ must be included using the same perturbative expansion in this limit, i.e., at the same scale and the same perturbative order, to ensure that the cancellations between them can take place and the proper full result is recovered.
This has important ramifications. First, since the full and nonsingular results are only known at fixed order, it is essential to turn off the resummation for $W_{77}^\sing$ for $k\sim\mb$ such that it also reduces to its fixed-order result. Second, the N$^n$LL$'$ resummation reduces to the fixed $\ord{\alpha_s^n}$ singular result, so consistently matching it to fixed order requires including the nonsingular to $\ord{\alpha_s^n}$.

Therefore, at N$^3$LL$'$ we need $W_{77}^\nons$ to $\ord{\as^3}$, which means we have to parametrize the unknown nonsingular function $w^{{\rm ns}(3)}_{77}(x)$. We do so by considering its required asymptotic behavior in the $x\to 0$ and $x\to 1$ limits. In particular, for $x\to 1$ we have to account for the singular-nonsingular cancellations at $\ord{\as^3}$, which basically implies that $w_{77}^{\nons(3)}$ and $w_{77}^{\sing(3)}$ are not independent.
In addition, we want to exploit the parameterization to estimate the perturbative uncertainty due to the missing $w^{\nons(3)}_{77}(x)$.

We begin by separating $w^{\nons(3)}_{77}(x)$ into a ``correlated'' and an ``uncorrelated'' piece,
\begin{equation}\label{eq:nons-three-loop}
  w^{{\rm ns}(3)}_{77}(x)= w^{{\rm ns}(3)}_{\rm cor}(x)+w^{{\rm ns}(3)}_{\rm uncor}(x)\,.
\end{equation}
The ``correlated'' term $w^{{\rm ns}(3)}_{\rm cor}(x)$ is designed to completely cancel the singular corrections in the $x\to 1$ limit without disturbing the hierarchy between singular and nonsingular contributions in the $x\to 0$ limit.
We define it as
\begin{equation}
  w^{{\rm ns}(3)}_{\rm cor}(x) = -(1-x)^3\,w^{{\rm s}(3)}_{77}(1)\,,
\end{equation}
where the \ThreeLoop singular function $w^{{\rm s}(3)}_{77}(x)$ is defined in \eq{singular-contribution}.
The overall factor $(1-x)^3$ simply cancels the overall $1/(1-x)^3$ in \eq{nonsingular-contribution}.
The remaining ``uncorrelated'' piece $w^{{\rm ns}(3)}_{\rm uncor}(x)$ can now be considered independent of the singular contribution, so we can parametrize it.
We do so by expanding it as
\begin{equation}\label{eq:ns-uncor-model}
  w^{{\rm ns}(3)}_{\rm uncor}(x)=(1-x)^3\,\sum_{k=0}^5 \,c_k^{\rm ns}\,L^k(x)
\qquad \text{with} \qquad
  L(x)=\frac{1}{4}\frac{w^{{\rm ns}(1)}_{77}(x)}{(1-x)^3}-\frac{9}{16}\,,
\end{equation}
where the function $L(x)$ is positive for $0<x<1$ and has the following asymptotics in the $x\to 0$ and $x\to 1$ limits,
\begin{align}\label{eq:ns-uncor-model-limits}
  L(x) & = -\frac{3}{2}-\ln x +\mathcal{O}(x)\,,
  &
  L(x) & = 0 + \mathcal{O}(1-x)\,.
\end{align}
We construct $L(x)$ using $w^{{\rm ns}(1)}_{77}(x)$
with the expectation that its powers provide a reasonable guess of the possible shape of the higher-order function $w^{{\rm ns}(3)}_{77}(x)$ in the intermediate region $0<x<1$.
Furthermore, \eq{ns-uncor-model} incorporates the knowledge that at $\ord{\as^3}$ the nonsingular in the limit $x\to 0$ is a degree-5 polynomial in $\ln x$.
Therefore we include up to five powers of $L(x)$ to ensure we are able to probe the complete logarithmic structure in the small-$x$ limit.

The parameters $c_{0\dots5}^{\rm ns}$ can be treated as theory nuisance
parameters with zero central values and the following variation magnitudes:
\begin{align}
  c^{\rm ns}_k = 0 \pm \delta c^{\rm ns}_k \, \qquad {\rm with} \qquad \delta c^{\rm ns}_{0\dots5}=(20, 100, 80, 10, 5, 1)
  \,.
\end{align}
To determine the range of variations for $c^{\rm ns}_{1\dots 5}$, we use the observation that in the limit $x\to 0$ the expression $4x \,w^{{\rm s}(2)}_{77}(x)$ provides a good estimate of the size of logarithmic terms in $w^{{\rm ns}(2)}_{77}(x)$ at one and two loops,
\begin{align} \label{eq:nonsnumericalestimate}
4x\,w^{{\rm s}(1)}_{77}(x)&=-7.00-4.00\ln x
\,,\nn\\
w^{{\rm ns}(1)}_{77}(x)&=\hphantom{-}3.75-4.00\ln x+\mathcal{O}(x)
\,,\nn\\
4x \,w^{{\rm s}(2)}_{77}(x)
&= 28.8+46.7\ln x+26.5\ln^2x+2.67\ln^3x
\,,\nn\\
w^{{\rm ns}(2)}_{77}(x)
&= 16.1+33.9\ln x+25.0\ln^2x+2.67\ln^3x +\mathcal{O}(x)
\,,\nn\\
4x\,w^{{\rm s}(3)}_{77}(x)
&= 406+142\ln x-113\ln^2x\hspace{1.3mm}-125\ln^3x\hspace{2mm}-21.7\ln^4x\hspace{1.3mm}-0.889\ln^5x
\,,\nn\\
&= 259+\hspace{1mm}95L(x)+189L^2(x)+15.5L^3(x)-15.0L^4(x)+0.889L^5(x)
\,.\end{align}
In particular, the highest power of $\ln x$ in $w^{{\rm ns}(1,2)}_{77}(x)$ is precisely determined by $4 x \,w^{{\rm s}(1,2)}_{77}$. The reason is that, similar to the leading-power case, it is expected that the universal cusp and a set of subleading noncusp anomalous dimensions govern the logarithmic structure at subleading power.
Thus, we similarly exploit $4x \,w^{{\rm s}(3)}_{77}(x)$ for estimating the typical size of $\delta c^{\rm ns}_{1\dots5}$.

To assess the uncertainty due to the missing $w_{77}^{\nons(3)}$ we separately vary each nuisance parameter in the ranges given above.
The different nuisance parameters are considered as independent such that the resulting individual uncertainties from varying them are added in quadrature. This is discussed in more detail in \sec{uncertainty}.
Of course, this means that the uncertainty necessarily increases by adding more parameters. Therefore, to obtain a realistic uncertainty estimate and avoid becoming overly conservative we also put the constraint that the total uncertainty estimate for the missing \ThreeLoop correction does not exceed the size of the \TwoLoop corrections. To satisfy this constraint, the values for $\delta c^{\rm ns}_{2,3,4}$ are chosen somewhat smaller than the corresponding coefficients of $4x \,w^{{\rm s}(3)}_{77}(x)$ in \eq{nonsnumericalestimate}.

It is easy to see from \eq{ns-uncor-model-limits} that all powers of $L(x)$ approach zero in the far tail of the spectrum.
Thus, in this limit the constant term $c_0$ dominates.
Therefore, we estimate the size of $\delta c_0^{\rm ns}$ using the Pad\'e approximation from the corresponding lower-order corrections,
\begin{align}
  w^{(k)}_{x\to 1}&=\lim_{x\to 1}\left[\frac{w^{{\rm ns}(k)}_{77}(x)}{(1-x)^3}+w^{{\rm s}(k)}_{77}(x)\right]\,,
  &
  w^{(3)}_{x\to 1}
  &=
  c_0^{\rm ns}
  \sim
  \frac{(w^{(2)}_{x\to 1})^2}{w^{(1)}_{x\to 1}}
  \approx
  \frac{2.92^2}{0.5}
  \sim 20\,.
\end{align}

Finally, we like to stress that the purpose of the parameterization given by
\eqs{nons-three-loop}{ns-uncor-model} is \emph{not} to construct an
approximation of the unknown function $w_{77}^{\nons(3)}(x)$.
Rather, the goal is first to enable a
consistent matching at N$^3$LL$'$, which is essentially achieved by the
separation in \eq{nons-three-loop}, and second to obtain a reliable estimate of
the perturbative uncertainty due to its unknown form. For this purpose, we only
need to estimate the \emph{typical size} of the theory nuisance parameters, say within a factor of a few, for which the above considerations are sufficient.
However, since \eq{nonsingular-contribution} includes all known $\ord{\as^3}$
contributions that are predictable from lower orders, and since the
parameterization of the remaining unknown $w_{77}^{\nons(3)}(x)$ does include
nontrivial information on its structure, we do expect some improvement in the
perturbative precision of our predictions beyond $\ord{\as^2}$, which will be
reflected in the size of the resulting uncertainty, as we will see in
\sec{results}.

\subsection{Profile scales}
\label{sec:profile}

The resummation of the singular contributions is determined by the choices
of the hard $\mu_H$, jet $\mu_J$, and soft $\mu_S$ scales in the factorization theorem in \eq{w77s}.
To achieve the proper resummation they must be chosen according to the
kinematics relevant in the different regions of the spectrum.
In addition, the nonsingular scale $\mu_{\rm ns}$ determines the
scale at which the fixed-order nonsingular terms in \eq{nonsingular-contribution} are evaluated.

For our scale choices we follow \refcite{Bernlochner:2020jlt}.
The $E_\gamma$ spectrum has three parametrically distinct kinematic regions:
\begin{itemize}
\item{\bf Shape function (nonperturbative) region:}
$\Lambda_{\rm QCD}\sim (m_B -2\, E_\gamma)\ll\mb$
\\
This corresponds to the peak of the spectrum where the full shape of the
shape function is relevant and the soft scale is fixed to the lowest
still-perturbative scale $\mu_S = \mu_0 \gtrsim \Lambda_{\rm QCD}$.

\item{\bf Shape function OPE region:}
$\Lambda_{\rm QCD}\ll(m_B-2E_\gamma)\ll\mb$
\\
This corresponds to the region left of the peak, i.e., the transition
region between the peak and the far tail.
Here the soft scale has canonical scaling $\mu_S \sim m_B-2E_\gamma$.

\item{\bf Local OPE (fixed-order) region:} $\Lambda_{\rm QCD}\ll(m_B-2E_\gamma)\sim\mb$
\\
This corresponds to the far tail of the spectrum, which is described by
fixed-order perturbation theory. Here, the distinction between singular and
nonsingular becomes meaningless and the resummation must be turned off to
ensure that singular and nonsingular contributions properly recombine into the
correct fixed-order result. This requires that all scales become equal
$\mu_S = \mu_J = \mu_H = \mu_\nons \equiv \mu_{\rm FO} \sim \mb$.
\end{itemize}
The canonical value for the hard scale in all regions is $\mu_H\sim\mb$. In the
first two regions the SCET resummation is applicable with the canonical scaling
$\mu_J \sim \sqrt{\mu_S\mu_H}$.

To account for these different scale hierarchies we use the common approach of
profile scales~\cite{Abbate:2010xh, Ligeti:2008ac}, where $\mu_S(E_\gamma)$, $\mu_J(E_\gamma)$, $\mu_\nons(E_\gamma)$ are taken as functions of
$E_\gamma$. The key advantage of using profile scales to connect the
descriptions in the different parametric regions is that they provide a smooth
transition between the regions that is solely implemented in terms of scale
choices, such that the ambiguity in the precise choices of the transition are
equivalent to scale ambiguities, which by construction are formally beyond the
order one is working and reduce as we go to higher order.

For our numerical analysis we employ the profile scales used in \refcite{Bernlochner:2020jlt}, given by
\begin{align}
\label{eq:profile-functions}
\mu_H&=e_H\mb
\,,\nn\\
\mu_S(E_\gamma)&=\mu_0+(\mu_H-\mu_0)\,f_{\theta}\biggl(\frac{E_1-E_{\gamma}}{E_1-E_2}\biggr)
\,,\nn\\
\mu_J(E_\gamma)&=\bigl[\mu_S(E_\gamma)\bigr]^{(1-e_J)/2}\,\mu_H^{(1+e_J)/2}
\,,\nn\\
\mu_{\rm ns}(E_\gamma)&=\bigl[\mu_S(E_\gamma)\bigr]^{(1-e_{\rm ns})/4}\mu_H^{(3+e_{\rm ns})/4}
\,,\end{align}
where the function $f_{\theta}(x)$ provides a smooth transition from
${f_{\theta}(x\leq0)=0}$ to ${f_{\theta}(x\geq1)=1}$,
\begin{equation}
f_{\theta}(x) =
\begin{cases}
    0 &\qquad x \leq0\,,\\
    2\,x^2 &\qquad 0<x\leq1/2\,,\\
    1-2\,(1-x)^2 &\quad 1/2<x\leq1\,,\\
    1 &\qquad 1<x
\,.\end{cases}
\end{equation}

\begin{figure}
\label{fig:singular-nonsingular}
\centering
\includegraphics[width=\WidthOneSubfig]{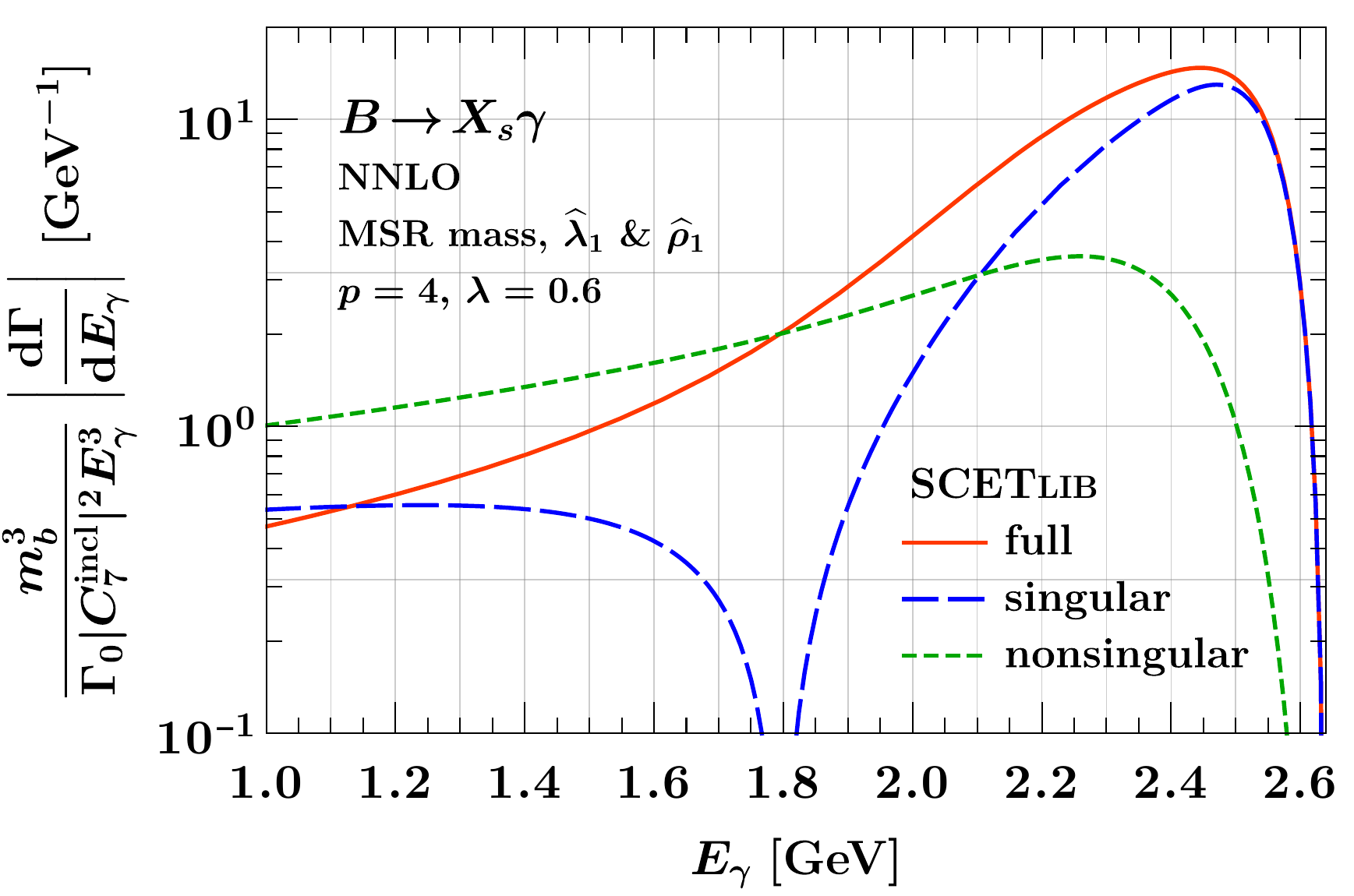}%
\caption{%
Comparison of the absolute values of the singular, nonsingular, and full contributions to the \btosgamma photon energy spectrum at fixed
NNLO in the MSR scheme.
}
\end{figure}

\begin{figure}\label{fig:profile-functions}
\centering
\includegraphics[width=\WidthOneSubfig]{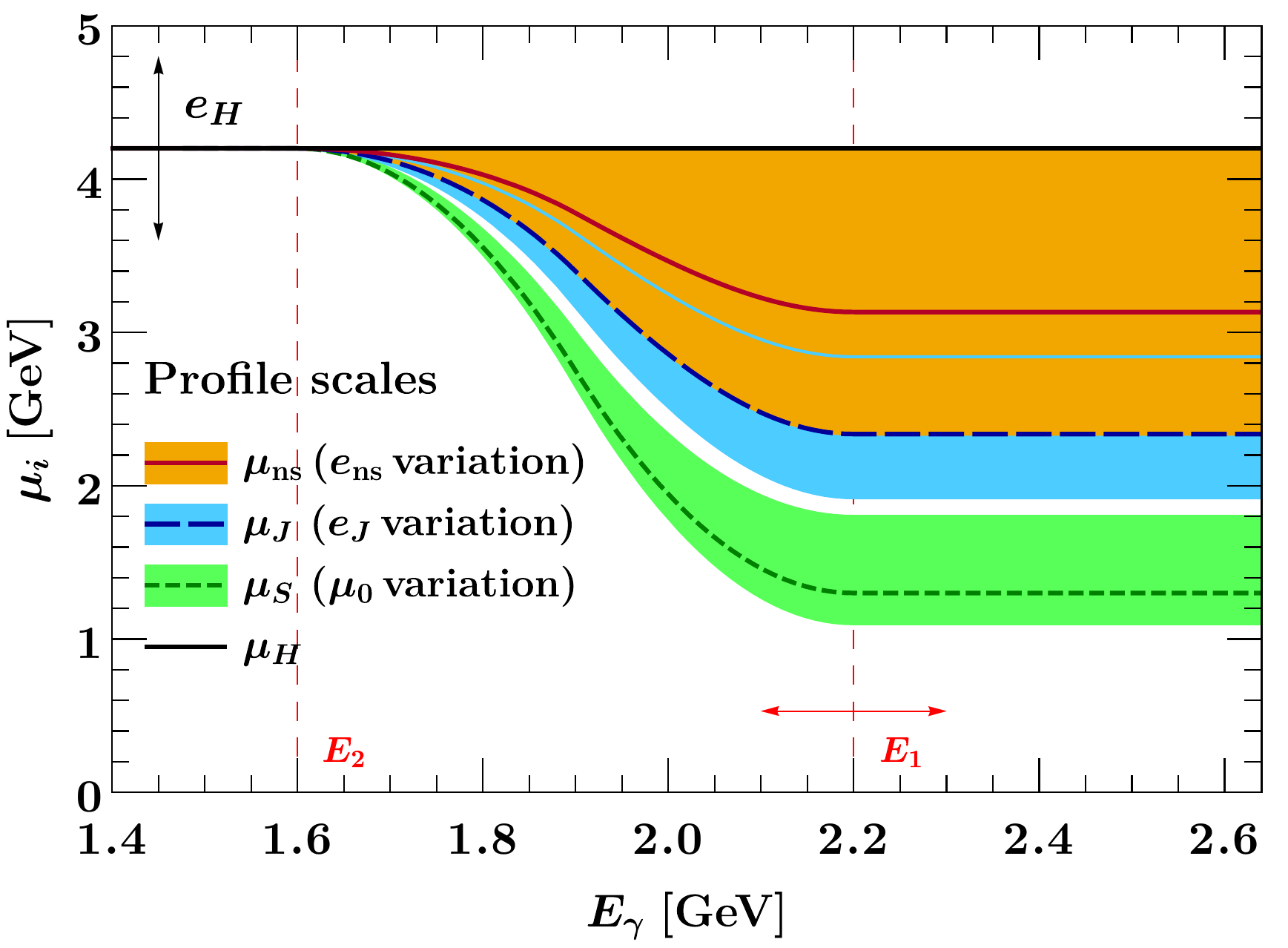}
\caption{%
Profile scales used for the hard, jet, soft, and nonsingular scales.
The bands show the individual ranges for the jet, soft, and nonsingular
scales. The black and red arrows indicate the variations for the parameter
$e_H$ and the transition point $E_1$.
}
\end{figure}

\Fig{singular-nonsingular} shows the absolute values of the singular and
nonsingular contributions as well as the full result at NNLO (i.e.\
without resummation), which is used to pick the transition points $E_1$ and
$E_2$ between the different parametric regions. For $E_\gamma\gtrsim E_1 =
2.2\GeV$, the singular contributions clearly dominate, which also corresponds to
the nonperturbative region. For $E_\gamma\lesssim E_2 = 1.6\GeV$, there are
large cancellations between the singular and nonsingular contributions. This
corresponds to the fixed-order region, where the separation into singular and
nonsingular is ill-defined and only their sum is meaningful. Therefore, the
resummation of the singular must be turned off to ensure that this cancellation
is not spoiled by it and the correct full result is recovered. Since there is
little space between $E_1$ and $E_2$, the transition region in between them
effectively coincides with the intermediate shape function OPE region.

To summarize we use the following values for the profile scale parameters~\cite{Bernlochner:2020jlt}:
\begin{align}\label{eq:scale-variation-ranges}
e_H &= \{1, 1/2, 2\}
\,,\quad &
\mu_0 &= \max(1,e_H)\times\{1.3, 1.1, 1.8\}\GeV
\,,\nn\\
e_J &= \{0,-1/3,+1/3\}
\,,\quad &
e_{\rm ns} &= \{0,-1,+1\}
\,,\nn\\
E_1&= \{2.2, 2.1, 2.3\}\GeV
\,,\quad &
E_2 &= 1.6\GeV
\,.\end{align}
For each parameter, the first value in the set is the central value and the next
two are the variations that we will use to assess perturbative uncertainties in
\sec{uncertainty}. The central values for the profile scales along with their
individual variation ranges are illustrated in \fig{profile-functions}.
Since the nonsingular contributions are treated
at fixed order, a priori there is no canonical scaling to guide the choice of
the nonsingular scale $\mu_\nons$ beyond the fixed-order region $E_\gamma \geq E_2$. In
practice, it is picked as the geometric mean of the hard and central jet
scales to account for the fact that the nonsingular terms have some sensitivity
to scales below $\mb$, and this choice is varied up to the hard and down to the
central jet scales as shown in \fig{profile-functions}.

Note that the individual scales are not independent of each other but are
parametrized in such a way that their relative hierarchies are preserved upon
varying the profile parameters. For example, they all depend on $\mu_H$, such
that varying $\mu_H$ up and down (by varying $e_H$) simultaneously moves the other scales up and down accordingly. In particular all scales always merge into
a common value for $E_\gamma \leq E_2$ to properly turn off the resummation. Similarly, $\mu_J$ and $\mu_\nons$ depend on $\mu_S$, such that varying $\mu_0$ not only varies $\mu_S$ but also moves $\mu_J$ and $\mu_\nons$ up and down
accordingly to preserve the hierarchy between them. The variations for $\mu_J$
and $\mu_\nons$ parametrized by $e_J$ and $e_\nons$ correspond to small deviations from the default hierarchy. This also means that the $\mu_0$, $e_J$, and $e_\nons$ variations smoothly turn off between $E_1$ and $E_2$ like the resummation itself, such that below $E_2$ only the overall $\mu_H$ variation remains corresponding to the usual fixed-order scale variation.

\subsection{Shape function}
\label{sec:shape-function}

The leading-power factorization theorem for \btosgamma in \eq{w77s} separates
all soft dynamics into the soft function
\begin{equation}\label{eq:hadronic-soft}
  \HadronicSoft(\omega,\mu)\defeq\langle B|\bar{b}_v\delta(iD_{+} - \delta + \omega)b_v|B\rangle
  \,,
\end{equation}
where $b_v$ is the HQET $b$-quark field, $\ket{B}$ is the full QCD $B$-meson state, and $\delta = m_B - \mbPole$. This definition of $S(\omega, \mu)$ is such that it has support for $\omega \geq 0$~\cite{Ligeti:2008ac}.
The soft function contains both perturbative soft radiation as well as the nonperturbative Fermi motion of the $b$ quark inside the $B$ meson.
Following \refcite{Ligeti:2008ac}, we further factorize it as
\begin{equation}\label{eq:soft-factorization}
\HadronicSoft(\omega,\mu)
= \int\!\df k\, \PartonicSoft(\omega-k,\mu)\,\ShapeFunction(k)
\,,\end{equation}
where the partonic soft function $\PartonicSoft$ can be calculated in
perturbation theory, while the shape function $\ShapeFunction(k)$ is a
nonperturbative object. It has support for $k\geq 0$ and peaks around
$k\sim\Lambda_{\rm QCD}$.

In the tail region where $\omega \gg \Lambda_{\rm QCD}$, the right-hand side of
\eq{soft-factorization} can be expanded in powers of $\Lambda_{\rm QCD}/\omega$,
\begin{equation}\label{eq:hadronic-soft-expansion}
  \HadronicSoft(\omega,\mu)=\sum_{n=0}^{\infty}\frac{(-1)^n}{n!}\frac{\df^n\PartonicSoft(\omega,\mu)}{\df\omega^n}M_n\,,
\end{equation}
where $M_n \sim \Lambda_{\rm QCD}^n$ are the moments
\begin{equation}
M_n\defeq\int\! \df k\, k^n\ShapeFunction(k)
\,.\end{equation}
Hence, in the tail region the leading nonperturbative corrections are encoded in the
first few nonperturbative moments of $\ShapeFunction(k)$, and its moment expansion recovers the local-OPE description of the spectrum.
On the other hand, in the peak region where $\omega\sim\Lambda_{\rm QCD}$, the
full function $\ShapeFunction(k)$ is needed.

As discussed in \refcite{Ligeti:2008ac}, an important feature of the
factorization in \eq{soft-factorization} is that it provides a common
description of the nonperturbative effects across these different kinematic
regions, incorporating all available perturbative information in the limit
$\omega \gg \Lambda_{\rm QCD}$ without having to explicitly carry out an
expansion in $\Lambda_{\rm QCD}/\omega$, whose precise region of validity would
be unclear. That is, all perturbative corrections to moments of $S(\omega, \mu)$
are encoded in the perturbative function $\PartonicSoft(\omega)$, while the
shape function $\ShapeFunction(k)$ is a genuinely nonperturbative function
that is formally independent of the perturbative order and can be extracted from experimental data.

The first few moments of $\ShapeFunction(k)$ are given in terms of $B$-meson matrix elements of local HQET operators,
\begin{align}\label{eq:shape-function-moments}
M_0 &= 1
\,,\nn \\
M_1 &= m_B - \mb + \dotsb
\,,\nn \\
M_2 &= (m_B - \mb)^2 -\frac{\lambdaOne}{3} + \dotsb
\nn \,,\\
M_3& = (m_B - \mb)^3 - \lambdaOne (m_B - \mb) +\frac{\rhoOne}{3}+ \dotsb
\,,\end{align}
The hadronic parameters $\lambdaOne$ and $\rhoOne$ are matrix elements of local HQET operators defined in a short-distance scheme as discussed in \sec{lambda1-rho1-short-distance}.
The ellipses denote terms that suppressed by relative powers of $\Lambda_{\rm QCD}/\mb$ arising from subleading shape functions that are typically absorbed into $\ShapeFunction(k)$~\cite{Bernlochner:2020jlt}, but which are not relevant for our purposes here.

A general method for parametrizing $\ShapeFunction(k)$ via a systematic
expansion around a given base model has been developed in
\refcite{Ligeti:2008ac}, which was used in \refcite{Bernlochner:2020jlt} to fit
$\ShapeFunction(k)$ from data. Since our primary interest in this paper are the
perturbative corrections, the precise form of $\ShapeFunction(k)$ is not
relevant here. We only need a reasonably realistic model for it in order to
illustrate our results numerically. For this purpose, we take the exponential
base model used in \refscite{Ligeti:2008ac, Bernlochner:2020jlt}, given by
\begin{equation}\label{eq:basis}
\ShapeFunction(k) =\frac{1}{\lambda}Y\Bigl(\frac{k}{\lambda}\Bigr)
\qquad\text{with}\qquad
Y(x) = \frac{(p+1)^{p+1}x^p}{p!}e^{-(p+1)x}
\,.\end{equation}
Its normalization and first moment are
\begin{equation}
M_0=\int\limits_0^\infty\!\df k\, \ShapeFunction(k) = 1
\,,\qquad
M_1=\int\limits_0^\infty\!\df k\, k\, \ShapeFunction(k) = \lambda
\,.\end{equation}
By picking $\lambda \approx m_B - \mb \approx 0.6\GeV$ this base model already
provides a good fit of the experimental measurements~\cite{Bernlochner:2020jlt}.

Note that evaluating the N$^n$LO soft function in a short-distance
scheme involves taking $n$ derivatives of $\ShapeFunction(k)$. Therefore, at
\NNNLLp, which needs the N$^3$LO soft function, and assuming integer $p$, we
require $p \geq 4$ to ensure that the soft function vanishes for $\omega\to 0$, which in turn is required for the $E_\gamma$ spectrum to vanish at the kinematic
endpoint $E_\gamma \to m_B/2$.

\section{Short-distance schemes}
\label{sec:short-distance-schemes}

A major challenge in $B$ physics is to parametrize nonperturbative effects in
such a way that their extraction from experimental measurements is stable with
(and ideally independent of) the perturbative order. Achieving this stability is
not trivial due to infrared sensitivity of the involved perturbative series and
the resulting ambiguity in the asymptotic series of perturbative QCD, which is known
as the renormalon problem. The renormalon problem manifests itself in practice
as poor or no convergence of the perturbative series even at low orders. Since
physical (measurable) quantities are independent of the perturbative order, the
large perturbative corrections at each order are compensated by corresponding
large changes in some associated nonperturbative parameter. In other words, the
renormalon ambiguity in the perturbative series is compensated order by order by
an equal and opposite renormalon ambiguity in the nonperturbative parameter.

Conceptually, to resolve this issue, the renormalon must be identified and
subtracted from both the perturbative quantity ($C$) and the associated
parameter ($p$), such that both become renormalon free and perturbatively
stable. To give a simple toy example,
\begin{equation}
C - p = (C - \delta p) - (p - \delta p) \equiv \widehat C - \widehat p
\,.\end{equation}
On the left-hand side, the renormalon only cancels between $C$ and $p$. On the
right-hand side, the so-called residual term $\delta p$ is a perturbative series
in $\as$ that contains the renormalon. Its specific choice defines a specific
so-called short-distance scheme. The renormalon then cancels within each of the
parenthesis defining the short-distance $\widehat C$ and $\widehat p$, which are
now separately free of the renormalon.

\subsection{Soft function}
\label{sec:short-distance-soft-function}

In reality, the structure is of course more complicated than the above simple
toy example.
In our case, the leading-power perturbative series that suffers from
renormalon ambiguities is that of the partonic soft function $C_0(\omega, \mu)$,
whose renormalons are cancelled by the nonperturbative object
$\ShapeFunction(k)$. Its leading renormalon ambiguity of
$\ord{\Lambda_{\rm QCD}}$ is due to the pole mass definition of the $b$ quark,
$\mbPole$, which explicitly enters in the definition of the soft function
$\HadronicSoft(\omega, \mu)$ in \eq{hadronic-soft} and henceforth shows up in
all the moments of $\ShapeFunction(k)$. (At subleading power, also the jet and
hard functions involve the pole-mass renormalon, which we come back to in
\sec{subleading-delta-mb-corrections}.) Furthermore, the hadronic parameter
$\lambda_1$, which first appears in the second moment of $\ShapeFunction(k)$,
has a subleading $\ord{\Lambda_{\rm QCD}^2}$ renormalon
ambiguity~\cite{Martinelli:1995zw}. Similarly, we expect the hadronic parameter
$\rho_1$, which first appears in the third moment, to have an
$\mathcal{O}(\Lambda_{\rm QCD}^3)$ renormalon.

We write the parameters in a generic short-distance scheme as
\begin{equation}\label{eq:mass-lambda-rho-def}
\mb = \mbPole- \delta m_b
\,,\qquad
\lambdaOne =\lambda_1 - \delta\lambda_1
\,, \qquad
\rhoOne = \rho_1 - \delta\rho_1
\,.\end{equation}
The residual terms $\delta m_b$, $\delta\lambda_1$, and $\delta\rho_1$
are defined to cancel the renormalon in their respective parameter such that
the short-distance parameters on the left-hand side are renormalon free.
The original%
\footnote{These are often referred to as defined in the ``pole scheme'' borrowing the language from the pole mass.} HQET parameters $\lambda_1$ and $\rho_1$ are defined in dimensional regularization as~\cite{Gremm:1996df}
\begin{equation}
\lambda_1 = \bra{B}\bar{b}_v(i D)^2b_v\ket{B}
\,,\qquad
\rho_1 = \bra{B}\bar{b}_v(iD_\mu)(v\cdot iD)(iD^\mu)b_v\ket{B}
\,.\end{equation}

The construction of the short-distance partonic soft function $\PartonicSoft(\omega,\mu)$ with the appropriate renormalon subtractions is
derived in detail in~\refcite{Ligeti:2008ac}. Up to N$^3$LO, we have
\begin{align}\label{eq:soft-function-short-distance}
\PartonicSoft(\omega,\mu)
&=\biggl[
   1
   -\frac{\delta\lambda_1}{6}\frac{\df^2}{\df\omega^2}
   -\frac{\delta\rho_1}{18}\frac{\df^3}{\df\omega^3}
   +\cdots
\biggr]
e^{\delta m_b\frac{\df}{\df\omega}}\PartonicSoftPole(\omega,\mu)
\nn \\
&=\biggl[
    1+\delta m_b\frac\df{\df\omega}
    +\frac{1}{2}\biggl(\delta m_b^2-\frac{\delta\lambda_1}{3}\biggr)\frac{\df^2}{\df\omega^2}
    \nonumber\\&\quad
    +\frac{1}{6}\biggl(\delta m_b^3-\delta m_b\delta\lambda_1-\frac{\delta\rho_1}{3}\biggr)\frac{\df^3}{\df\omega^3}
    +\cdots
  \biggr]\PartonicSoftPole(\omega,\mu)
\,,\end{align}
where the original pole-scheme $\PartonicSoftPole(\omega, \mu)$
is defined and given in \app{soft}.
Importantly, for the renormalons to cancel on the right-hand side, it must always be fully expanded to a given fixed order in $\alpha_s$, including
the $\delta m_b$, $\delta\lambda_1$, $\delta\rho_1$ and their products with each other and with $C_0$.
With these subtractions both
$\PartonicSoft(\omega, \mu)$ and $\ShapeFunction(k)$ become renormalon free, up to
yet higher-order renormalons. In particular, the moments of $\ShapeFunction(k)$ are
then given by the short-distance parameters $\mb$, $\lambdaOne$, $\rhoOne$, as
shown in \eq{shape-function-moments}.
[Note that $\ShapeFunction(k)
\equiv\widehat\ShapeFunction(k)$ in \eq{soft-factorization} is already the
short-distance parameter.]
To evaluate the convolution integral $\PartonicSoft \otimes\ShapeFunction$,
we use integration by parts to move all the derivatives in
\eq{soft-function-short-distance} to act on
$\ShapeFunction$~\cite{Ligeti:2008ac}.
The renormalon subtractions significantly improve the perturbative convergence
of the soft function compared to the pole scheme, which we demonstrate
numerically in \sec{discussion-impact-short-distance}.

In general, the residual terms are scale dependent, leading to a similar scale
dependence of the short-distance parameter, which we suppress for simplicity in
our generic notation. The scale dependence can be explicit, as e.g.\ for the
\MSbar mass, in which case it is usually governed by an associated RGE. It can
also be only internal, as e.g.\ for the \OneS mass or the MSR mass, in which
case $\delta m_b$ (and also $\mb$) is formally scale independent (with only the
usual scale dependence from truncating the perturbative series that is cancelled
by higher orders). In either case, the residual terms must be expanded at the
same scale $\mu$, i.e., in terms of the same $\as(\mu)$, that is used for the
perturbative series whose renormalon is supposed to be subtracted to ensure that
the renormalon actually cancels. In our case this is the soft scale $\mu_S$ at
which we evaluate the fixed-order boundary condition $\PartonicSoft(\omega,
\mu_S)$ in the factorization theorem in \eq{w77s}.

In this context, the power counting of the HQET Lagrangian provides a powerful constraint on the
generic size of the residual mass term $\delta m_b$,
\begin{equation}
\mathcal{L}_{\rm HQET} = \bar{b}_v\,(i v\cdot D - \delta m_b)\,{b}_v
\,.\end{equation}
A suitable short-distance scheme for the bottom-quark mass should respect the
power counting of residual soft momentum $\delta m_b \sim \as \,v\cdot D \sim
\as \, k\sim \as\Lambda_{\rm QCD}$~\cite{Falk:1992fm}. Note that the
perturbative series for $\delta m_b$ starts at NLO and therefore scales like
$\as$. As stated in the previous section, in \btosgamma the residual momentum of
the bottom quark in the $B$ meson scales like $k\sim m_B-2E_\gamma$, which in
the peak region scales like $\Lambda_{\rm QCD}$. Thus, in the peak region one
expects that a low-scale mass scheme, such as the \OneS
scheme~\cite{Hoang:1998ng, Hoang:1998hm, Hoang:1999zc} or the MSR
scheme~\cite{Hoang:2008yj} with $R\sim\Lambda_{\rm QCD}$, are applicable. In the
context of \btosgamma, the \OneS mass scheme has been extensively discussed
in \refcite{Ligeti:2008ac}, and we remind the reader of the main features of the
MSR scheme in the following section.

\subsection{MSR mass scheme}

The MSR mass scheme is a short-distance mass scheme designed to subtract the pole-mass renormalon by introducing an infrared cutoff scale $R$ as follows,
\begin{equation}\label{eq:MSR-mass-residual}
  \delta m_b(R)\defeq\mbPole-m_b^{\rm MSR}(R)=R\sum_{n=1}^{\infty}a_n^{\rm MSR}\Bigl[\frac{\as(R)}{4\pi}\Bigr]^n
\,,\end{equation}
where $a_n^{\rm MSR}$ coefficients are determined from matching the MSR mass scheme onto the $\rm \overline{MS}$ mass scheme at $\overline{m}_b(\overline{m}_b)$.
The infrared scale $R$ controls the size of self-energy contributions which are absorbed into the mass definition \cite{Hoang:2008yj, Hoang:2014oea}.
In our analysis we use the so-called ``natural'' MSR mass definition, where the series coefficients $a_n^{\rm MSR}=a_n^{\MSbar}(n_l, n_h=0)$\,, see \refcite{Hoang:2017suc} for more details.

The MSR mass is a natural extension of the \MSbar mass for $R \leq \overline{m}_b(\overline{m}_b)$, which interpolates between all short-distance schemes with residual power counting $\delta m_b(R) \sim R \,\as$.
In the limit where $m_b^{\rm MSR} (m^{\rm MSR}_b) \to \overline{m}_b(\overline{m}_b)$ one approaches the \MSbar mass, whereas in the opposite limit, $R\to 0$\,, the MSR mass formally approaches the pole mass. In practice, when taking this limit one encounters the Landau pole of the coupling constant. This issue is deeply related to the pole mass renormalon and cannot be addressed unambiguously.

Values of the MSR mass at different $R$ scales are related by the so-called $R$-evolution equation~\cite{Hoang:2008yj}, whose solution resums logarithms $\ln(R_1/R_0)$ in the perturbative correction between $m_b^{\rm MSR}(R_1)$ and $m_b^{\rm MSR}(R_0)$.
The $R$ evolution can be used to obtain the MSR mass value at a low $R$ from the \MSbar mass and vice versa.

In contrast to the \OneS scheme, the infrared scale $R$ of the MSR scheme is an external parameter. We exploit this feature and pick $R = 1\GeV$ to be of the same order as the soft scale in the peak region.
In \sec{discussion-impact-short-distance} we demonstrate that this ensures a proper cancellation of the renormalon in the soft function.
In principle, it is possible to pick a different value for the $R$ scale in different regions of the spectrum by introducing a profile function $R(E_\gamma)$~\cite{Butenschoen:2016lpz}, as long as the $R$ evolution is consistently used to relate the shape functions at different $R$ scales.
This can be used to enforce $R\sim\mu_S$ over the whole spectrum and to eliminate logarithms $\ln(\mu_S/R)$ in the series of $\delta m_b(R)$.
\Refcite{Abbate:2010xh} implements this $R$-evolution setup for an analogous soft function for the thrust distribution in jet production.
In our case the use of $R$ evolution does not lead to a significant improvement in convergence, so for simplicity in our numerical analysis we use a fixed value of $R$.

\subsection{Short-distance schemes for \texorpdfstring{$\lambda_1$}{lambda\_1} and \texorpdfstring{$\rho_1$}{rho\_1}}
\label{sec:lambda1-rho1-short-distance}

To express the $B$-meson matrix elements $\lambda_1$ and $\rho_1$ in a
short-distance scheme, we use analogous schemes to the ``invisible'' scheme for
$\lambda_1$~\cite{Ligeti:2008ac}, where $\delta\lambda_1 \propto \alpha_s^2$.
The reason for this $\alpha_s^2$ scaling in the invisible scheme compared to the
kinetic scheme~\cite{Czarnecki:1997wy, Czarnecki:1997sz}, where $\delta\lambda_1
\propto \alpha_s$, is that in the invisible scheme one employs Lorentz-invariant
UV regulators for regularizing the kinetic energy
operator~\cite{Martinelli:1995zw, Neubert:1996zy}, while in the kinetic scheme
the regulator is not Lorentz-invariant. Indeed it has been shown in
\refcite{Ligeti:2008ac} that using the kinetic scheme for $\lambda_1$ leads to
an over-subtraction of the $u=1$ renormalon in the soft function. It is worth to
note here that there is no analogous study for the $u=3/2$ renormalon present in
$\rho_1$. Following these features of the invisible scheme for $\lambda_1$, we
write
\begin{align}\label{eq:lambda1-rho1-short-distance}
\delta\lambda_1(R_\lambda,\mu)
&= R_{\lambda}^2\, \frac{\as^2(\mu)}{\pi^2} \biggl[
  \delta\lambda_1^{(2)}
  + \frac{\as(\mu)}{\pi} \Bigl(\delta\lambda_1^{(3)} + \delta \lambda_1^{(2)}\,\beta_0\,\ln\frac{\mu}{R_\lambda} \Bigr)\biggr]
  +\mathcal{O}(\as^4)
\,,\nonumber\\
\delta\rho_1(R_\rho,\mu)
&= R_{\rho}^3 \,\frac{\as^3(\mu)}{\pi^3} \, \delta\rho_1^{(3)}
  +\mathcal{O}(\as^4)
\,,\end{align}
where we set $R_\lambda=R_\rho=1\GeV$ by default.
Note that $\lambda_1$ and $\rho_1$ appear in the second and third moments of the shape function and contain an $\mathcal{O}(\Lambda_{\rm QCD}^2)$ and $\mathcal{O}(\Lambda_{\rm QCD}^3)$ renormalon ambiguities, respectively.
Therefore by dimensional analysis they must scale like $\delta\lambda_1 \propto R_{\lambda}^2$ and $\delta\rho_1 \propto R_{\rho}^3$.
Furthermore, we take $\delta\rho_1 \propto \alpha_s^3$ to impose a plausible ``invisibility'' of our scheme choice and to avoid diluting the lower-order corrections arising from $\delta m_b$ and $\delta \lambda_1$.
For $\delta\lambda_1^{(2)}$ we use the value given in~\refcite{Ligeti:2008ac} for the invisible scheme, $\delta\lambda_1^{(2)}=\pi^2/3 -1$.
Values of $\delta\lambda_1^{(3)}$ and $\delta\rho_1^{(3)}$ are not available so far.

Usually, one defines a short-distance scheme to all orders by exploiting the perturbative series of some physical and thus renormalon-free quantity. However, this is not strictly necessary, since after all the main goal of the renormalon subtractions is to obtain a stable perturbative result. Thus, here we take a pragmatic approach and simply define our ``invisible'' scheme for $\widehat\lambda_1$ and $\widehat\rho_1$ at $\ord{\as^3}$ by choosing numerical values for $\delta\lambda_1^{(3)}$ and $\delta\rho_1^{(3)}$ such that the resulting soft function at different perturbative orders manifests good convergence, i.e.\ that the size of scale variations reduces when including higher-order corrections, that the resulting uncertainty bands at different orders have reasonable overlaps, that the peak position for the soft function remains stable, and finally that it remains positive at small $k$ and approaches zero with similar slopes at different orders.
Following this procedure, we find a satisfactory convergence for the soft function, see \fig{discussion-MSR}, by taking
\begin{equation}
\delta\lambda_1^{(3)}=16
\,,\qquad
\delta\rho_1^{(3)}=-3
\,.\end{equation}

In principle, we could also consider the perturbative convergence of the final spectrum. However, since it also receives contributions from the jet and hard functions, the dependence on the soft function is washed out in the spectrum.
Therefore, we use the convergence of the soft function to define the short-distance scheme. This is also the most natural, since the soft function is the object containing the renormalons to be subtracted. This is somewhat similar to the ``shape-function'' scheme in \refcite{Bosch:2004th}, where the second and third moments of the perturbative shape function with some, largely arbitrary, hard cutoff is used to define the short-distance parameters.
(Similarly, the short-distance $b$-quark mass is defined based on the first moment.)
The disadvantage of that approach is that it yields $\delta\lambda, \delta\rho_1 \sim \as$, which leads to massive oversubtraction similar to the kinetic scheme.

\subsection{Subleading \texorpdfstring{$\delta m_b$}{delta mb} corrections}
\label{sec:subleading-delta-mb-corrections}

As discussed in \sec{short-distance-soft-function}, the leading renormalon at
leading power comes from the $b$-quark mass that enters via the argument of the
soft function. In addition to the soft function, the $b$-quark mass also enters
in the hard and jet functions through the SCET label momentum $p^- \sim m_b$. By
default, label momentum conservation sets $p^- = \mbPole$. Formally, choosing a
different label $p^- = \mb$ amounts to a power-suppressed effect. For this
reason, in \refcite{Bernlochner:2020jlt} the resulting corrections from changing
to the $\mb$ scheme could effectively be absorbed into the nonsingular
corrections. As we will see, at \NNNLLp this is no longer viable, so
instead we will explicitly switch both hard and jet functions to a
short-distance mass scheme.

To derive the scheme change, we consider the partonic function
$W(k)$ appearing in \eq{perturbative-contributions}, which can be either the singular, the nonsingular, or the full contribution. It has mass dimension $-1$ and depends on two dimensionful quantities, $k$ and $m_b$. Therefore, by dimensional analysis it must have the form
\begin{equation} \label{eq:Wk_structure}
W(k)
= \frac{1}{m_b}w\Bigl[\frac{k}{m_b}, \as(\mu), \ln\frac{\mu}{m_b}\Bigr]
= \frac{1}{m_b}w\Bigl[\frac{k}{m_b}, \as(m_b)\Bigr]
\,,\end{equation}
where all dependence on $m_b$ is made explicit on the right-hand side and $w(x, \as, L)$ is a scaleless function of its arguments.
Since $W(k)$ is defined to be $\mu$ independent, the $\mu$ dependence on the
right-hand is only the internal $\mu$ dependence from $\as(\mu)$ which cancels
order by order. Therefore, we can pick $\mu = m_b$, which eliminates all logarithms and allows us to track the associated $m_b$ dependence via the dependence on $\as(m_b)$. After switching the scheme, we can easily reintroduce the $\mu$ dependence by reexpanding $\as(m_b)$ in terms of $\as(\mu)$.

The partonic rate in the pole scheme is given by \eq{Wk_structure} evaluated at $k = \mbPole - 2E_\gamma$ and $m_b = \mbPole$. To switch to a short-distance scheme, we thus have to replace $k \to k + \delta m_b$ and $m_b \to \mbPole = \mb + \delta m_b$ in \eq{Wk_structure} and expand in $\delta m_b$. This gives
\begin{align} \label{eq:fixed-order-delta-mb-corrections}
\mb W(\mb x)
&= \frac{1}{1 +\delta m_b/\mb}w\Bigl(\frac{x + \delta m_b/\mb}{1 + \delta m_b/\mb}, \as(\mb + \delta m_b) \Bigr)
\nn \\
&= \biggl\{
    1
    +\frac{\delta m_b}{\mb}\frac{\df}{\df x}(1-x)
    +\frac{1}{2}\frac{\delta m_b^2}{\mb^2}\frac{\df^2}{\df x^2}(1-x)^2
    \nonumber\\&\quad
    +\frac{1}{3!}\frac{\delta m_b^3}{\mb^3}\frac{\df^3}{\df x^3}(1-x)^3
    +\frac{\delta m_b}{\mb}\beta[\as(\mb)]\frac{\df}{\df\as}
    +\ord{\as^4}
  \biggr\} w[x, \as(\mb)]
\,,\end{align}
where for convenience we multiplied by $\mb$ and switched variables, $k = \mb x$,
such that $\mb W(\mb x)$ is a dimensionless function of $x$. In the second step,
we expanded in $\delta m_b$ keeping only terms that contribute up to
$\ord{\as^3}$, recalling that $\delta m_b \sim \ord{\as}$ and $\beta(\as)\sim \as^2$. The derivatives act on everything to their right.

Substituting the explicit $\as$ expansions for $\delta m_b/\mb$, $\beta(\as)$, and $w(x, \as)$, it is straightforward to derive the correction terms $\Delta w_{77}^{\sing(n)}$ and $\Delta w_{77}^{\nons(n)}$ appearing in \eqs{singular-contribution}{nonsingular-contribution}. Writing the $\as$ expansion of $\delta m_b/\mb$ as
\begin{equation}\label{eq:deltam-definition}
\frac{\delta m_b}{\mb}
= \frac{\as(\mu)}{4\pi}\, \deltam^{(1)}(\mu)
  + \Bigl[\frac{\as(\mu)}{4\pi}\Bigr]^2\deltam^{(2)}(\mu)
  + \Bigl[\frac{\as(\mu)}{4\pi}\Bigr]^3\deltam^{(3)}(\mu)
  + \ord{\as^4}
\,,\end{equation}
we find for the singular
\begin{align}\label{eq:singular-delta-mb-corrections}
\Delta w_{77}^{\sing(1)}(\mu, x)
&=\frac{1}{4}\delta_m^{(1)}(\mu)\frac{\df}{\df x}w_{77}^{\sing(0)}(x)
\,,\nn \\
\Delta w_{77}^{\sing(2)}(\mu, x)
&= \frac{1}{16} \biggl\{
    \delta_m^{(2)}(\mu)\frac{\df}{\df x}
   +\frac{1}{2} \bigl[\delta_m^{(1)}(\mu)\bigr]^2\frac{\df^2}{\df x^2}
 \biggr\}w_{77}^{\sing(0)}(x)
 \nn\\&\quad
 + \frac{1}{4} \delta_m^{(1)}(\mu)\frac{\df}{\df x}\bigl[(1-x)w_{77}^{\sing(1)}(x)\bigr]
\,,\nn \\
\Delta w_{77}^{\sing(3)}(\mu, x)&=
 \frac{1}{64}\biggl\{
    \delta_m^{(3)}(\mu)\frac{\df}{\df x}
   +\delta_m^{(1)}(\mu)\delta_m^{(2)}(\mu)\frac{\df^2}{\df x^2}
   +\bigl[\delta_m^{(1)}(\mu)\bigr]^3\frac{1}{3!}\frac{\df^3}{\df x^3}
 \biggr\} w_{77}^{\sing(0)}(x)
 \nn\\&\quad
 +\frac{1}{16}\biggl\{
   -2\beta_0\delta_m^{(1)}(\mu)
   +\Bigl[
      \delta_m^{(2)}(\mu)
     +2\beta_0\delta_m^{(1)}(\mu)\ln\frac{\mu}{\mb}
   \Bigr]\frac{\df}{\df x}(1-x)
   \nn\\&\qquad\qquad
   +\frac{1}{2}\bigl[\delta_m^{(1)}(\mu)\bigr]^2\frac{\df^2}{\df x^2}(1-x)^2
 \biggr\} w_{77}^{\sing(1)}(x)
 \nn\\&\quad
 +\frac{1}{4}\delta_m^{(1)}(\mu)\frac{\df}{\df x}(1-x)w_{77}^{\sing(2)}(x)
\,,\end{align}
and the nonsingular
\begin{align}\label{eq:nonsingular-delta-mb-corrections}
\frac{\Delta w^{\nons(2)}_{77}(\mu_\nons, x)}{(1-x)^3}
&= \frac{1}{4} \deltam^{(1)}(\mu_\nons)\frac{\df}{\df x}\biggl[\frac{w_{77}^{\nons(1)}(x)}{(1-x)^2}\biggr]
\,, \nn \\
\frac{\Delta w^{\nons(3)}_{77}(\mu_\nons, x)}{(1-x)^3}&=
\frac{1}{16} \biggl\{
 -2\beta_0\deltam^{(1)}(\mu_\nons)
 +\Bigl[
    \deltam^{(2)}(\mu_\nons)
   +2\beta_0\deltam^{(1)}(\mu_\nons)\ln\frac{\mu_\nons}{\mb}
 \Bigr]\frac{\df}{\df x}(1-x)
 \nn\\&\qquad\quad
 +\frac{1}{2}\bigl[\deltam^{(1)}(\mu_\nons)\bigr]^2\frac{\df^2}{\df x^2}(1-x)^2
\biggr\}\frac{w_{77}^{\nons(1)}(x)}{(1-x)^3}
\nn\\&\quad
+\frac{1}{4}\deltam^{(1)}(\mu_\nons)\frac{\df}{\df x}\frac{w_{77}^{\nons(2)}(x)}{(1-x)^2}
\,.\end{align}

As we have seen above, there are three sources of $\delta m_b$ corrections:
\begin{enumerate}
\item Shifting the argument $k\to k+\delta m_b$.
\item Changing to $\mb$ in the argument $k/\mb$, which yields the rescaling $x \to x/(1 + \delta m_b/\mb)$.
\item Changing to $\mb$ in the $\mu$ dependence of $\as(\mb)$.
\end{enumerate}
Considering just the singular contributions and only keeping the
first and neglecting the latter two corrections amounts to only keeping the
leading-power terms in \eq{fixed-order-delta-mb-corrections},
\begin{align} \label{eq:delta-mb-corrections-leading-power}
\mb W^\sing(\mb x)
&= \biggl[
    1
    +\frac{\delta m_b}{\mb}\frac{\df}{\df x}
    +\frac{1}{2}\frac{\delta m_b^2}{\mb^2}\frac{\df^2}{\df x^2}
    +\frac{1}{3!}\frac{\delta m_b^3}{\mb^3}\frac{\df^3}{\df x^3}
  \biggr] w^\sing[x, \as(\mb)]
\,.\end{align}
The relevant formal power counting here is $\delta m_b/\mb\sim\lambda\ll1$, $\df/ \df x\sim 1/x\sim\lambda^{-1}$, so all terms on the right-hand side are leading power.
These terms are exactly reproduced by the factorized result at fixed order
by changing the soft function to the $\mb$ scheme, which involves the analogous shift of its argument.
In the numerical implementation, the derivatives $\df/\df (\mb x) = \df/\df k$
are moved via integration by parts to act on the shape function $\ShapeFunction(k)$
so they count as $1/\Lambda_{\rm QCD}$. Since $\delta m_b\sim\Lambda_{\rm QCD}$,
we see again that all terms in \eq{delta-mb-corrections-leading-power} are leading power, counting as $(\delta m_b\, \df/\df k)^n \sim 1$.

All terms $\sim x^n w^\sing(x, \as)$ are thus induced by the second source.
By moving the derivatives to act onto the shape function, we see that they
are explicitly power-suppressed by $x$ and hence nonsingular. For this reason,
they could be included as part of the nonsingular correction terms
$\Delta w_{77}^\nons$, as was done in \refcite{Bernlochner:2020jlt}.
In the singular contributions, the $k/m_b$ dependence only
appears in logarithms which are factorized into the soft, jet, and hard functions,
where $m_b$ corresponds to the large $p^-$ label momentum, which only appears
in the hard and jet functions, while the soft function only depends on the small
momentum $k$. Therefore, the associated correction terms can be reproduced
by the leading-power factorized result by changing the $m_b$ dependence in the
hard and jet functions to the short-distance $\mb$.

Finally, the third source produces the last term in \eq{fixed-order-delta-mb-corrections},
\begin{equation}
\frac{\delta m_b}{\mb}\beta(\as)\frac{\df}{\df\as} [w(x, \as(\mb)]
= -\frac{\as^3 C_F}{8\pi^3}\,\delta_m^{(1)}\,\beta_0\,w_{77}^{\sing(1)}(x)
+ \ord{\as^4}
\,.\end{equation}
Since it starts at $\ord{\as^3}$ it first appears at \NNNLLp. Formally, this
term is also subleading power, because $\delta m_b/\mb\sim\lambda$. However,
since there is no explicit kinematic suppression by $x$, the $x$ dependence
itself is still singular $\sim 1/x$, involving $\delta(x)$ and logarithmic plus
distributions. Hence, this term cannot simply be absorbed into the nonsingular
contributions but must be properly accounted for in the resummed singular
contribution. The $\mb$ dependence in the corresponding fixed-order
$\ln(\mu/\mb)$ also corresponds to the $p^-$ label momentum. Therefore, to
account for the distributional structure of this contribution and to resum it,
we consistently switch the hard and jet functions to the short-distance mass
$\mb$. The details of this procedure are discussed in the following two
subsections.

\subsection{Hard function}
\label{sec:hard-short-distance-scheme}

The hard function in a short-distance scheme is obtained by writing the
$b$-quark mass in the pole-scheme hard function in \eq{hard} in terms of a
short-distance mass and reexpanding the result strictly in powers of
$\as(\mu)$. At $\rm N^3LO$ we obtain
\begin{equation} \label{eq:short-distance-hard}
  \Hard(\mb,\mu)=\HardPole(\mb,\mu)-\sum_{n=2}^\infty\sum_{m=0}^{2n-3}\Delta H_m^{(n)}(\mu)\biggl[\frac{\as(\mu)}{4\pi}\biggr]^n\ln^m\!\frac{\mu}{\mb}
  \,,\end{equation}
where the coefficients $\Delta H_m^{(n)}$ are given by
\begin{align}
  \Delta H_0^{(2)}(\mu)&=\deltam^{(1)}(\mu)H_1^{(1)}\,,
  \nonumber\\
  \Delta H_1^{(2)}(\mu)&=2\deltam^{(1)}(\mu)H_2^{(1)}\,,
  \nonumber\\
  \Delta H_0^{(3)}(\mu)&=\deltam^{(2)}(\mu)H_1^{(1)}+\deltam^{(1)}(\mu)H_1^{(2)}-\frac{1}{2}\bigl(\deltam^{(1)}(\mu)\bigr)^2\Bigl(H_1^{(1)}+2 H_2^{(1)}\Bigr)\,,
  \nonumber\\
  \Delta H_1^{(3)}(\mu)&=2\deltam^{(2)}(\mu)H_2^{(1)}+2\deltam^{(1)}(\mu)H_2^{(2)}-\bigl(\deltam^{(1)}(\mu)\bigr)^2H_2^{(1)}\,,
  \nonumber\\
  \Delta H_2^{(3)}(\mu)&=3\deltam^{(1)}(\mu)H_3^{(2)}\,,
  \nonumber\\
  \Delta H_3^{(3)}(\mu)&=4\deltam^{(1)}(\mu)H_4^{(2)}\,.
\end{align}
Here $H_m^{(n)}$ are the coefficients of the pole-scheme hard function defined in \eq{hard}, and $\deltam^{(n)}(\mu)$ are defined in \eq{deltam-definition}.

Similarly, the RGE for the hard function in a short-distance mass scheme is derived by rewriting the pole mass in \eq{hard-RGE} in terms of a short-distance mass,
\begin{equation}\label{eq:short-distance-hard-RGE}
\frac{\df\Hard(\mb,\mu)}{\df\ln\mu}
= \biggl\{\Gamma^H[\as(\mu)]\ln\frac{\mu}{\mb} + \gamma^H[\as(\mu)]+\Delta\gamma^H(\mu)\biggr\}\Hard(\mb,\mu)
\,,\end{equation}
where
\begin{align}
  \Delta\gamma^H(\mu)
  &= \Gamma^H[\as(\mu)]\ln\frac{\mb}{\mbPole}
  = -\Gamma^H[\as(\mu)]\ln\Bigl(1+\frac{\delta m_b}{\mb}\Bigr)
\nn\\&
  = -\Bigl[\frac{\as(\mu)}{4\pi}\Bigr]^2\Gamma^H_0\deltam^{(1)}(\mu)
\nn\\&\quad
    -\Bigl[\frac{\as(\mu)}{4\pi}\Bigr]^3\biggl\{
      \Gamma^H_1\deltam^{(1)}(\mu)
      +\Gamma^H_0\biggl[
        \deltam^{(2)}(\mu)
        -\frac{1}{2}(\deltam^{(1)}(\mu))^2
      \biggr]
    \biggr\}
  +\mathcal{O}(\as^4)
  \,,
\end{align}
and the hard cusp anomalous dimension coefficients $\Gamma_n^H=-2\Gamma_n$ are
defined in \eq{anomdimcoeffs}.
Note that the reexpansion in \eqs{short-distance-hard}{short-distance-hard-RGE}
must be performed in terms of the same $\alpha_s(\mu)$, such that $\Hard(\mb, \mu)$
in \eq{short-distance-hard} indeed satisfies the RGE in \eq{short-distance-hard-RGE}.

We expect that the renormalon associated with the pole mass cancels in the
perturbative series of the hard anomalous dimension.
The cusp anomalous dimension is universal and arises in the evolution of
many perturbative objects with Sudakov double logarithms that do not involve the
$b$-quark mass at all.
Therefore, the $\Gamma^H$ series cannot know about the pole-mass renormalon,
so the cancellation must happen between $\gamma^H$ and $\Delta\gamma^H$.
For this reason, it is important to consistently expand $\Delta\gamma^H$ in powers of $\as(\mu)$ to the same order as $\gamma^H$.

The all-order solution to the differential equation (\ref{eq:short-distance-hard-RGE}) can be written as
\begin{equation}
  \HardEvolution(\mb,\mu_H,\mu)
  =
  \HardEvolutionPole(\mb,\mu_H,\mu)\times \Delta U_H(\mu_H,\mu)
  \,,
\end{equation}
where $U_H$ is the usual hard evolution factor given in \eq{hard-RGE-solution},
and the correction factor $\Delta U_H$ is given by
\begin{equation}
\Delta U_H(\mu_H,\mu) = \exp\biggl[\int_{\mu_H}^\mu\!\df\ln\mu\, \Delta\gamma^H(\mu)\biggr]
\,.\end{equation}
Note that in general, the $\mu$ dependence of $\Delta\gamma^H(\mu)$ coming from $\delta_m^{(n)}(\mu)$ can be more involved than for the usual anomalous dimension,
such that the $\mu$ integral may have to be performed numerically.
Using the MSR mass and up to N$^3$LL, we can still perform the integral by employing an analytic approximation analogous to the one used for the $K$ and $\eta$ integrals in \eq{define-eta-K}.
We find up to N$^3$LL
\begin{align}\label{eq:delta-ln-UH}
\ln \Delta U_H(\mu_H,\mu)
&= \frac{\as(\mu_H)}{4\pi}\frac{\Gamma^H_0}{2\beta_0}\delta_{\rm MSR}^{(1)}(r-1)
  \nn\\&\quad
  +\frac{\as(\mu_H)}{4\pi}\frac{\Gamma^H_0}{4\beta_0}\biggl\{
    \frac{\as(\mu_H)}{4\pi}\biggl[
      \left(
         \delta_{\rm MSR}^{(2)}
        -\frac{1}{2}(\delta_{\rm MSR}^{(1)})^2
        +2\delta_{\rm MSR}^{(1)}\beta_0\ln\frac{\mu_H}{R}
      \right)
      \nn\\&\quad
      +\delta_{\rm MSR}^{(1)}\left(\frac{\Gamma^H_1}{\Gamma^H_0}-\frac{\beta_1}{\beta_0}\right)
    \biggr](r^2-1)
    -\delta_{\rm MSR}^{(1)}(r-1)^2
  \biggr\}
  \,,
\end{align}
where $r = \as(\mu)/\as(\mu_H)$ and
\begin{align}
  \delta_{\rm MSR}^{(n)}&\defeq\frac{R}{m_b^{\rm MSR}(R)}\, a_n^{\rm MSR}
\,.\end{align}
At NNLL only the first line on the right-hand side of \eq{delta-ln-UH} is kept.

\subsection{Jet function}
\label{sec:jet-short-distance-scheme}

Similar to the hard function in the previous section, we define the jet function in a short-distance scheme by expressing $\mbPole$ in terms of a short-distance mass $\mb$.
To this end, we start with the jet function in the pole scheme given in \eq{jet} and set $s=\mbPole\,\omega$. Then we write $\mbPole$ in terms of $\mb$ and reexpand the result strictly in powers of $\as(\mu)$,
such that $\mbPole\JetPole(\mbPole\omega,\mu_J)=\mb\Jet(\mb\omega,\mu_J)$ order by order.
This yields up to ${\rm N^3LO}$
\begin{equation}\label{eq:jet-short-distance}
  \Jet(\mb\omega,\mu)=\JetPole(\mb\omega,\mu)+\sum_{n=2}^3\sum_{m=-1}^{2n-4} \Delta J_m^{(n)} \biggl[\frac{\as(\mu)}{4\pi}\biggr]^n\frac{1}{\mu^2}\mathcal{L}_m\Bigl(\frac{\mb\omega}{\mu^2}\Bigr)
  \,.
\end{equation}
The expansion coefficients $\Delta J_m^{(n)}$ read
\begin{align}
  \Delta J_{-1}^{(2)}(\mu)&=\deltam^{(1)}(\mu)J_0^{(1)}\,,
  \nonumber\\
  \Delta J_0^{(2)}(\mu)&=\deltam^{(1)}(\mu)J_1^{(1)}\,,
  \nonumber\\
  \Delta J_{-1}^{(3)}(\mu)&=\deltam^{(2)}(\mu)J_0^{(1)}+\deltam^{(1)}(\mu) J_0^{(2)}-\frac{1}{2}\bigl(\deltam^{(1)}(\mu)\bigr)^2\Bigl(J_0^{(1)} - J_1^{(1)}\Bigr)\,,
  \nonumber\\
  \Delta J_0^{(3)}(\mu)&=\deltam^{(2)}(\mu)J_1^{(1)}+\deltam^{(1)}(\mu)J_1^{(2)}-\frac{1}{2}\bigl(\deltam^{(1)}(\mu)\bigr)^2J_1^{(1)}\,,
  \nonumber\\
  \Delta J_1^{(3)}(\mu)&=2\deltam^{(1)}(\mu)J_2^{(2)}\,,
  \nonumber \\
  \Delta J_2^{(3)}(\mu)&=3\deltam^{(1)}(\mu)J_3^{(2)}\,,
\end{align}
where the $\deltam^{(n)}$ are defined in \eq{deltam-definition} and the
$J_m^{(n)}$ coefficients are those of the original pole-scheme jet function
as defined in \eq{jet}.

\section{Perturbative uncertainties}
\label{sec:uncertainty}

For our predictions we can distinguish perturbative and parametric
uncertainties. Parametric uncertainties arise from the uncertainty in input
parameters, such as $\CSevenInclusive$, $|V_{\rm tb}\,V_{ts}^*|^2$, $m_b$,
$\ShapeFunction(k)$. These are not considered in the following. We normalize our
numerical results by dividing out the overall prefactor $\Gamma_0
\CSevenInclusiveSquared$, so the associated uncertainties drop out. We also
ignore the parametric uncertainties due to the shape function $\ShapeFunction(k)$
and $m_b$, which do affect the shape of the spectrum, because we do not compare
with experimental \btosgamma measurements, in which case they would be determined
by the fit to the data.

Our primary focus is on the perturbative results and their perturbative uncertainties
arising from missing higher-order corrections. For these there are various
different sources, which fall into two categories:
\begin{itemize}
\item{\bf Profile scale variations:}
We identify three sources of perturbative uncertainties that are estimated by a
suitable set of variations of the profile scales discussed in \sec{profile}. The
resummation uncertainty $\DeltaResummation$ is obtained by taking the maximum
envelope of all $27$ simultaneous variations of the profile function parameters
$e_H$, $e_J$, $\mu_0$, which corresponds to scale variations in the resummed
singular contributions, together with corresponding correlated variations in the
nonsingular contributions. Note also that despite its name, $\Delta_{\rm resum}$
reduces to the overall fixed-order scale variation in the fixed-order region.
The nonsingular uncertainty $\DeltaNonsingular$ is determined by varying the
$e_{\rm ns}$ parameter, which determines the central value of the nonsingular scale in the resummation regions. The matching
uncertainty $\DeltaMatching$ comes from varying the transition point $E_1$,
which marks the start of the transition region. These three sources are
considered independent and are thus added in quadrature,
\begin{equation}
\DeltaScalevar = \DeltaResummation \oplus\DeltaNonsingular \oplus\DeltaMatching
\,.\end{equation}
For notational convenience, we use the symbol $\oplus$ to denote addition in
quadrature, $x\oplus y\defeq\sqrt{x^2+y^2}$.

\item{\bf Theory nuisance parameter variations:}
The uncertainties $\DeltaHardNP$ and $\Delta_{c^{\rm ns}}$ are estimated by varying the nuisance parameters $h_3$ and $c^{\rm ns}_k$, respectively, within the ranges given in \secs{singular}{nonsingular}, where we abbreviate
$\Delta_{c^{\rm ns}}=(\sum_{k=0}^5\Delta_{c^{\rm ns}_k}^2)^{1/2}$.
Note that by definition the central values of all nuisance parameters are zero.

The nuisance parameters at \NNNLLpMatched are introduced in such a way that the scale dependence cancels at this order, i.e., all terms that are predicted by scale
dependence are correctly included.
This means that at \NNNLLpMatched the profile scale variations estimate the
uncertainty due to the missing next order N$^4$LL$'+$N$^4$LO,
while the nuisance parameters capture the uncertainty due to the missing
$\mathcal{O}(\as^3)$ ingredients at \NNNLLpMatched.
\end{itemize}

The total perturbative uncertainty is obtained by adding all sources in quadrature,
\begin{equation} \label{eq:total-uncertainty}
\Delta_{\rm total} = \DeltaScalevar \oplus\DeltaHardNP \oplus\DeltaNonsingularNP
\,.\end{equation}
The nuisance parameter uncertainties only contribute at the highest order \NNNLLpMatched, while at lower orders we simply have $\Delta_{\rm total}=\DeltaScalevar$.

Note that in \refcite{Bernlochner:2020jlt} the perturbative uncertainty was
estimated from profile scale variations by taking the maximum envelope of $3^5 = 243$ profile scale variations corresponding to simultaneous variations of the above
five profile scale parameters.
Here we have refined the estimation procedure by separating conceptually
different sources of perturbative uncertainties, which leads to an overall more consistent picture of the resulting uncertainties when including the new highest
order at \NNNLLp. In part, this becomes possible
because we are now able to reexpand the fixed-order hard function against the
product of the fixed-order jet and soft functions.
We have also checked that the total $\DeltaScalevar$ estimated as described
above is comparable in size to what we obtain in our setup by taking the maximum
envelope of all $3^5 = 243$ variations excluding a small set of obvious outliers.

\section{Results}
\label{sec:results}

In this section, we present our numerical results for the photon energy spectrum.
All numerical results are implemented and obtained with the
\scetlib{}~\cite{scetlib} library.

Our default numerical setup is as follows.
The values used for input parameters are summarized in \tab{theory-parameters}.
The spectrum is always divided by the overall normalization factor
$\Gamma_0\CSevenInclusiveSquared$, so its numerical value is not needed.
For the shape-function model in \eq{basis} we use $\lambda=0.6\GeV$
and $p=4$ as the default settings. We illustrate the impact of changing
$p$ and $\lambda$ later in this section.
We neglect finite-charm-mass corrections and work in QCD with $n_f=4$ massless quarks.
We always use the 4-loop running of $\as$, which is sufficient for resummation at $\rm N^3LL$.
We use the MSR scheme for the $b$-quark mass and adopt short-distance schemes
for $\lambda_1$ and $\rho_1$ as discussed in \sec{lambda1-rho1-short-distance}.
The impact of the short-distance mass scheme is discussed in \sec{discussion-impact-short-distance}.
Throughout this section, the colored bands always show the perturbative uncertainties obtained from profile scale variations $\DeltaScalevar$.

\begin{table}[t!]
\centering
\begin{tabular}{c|c}
\hline\hline
Parameter&Value\\\hline
$m_B$&$5.279\GeV$\\
$\as(\mu=4.7\GeV)$&$0.2155$\\
$m_b^{\rm MSR}(R=1\GeV)$&$4.7\GeV$\\
$\lambda$&$0.6\GeV$\\
$p$&$4$\\
\hline\hline
\end{tabular}
\caption{%
Numerical values of required input parameters.
}
\label{tab:theory-parameters}
\end{table}

\subsection{Main results}
\label{sec:main-results}

\begin{figure}\label{fig:resummed-vs-full}
\centering
\includegraphics[width=\WidthTwoSubfigs]{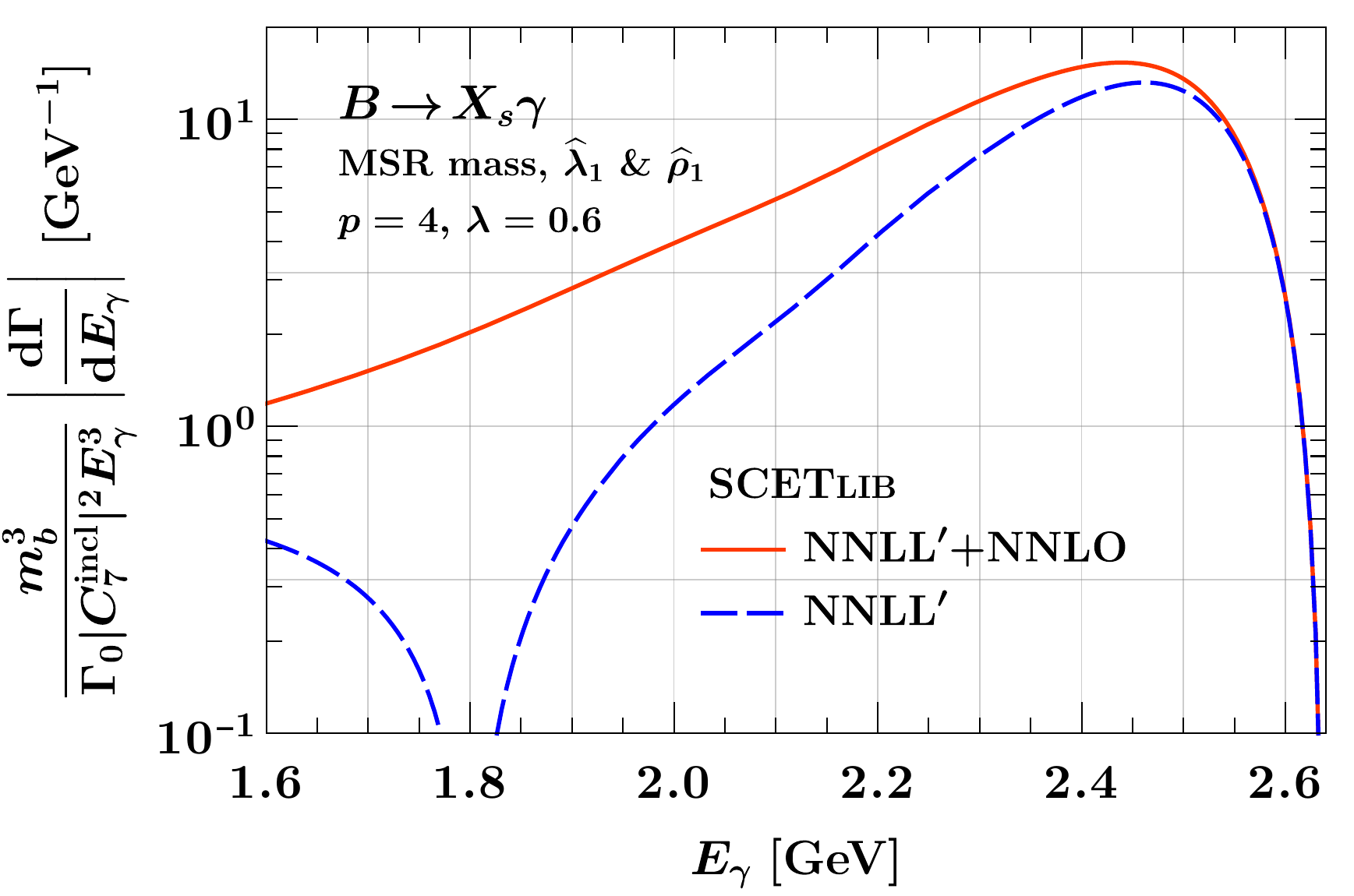}%
\hfill%
\includegraphics[width=\WidthTwoSubfigs]{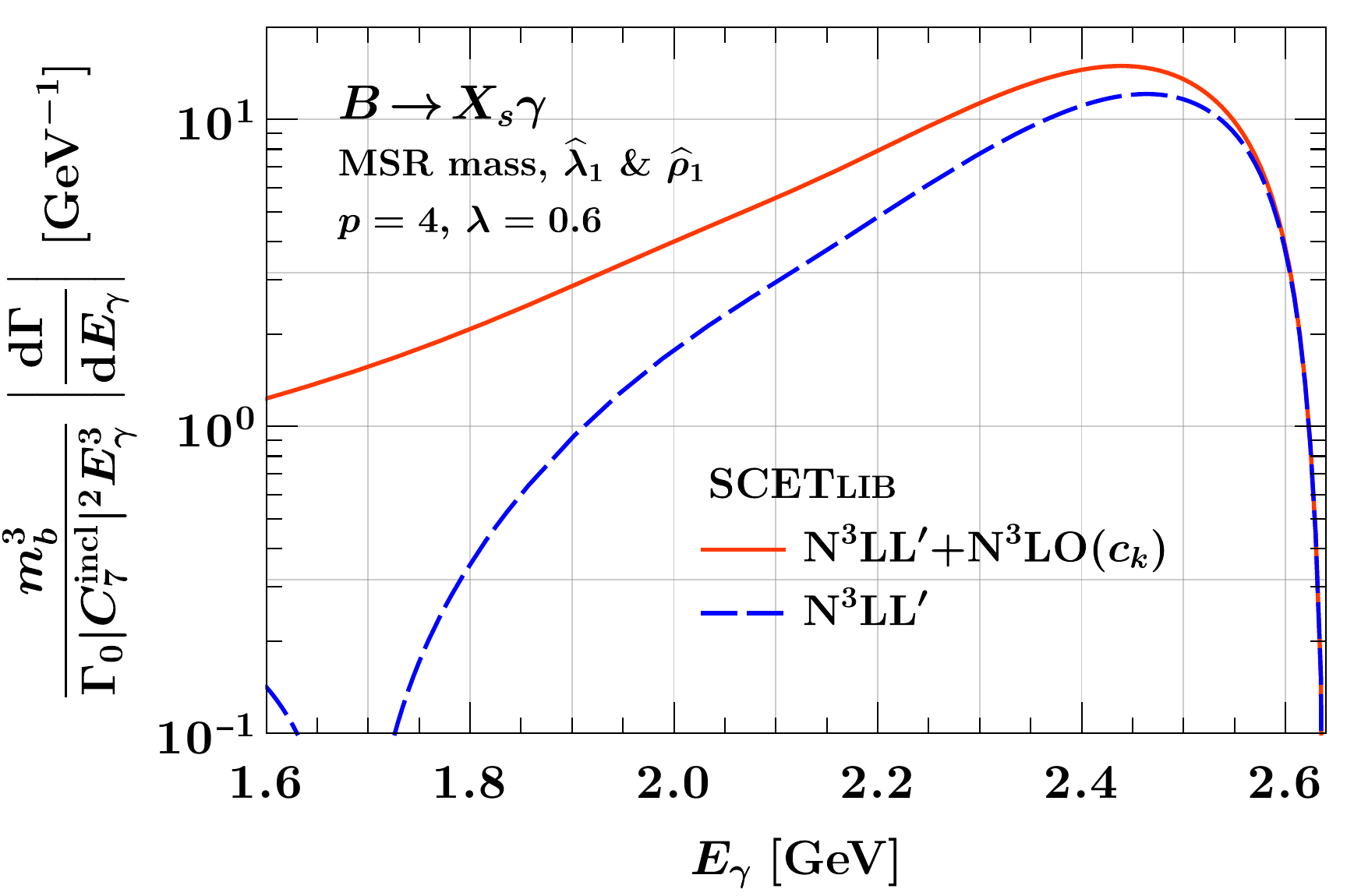}%
\caption{%
Comparison of the absolute value of resummed contribution to the full \btosgamma photon energy spectrum in the MSR mass scheme.
The left panel shows the \TwoLoop predictions, the right panel shows the \ThreeLoop predictions.
The overall factor $\Gamma_0\CSevenInclusiveSquared(E_\gamma/\mb)^3$ is divided out (see \eq{spectrum}).
}
\end{figure}

To begin, in \fig{resummed-vs-full}, we show the contribution of the resummed singular corrections to the full result at \NNLLp (left panel) and \NNNLLp (right panel).
The resummed contribution is indeed dominant across the peak of the spectrum,
while it decreases rapidly in the tail and eventually changes sign at $E_\gamma\lesssim 1.8\GeV$, where the resummation is getting turned off. Here, only the full matched result is meaningful, which remains positive and slowly approaches zero in the far tail.

\begin{figure}[t!]
\centering
\includegraphics[width=0.484\textwidth]{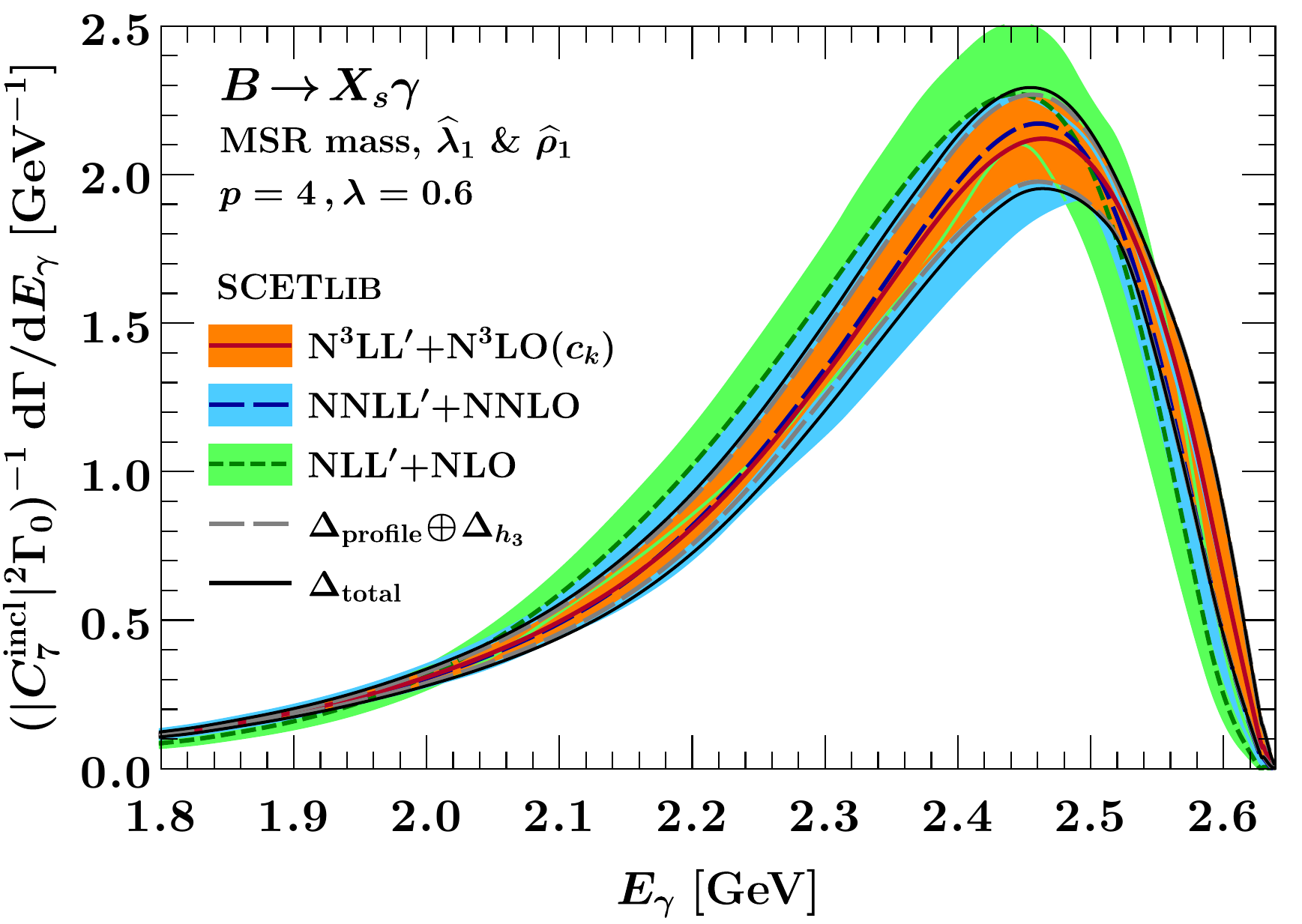}%
\hfill%
\includegraphics[width=0.50\textwidth]{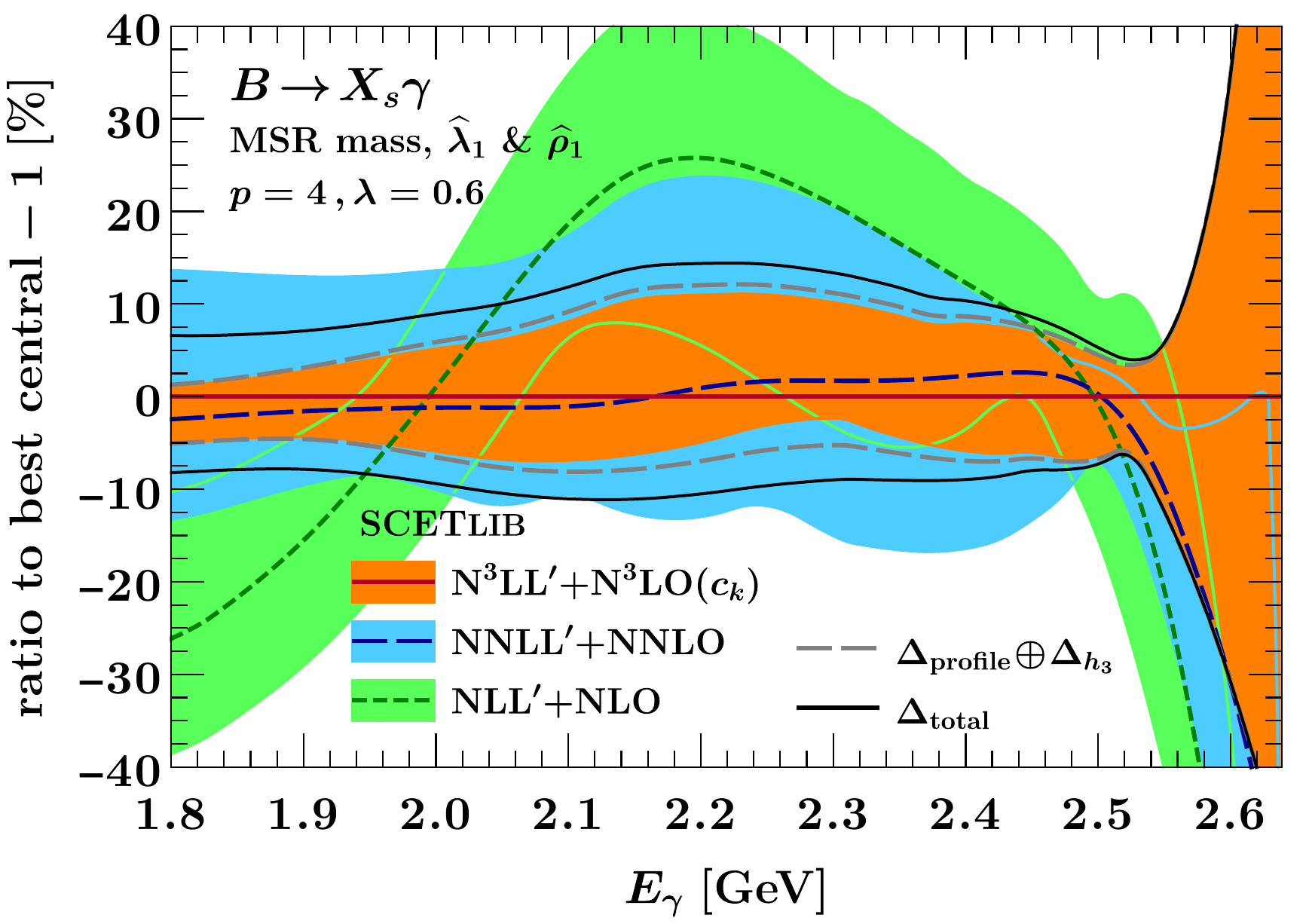}%
\\
\includegraphics[width=0.484\textwidth]{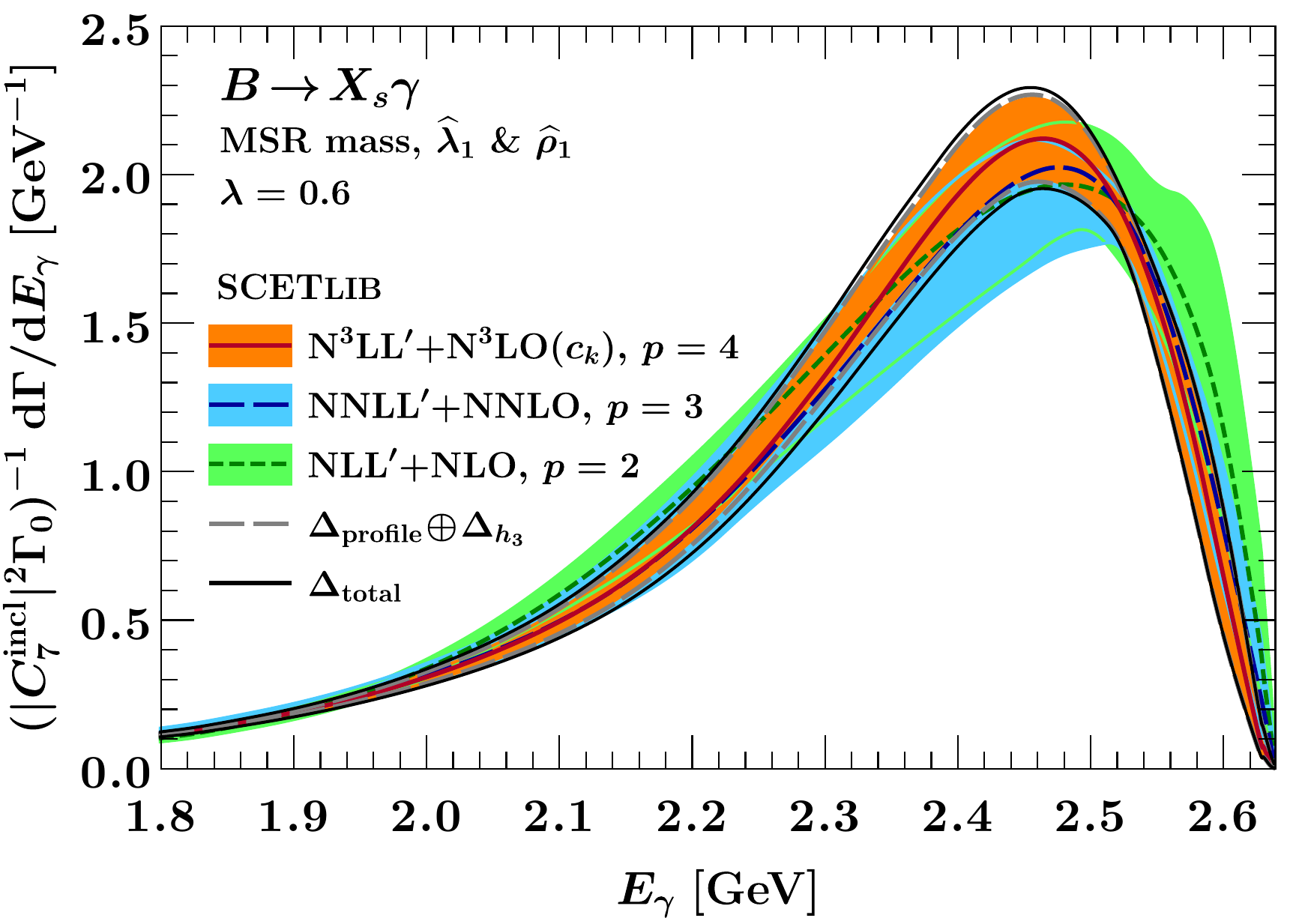}%
\hfill%
\includegraphics[width=0.50\textwidth]{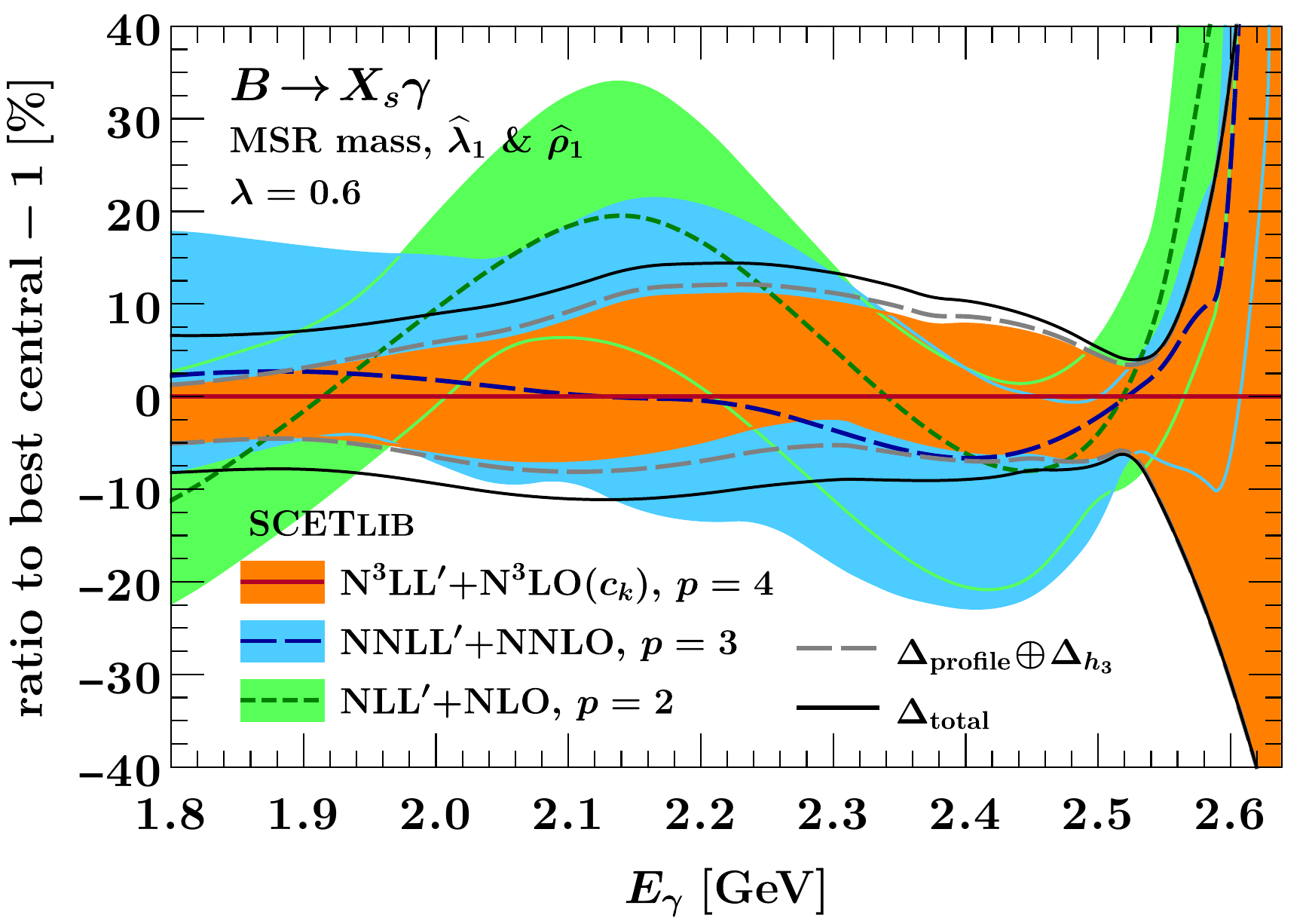}
\label{fig:spectrum-main}%
\caption{%
The \btosgamma spectrum at different perturbative orders. The $E_\gamma$
spectrum itself is shown on the left, while the relative differences to the
highest-order central value are shown on the right. In the top row, we use the
same value $p = 4$ at each order, while in the bottom row we use successive
values at each order. The colored bands show the perturbative uncertainty
estimated by just profile scale variations, $\DeltaScalevar$. The dashed gray
line includes in addition the uncertainty due to $h_3$. The solid black line
further includes in addition the nonsingular nuisance parameters, corresponding
to the total perturbative uncertainty at \NNNLLpMatched.
}
\end{figure}

\begin{figure}[t!]\label{fig:uncertainty-components}
\centering
\includegraphics[width=\WidthTwoSubfigs]{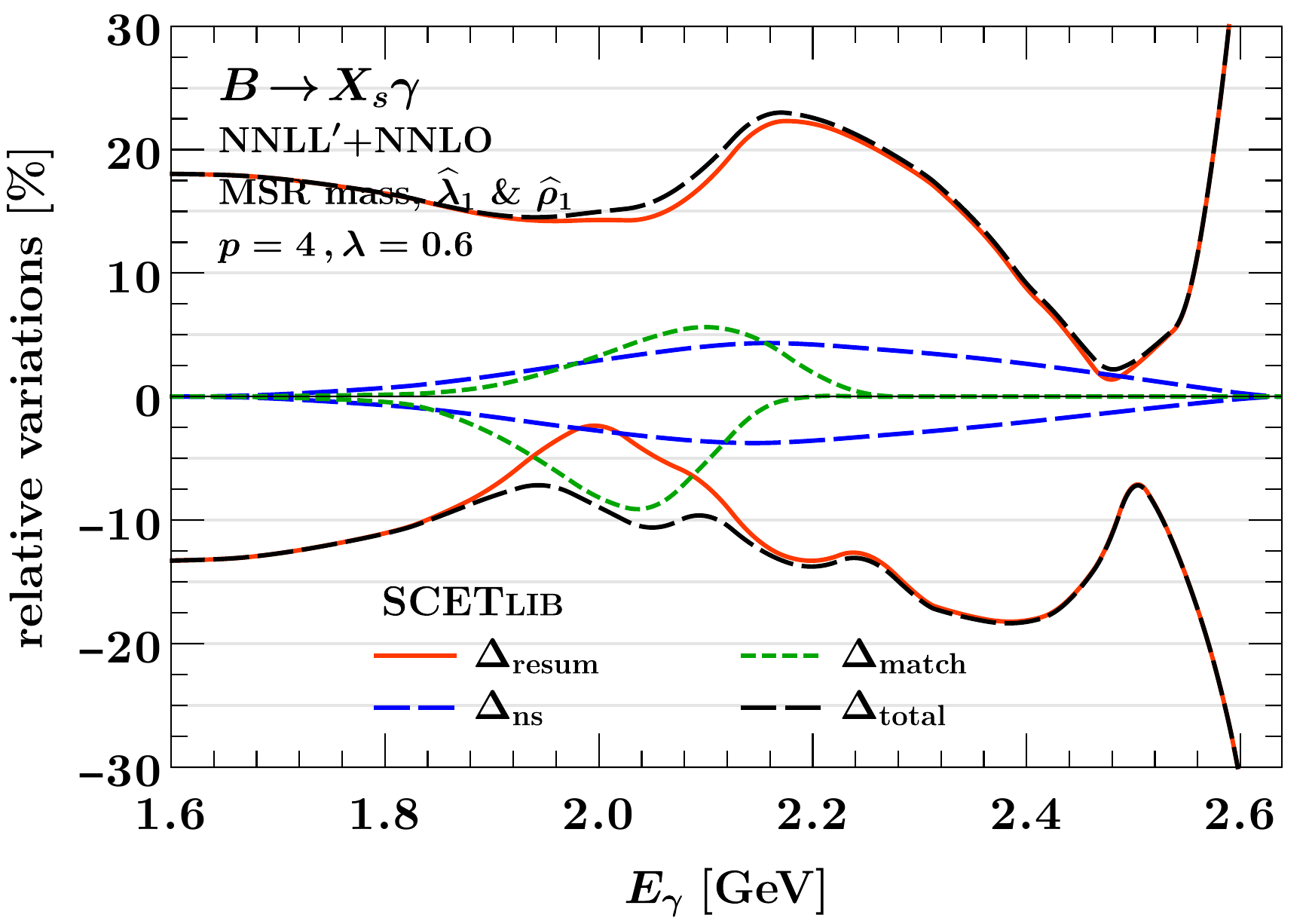}%
\hfill%
\includegraphics[width=\WidthTwoSubfigs]{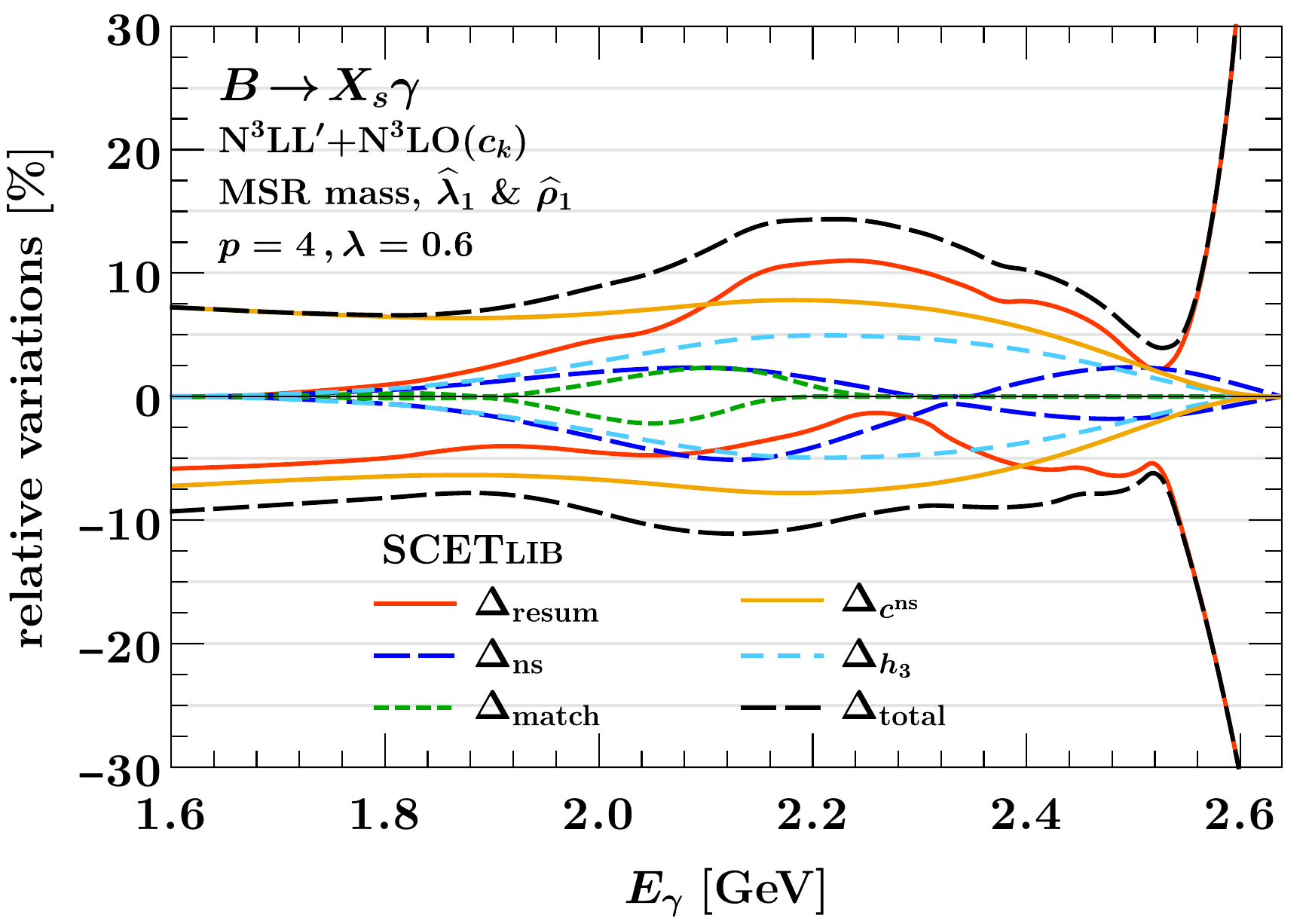}
\caption{%
Breakdown of the perturbative uncertainty for the \btosgamma spectrum into its components at \NNLLpMatched (left panel) and \NNNLLpMatched (right panel).
}
\end{figure}

Our main predictions for the \btosgamma photon energy spectrum at different
perturbative orders are presented in \fig{spectrum-main}.
In addition to the colored bands showing $\DeltaScalevar$,
the gray dashed line shows $\DeltaScalevar\oplus\DeltaHardNP$, and the black solid line shows $\Delta_{\rm total}$ defined in \eq{total-uncertainty}.
The first column shows the
results for the spectrum, and the second column shows the relative difference to
the central value at the highest order, i.e.\ to the red solid line on the left. In the first row, we use the same value for the
shape-function parameter $p = 4$ for all orders, whereas in the second row, we
use different values for $p$.
In \fig{uncertainty-components} we show the breakdown of the relative perturbative uncertainties into the individual contributions at \NNLLpMatched and \NNNLLpMatched.

We remind the reader that the choice $p=4$ was needed to ensure that the
spectrum at the highest order \NNNLLp still vanishes in the limit $E_\gamma \to
m_B/2$. In practice, when fitting to the experimental data, the fit will always
fix the precise shape near the endpoint to that of the data irrespective of the
perturbative order, while the perturbative differences get moved (at least
partially) into the fit result for $\ShapeFunction$. By using the same fixed
model for $\ShapeFunction$ at each order, we can directly assess the
perturbative convergence in the spectrum. However, by using a common $p$, the
spectrum at lower orders vanishes correspondingly faster, i.e., quadratically at
\NNLLp and cubically at \NLLp, which also affects to some extent the shape of
the spectrum into the peak. Therefore, in the lower panels of
\fig{spectrum-main} we also show an alternative order comparison, where we do
change the model at each order by using successively lower values for $p$ at the
lower orders, such that the spectrum vanishes linearly at each order.

The results in \figs{spectrum-main}{uncertainty-components} manifest a good
perturbative convergence, especially from \NNLLpMatched to \NNNLLpMatched.
The relative uncertainties are under
good control over the entire $E_\gamma$ range, except at the very endpoint
where the spectrum vanishes so the relative uncertainties necessarily blow up.
Apart from the very endpoint, the total uncertainty at \NNNLLpMatched is at
most 15\% and in most of the $E_\gamma$ range well below that.

The uncertainties at \NNNLLpMatched are substantially reduced compared to
\linebreak \NNLLpMatched, even accounting for the fact that not all \ThreeLoop
perturbative ingredients are known. As expected, the higher-order uncertainties
estimated from profile scale variations ($\DeltaResummation$,
$\DeltaNonsingular$, $\DeltaMatching$) are much reduced. Also recall that in the
fixed-order tail $\DeltaResummation$ turns into the usual fixed-order scale
variation. The uncertainty
$\DeltaHardNP$ is visible but subdominant. Across the entire peak of the
spectrum, the uncertainty $\DeltaNonsingularNP$ from the missing \ThreeLoop
nonsingular corrections is at most comparable to the other sources thanks to the power suppression of the nonsingular. As expected,
in the tail below $E_\gamma\lesssim E_1 = 2.1\GeV$ it starts to take over and
becomes the dominant uncertainty. This demonstrates that in our approach
we are able to substantially benefit from the increased precision of the
N$^3$LL$'$ resummation even in the absence of the full N$^3$LO result.
Furthermore, in the fixed-order tail
the total uncertainty dominated by $\DeltaNonsingularNP$ is still reduced
compared to the scale-variation based estimate at \NNLLpMatched.
As discussed at the end of \sec{nonsingular}, this is anticipated and justified
because of the additional nontrivial perturbative information included at
N$^3$LO$(c_k)$.

\begin{figure}[t!]\label{fig:model-variation-MSR}
\centering
\includegraphics[width=\WidthTwoSubfigs]{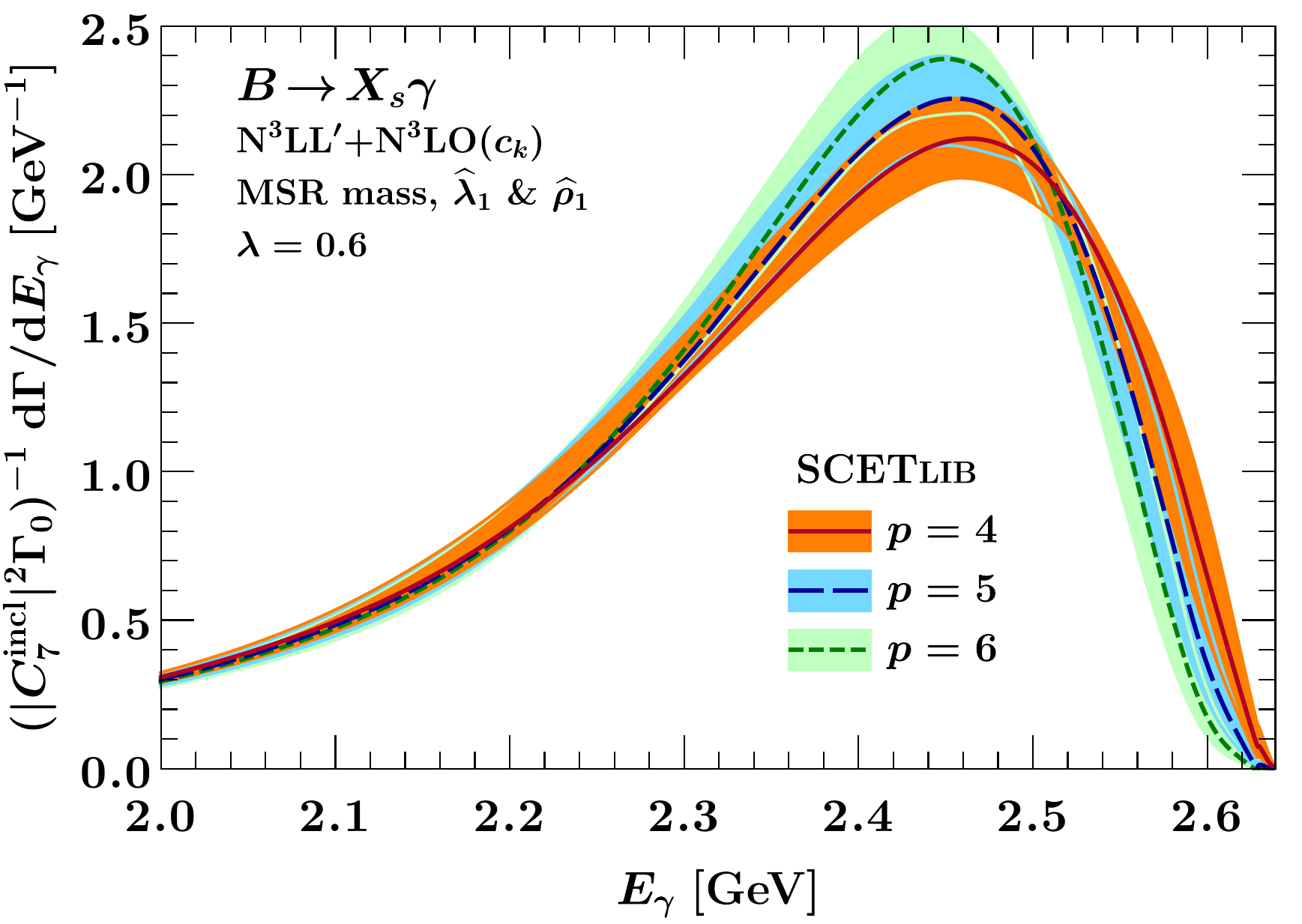}
\includegraphics[width=\WidthTwoSubfigs]{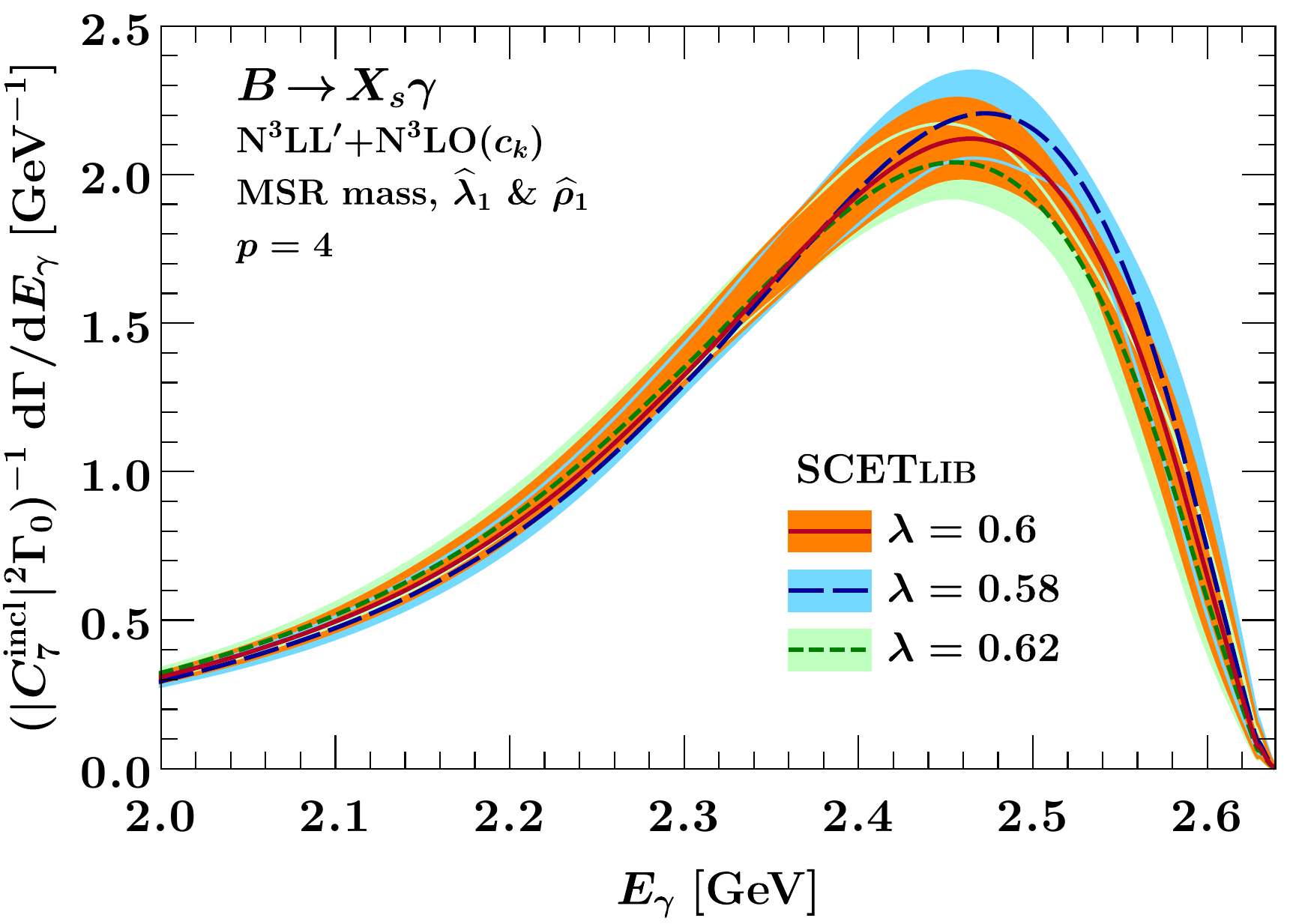}
\caption{%
The \btosgamma spectrum using different values for the shape-function parameters $p$ (left panel) and $\lambda$ (right panel) at \NNNLLpMatched.
 The remaining theory parameters are set to their default values.
}
\end{figure}

The effect of varying the shape function model parameters $p$ and $\lambda$ is illustrated in \fig{model-variation-MSR}. As expected, the position and height of the peak depend on the model parameters.
We can clearly see how the value of $p$ controls how fast the spectrum vanishes toward the endpoint $E_{\gamma}\to m_B/2$.
The value of $\lambda$ determines the width of shape function, which as a result controls the width of the peak.
We stress that these results are just meant as an additional illustration.
In particular, the differences seen in \fig{model-variation-MSR} are not to be interpreted as an additional theoretical uncertainty in the predictions.
As mentioned before, the actual shape of $\ShapeFunction$ will be determined by fitting to the experimental data.

\subsection{Different treatments of higher-order singular cross terms}
\label{sec:discussion-perturbative-treatment}

When evaluating the singular correction using the factorization theorem in \eq{w77s} by default we expand the fixed-order boundary conditions of the hard, jet, and soft functions against each other. This is usually done, because it ensures that when switching off the resummation in the far-tail region
and adding the fixed-order nonsingular correction, the fixed-order results are exactly reproduced.
In \refcite{Bernlochner:2020jlt} it was observed that in the \OneS scheme and up to
NNLL$'+$NNLO the perturbative convergence in the resummation region is substantially improved by keeping the hard function unexpanded as an overall factor. The main reason is that this ensures that the hard function does not affect the shape of the spectrum,
which receives relatively large corrections from the jet and soft functions.
The disadvantage is that this has the danger of generating unphysically large higher-order corrections in the fixed-order limit. In \refcite{Bernlochner:2020jlt} it
was checked that this does not happen in the region of interest.

In this section, we study the differences in our perturbative setup between
expanding the hard function against soft and jet functions (our default) vs.
keeping it unexpanded.

\begin{figure}[t!]
\centering
\includegraphics[width=0.484\textwidth]{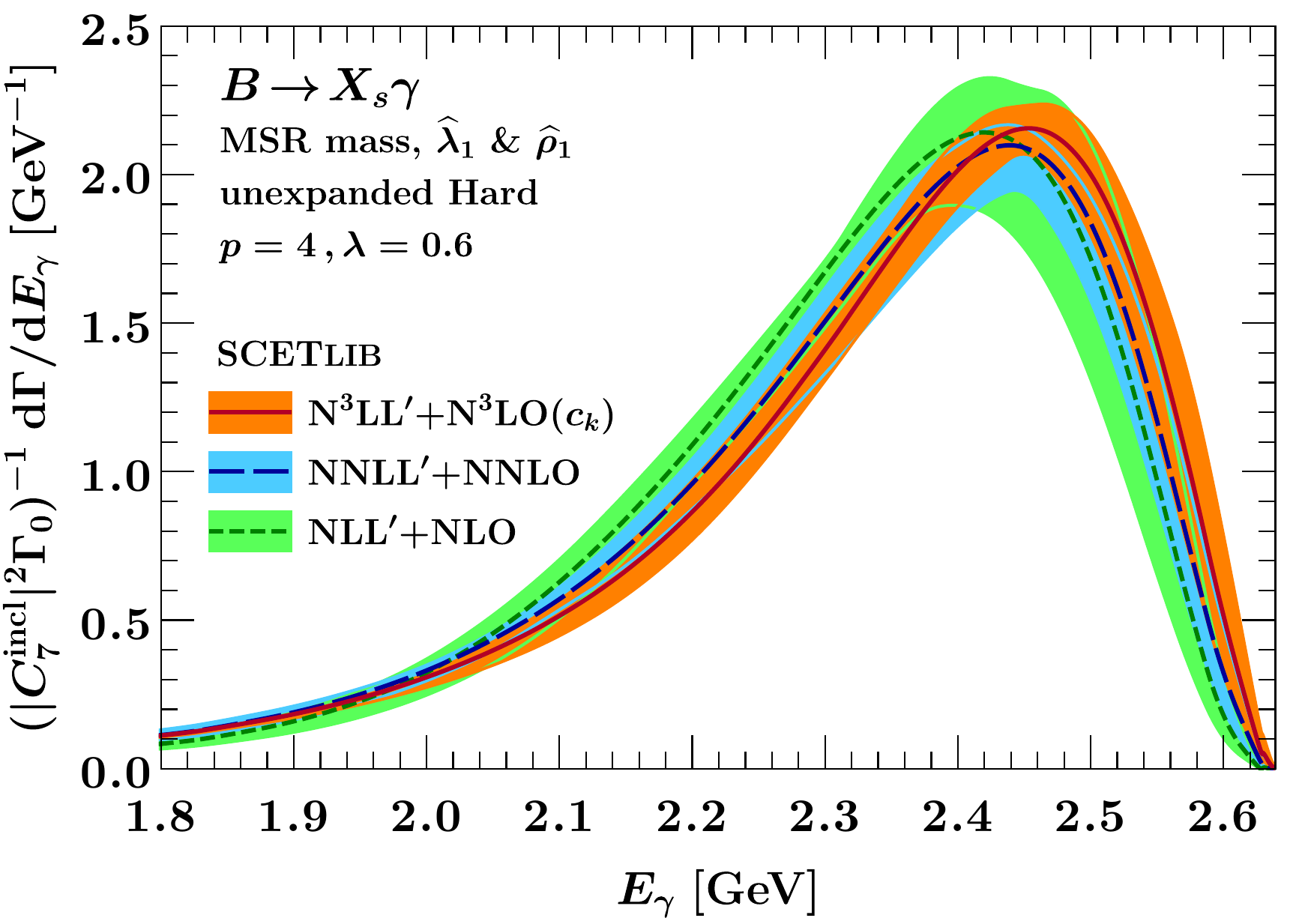}
\includegraphics[width=0.50\textwidth]{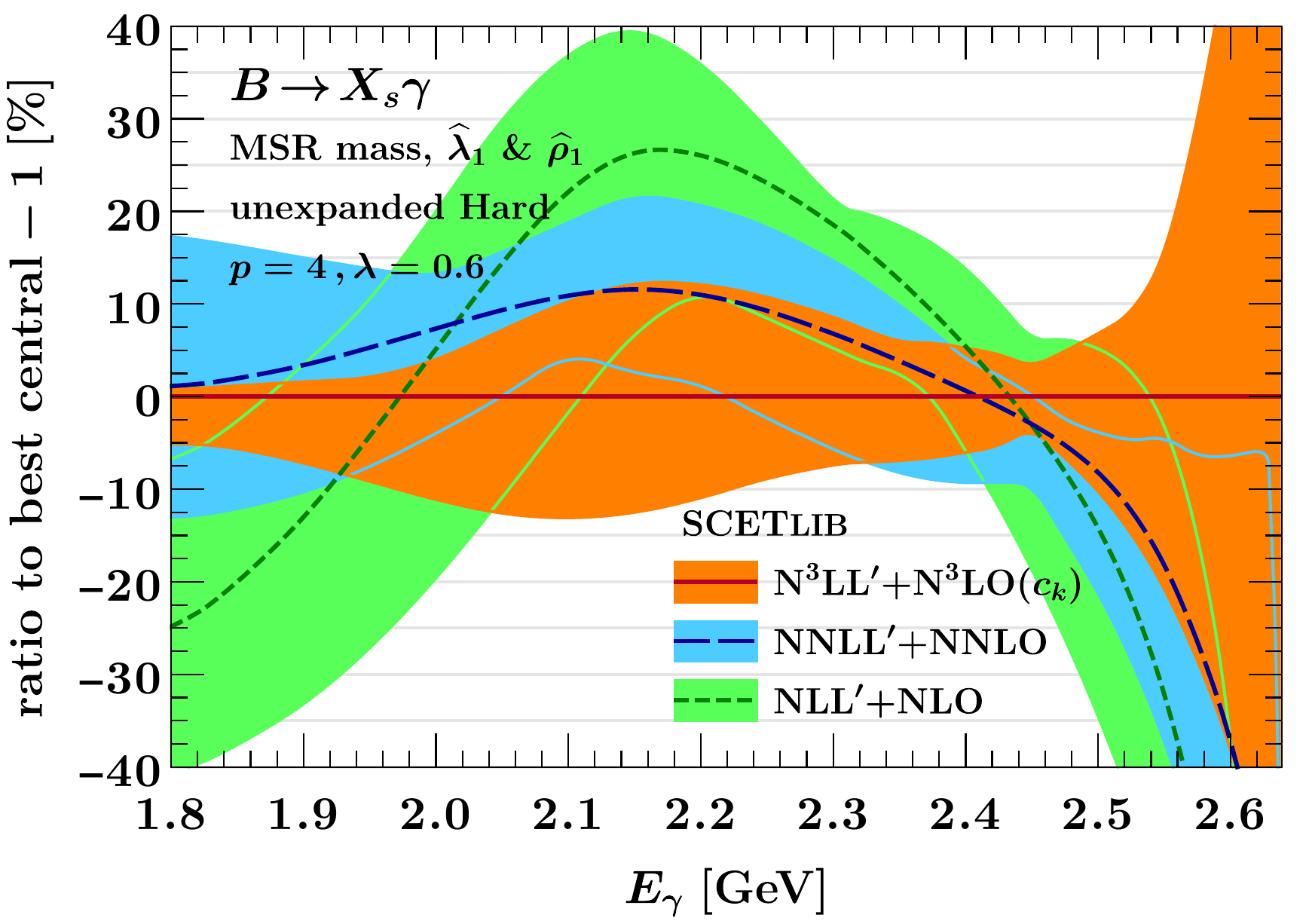}
\caption{\btosgamma spectrum with unexpanded hard function.}\label{fig:decay-rate-spectrum-unexpandedH}
\end{figure}

A priori, the additional higher-order terms induced by keeping the hard
function unexpanded can easily spoil the delicate cancellation between singular and
nonsingular contributions in the far tail of the spectrum.
To fix this problem, we compensate for these higher-order terms by adding a
constant term that cancels them in the fixed-order region but is only a power-suppressed correction in the peak region,
\begin{equation}\label{eq:spectrum-unexpanded-hard}
  \Bigl[W_{77}^{\rm s}(k) + W_{77}^{\rm ns}(k) \Bigr] \to
  \Bigl[W_{77}^{\rm s,\unexpandedHard}(k) - W_{77, {\rm FO}}^{\rm s,\unexpandedHard}(\mb)\Bigr] + \Bigl[W_{77}^{\rm ns}(k) + W_{77, {\rm FO}}^{\rm s,\expandedHard}(\mb) \Bigr]
\,,\end{equation}
where we subtract the singular with unexpanded hard function, $W_{77, {\rm FO}}^{\rm s,\unexpandedHard}(\mb)$ and add it back with expanded out hard function, $W_{77, {\rm FO}}^{\rm s,\expandedHard}(\mb)$.
Both of these terms are evaluated at fixed order and at $k=\mb$, corresponding to the tail limit $x=1$, so in the peak region they are indeed power suppressed.
This prescription allows us to take advantage of keeping the hard function unexpanded in the peak region while avoiding unphysically large corrections from higher-order cross terms in the fixed-order region, since they are explicitly removed in the $x\to 1$ limit.

\begin{figure}[t!]
\centering
\includegraphics[width=0.488\textwidth]{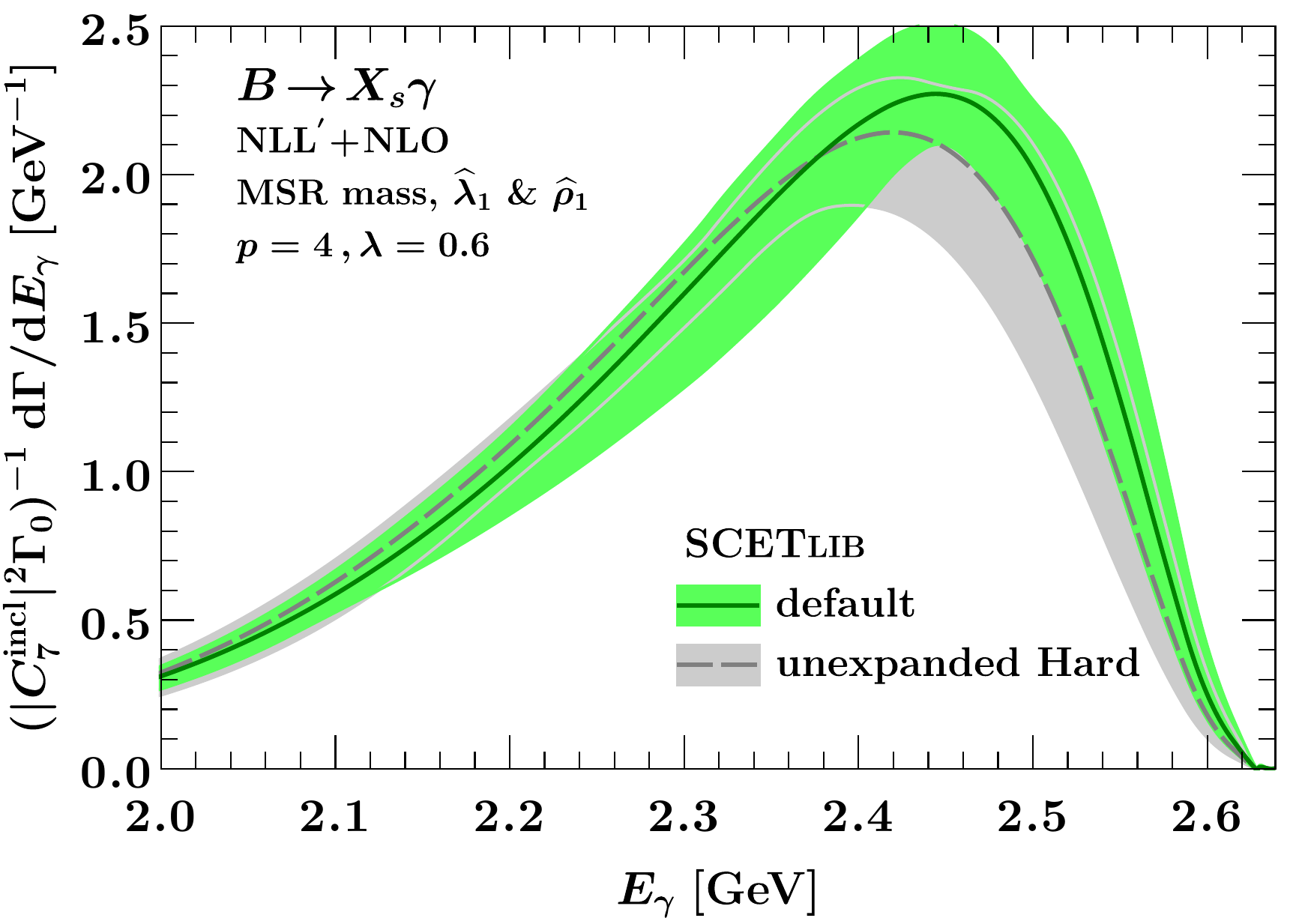}
\includegraphics[width=0.50\textwidth]{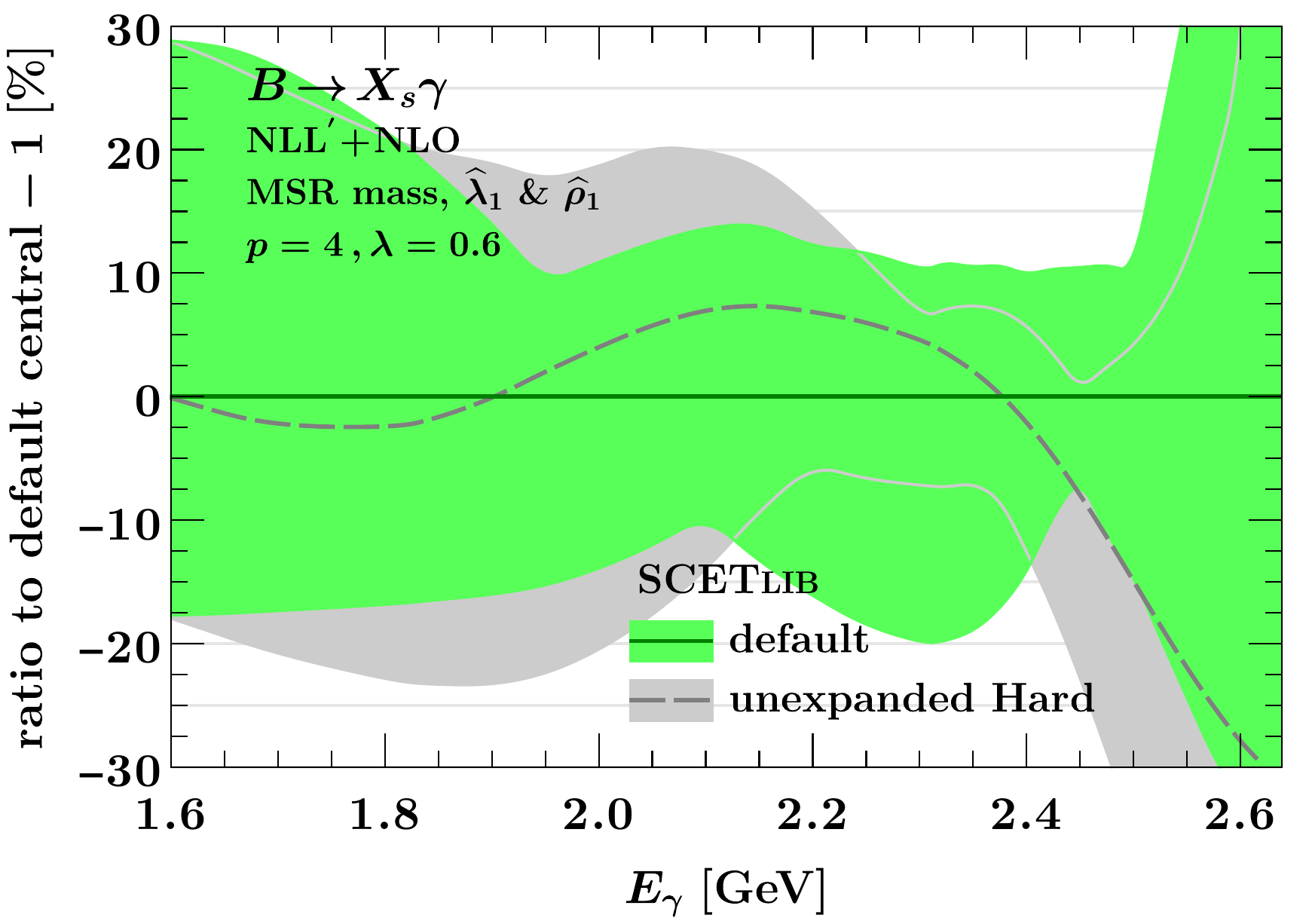}
\\
\includegraphics[width=0.488\textwidth]{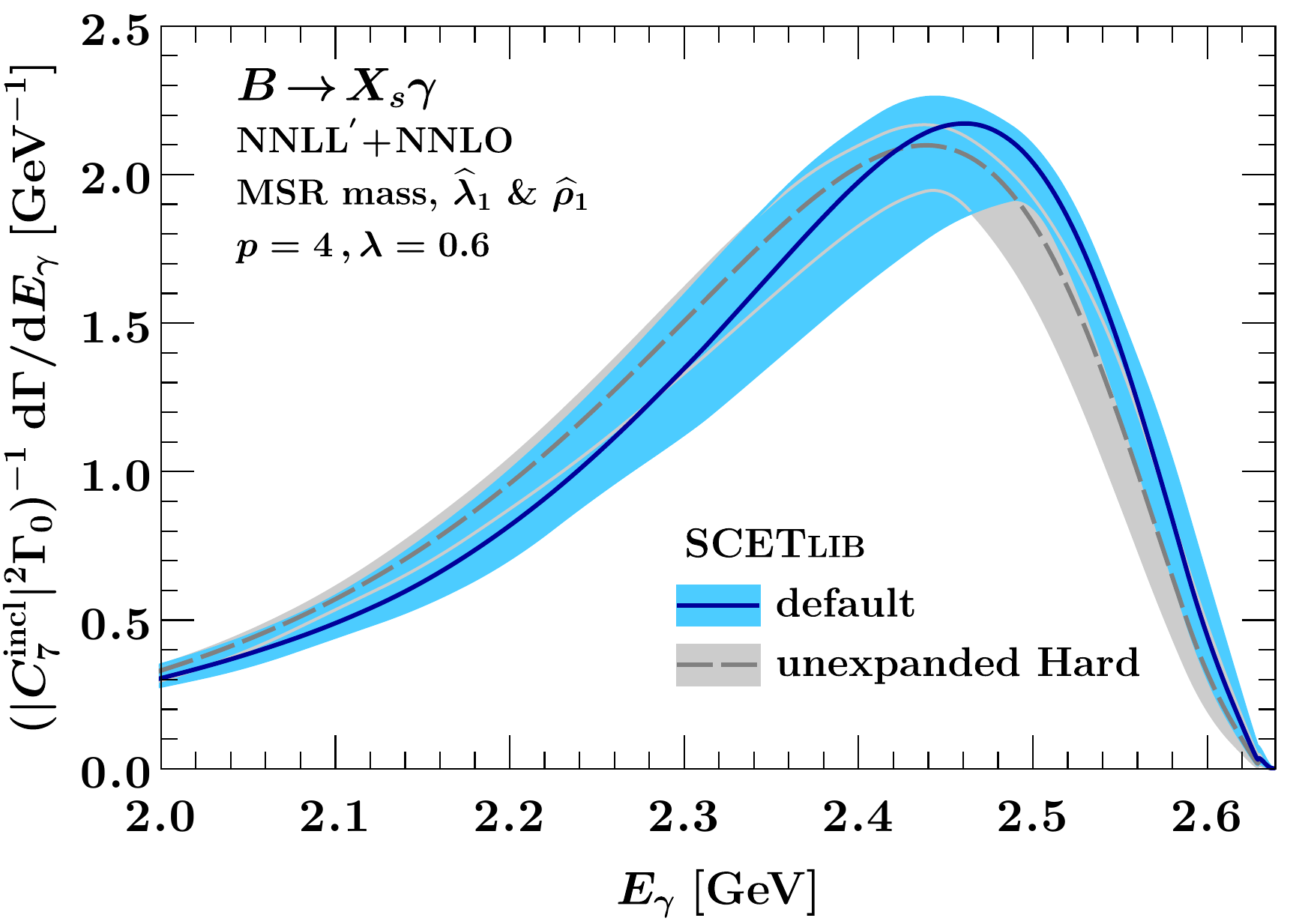}
\includegraphics[width=0.50\textwidth]{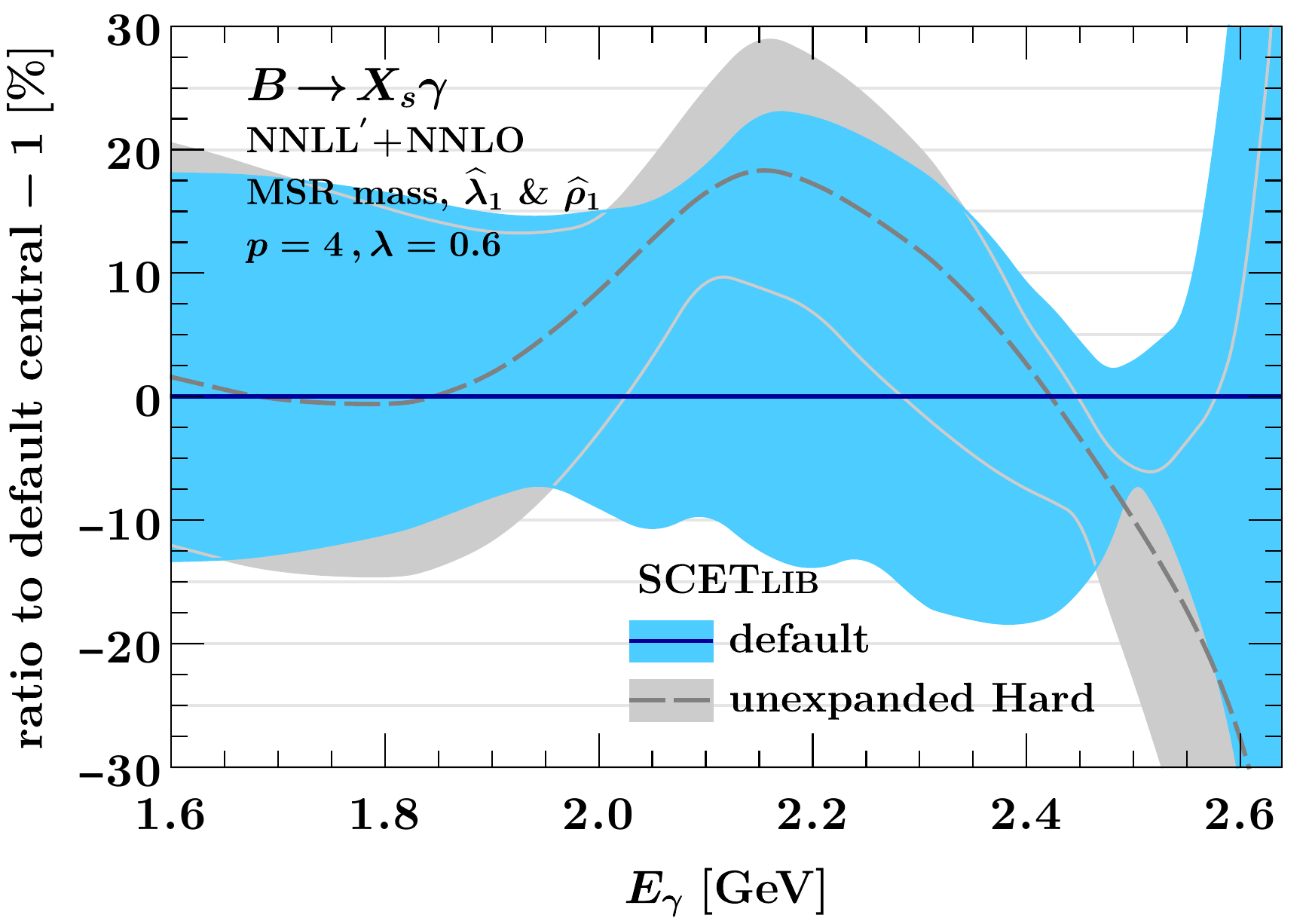}
\\
\includegraphics[width=0.488\textwidth]{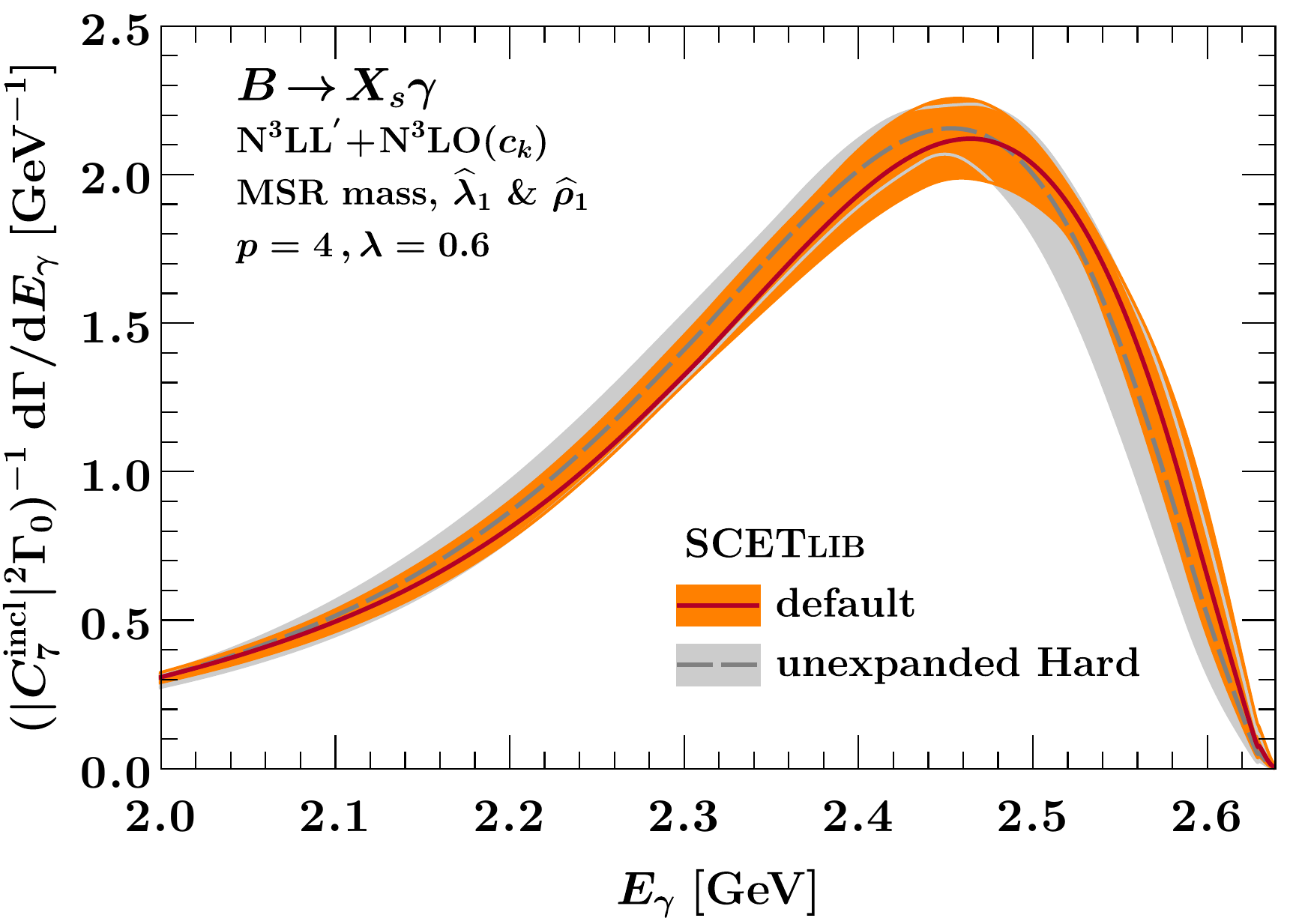}
\includegraphics[width=0.50\textwidth]{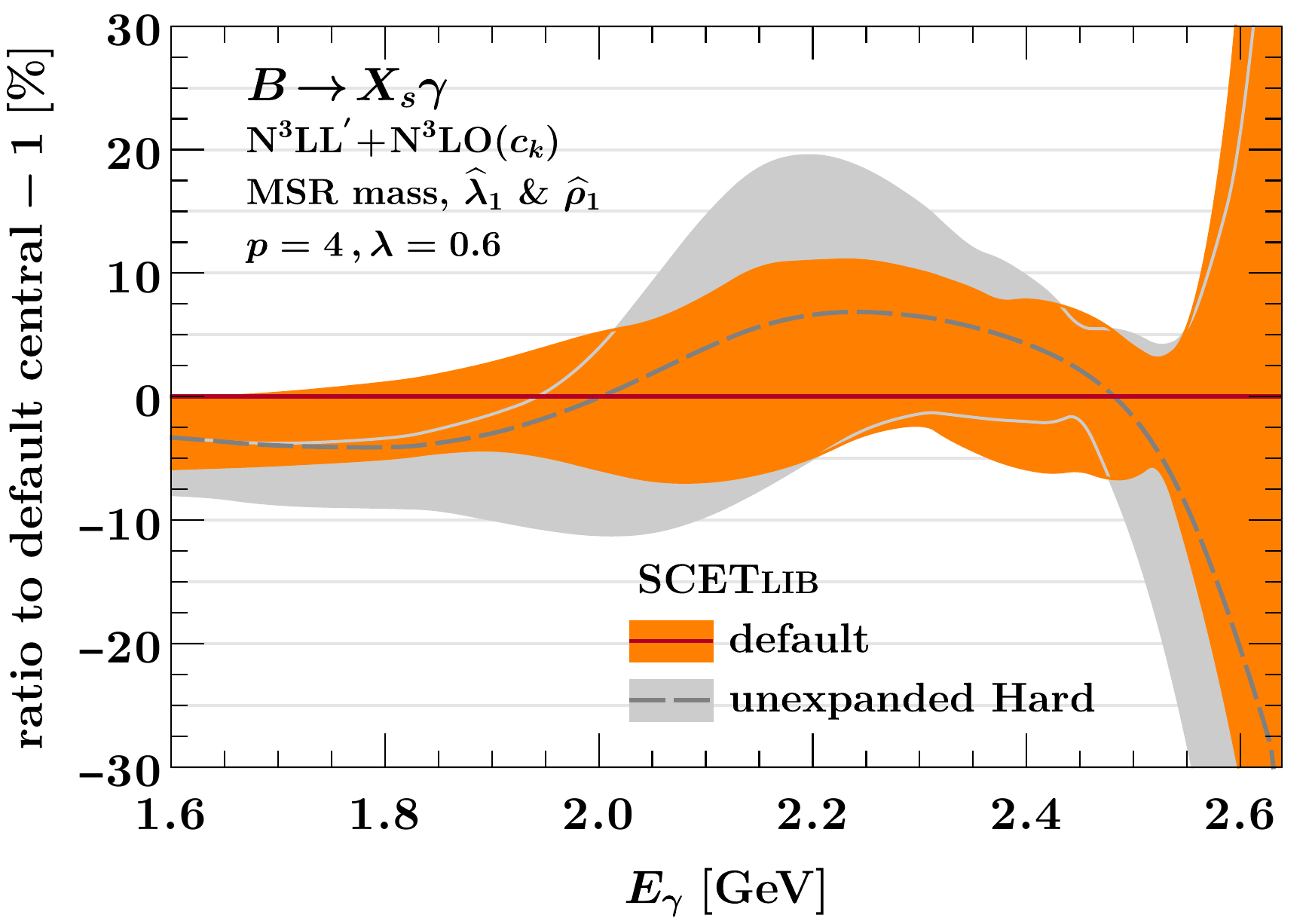}
\caption{%
  Comparison of \btosgamma spectrum with expanded and unexpanded hard function.
  The colored and gray bands at each order display the spectrum with expanded and unexpanded hard function, respectively.
}\label{fig:decay-rate-spectrum-total-vs-expand-H}
\end{figure}

The numerical results using this prescription are shown in \fig{decay-rate-spectrum-unexpandedH}, where we adopt the MSR scheme with short-distance schemes for $\widehat\lambda_1$ and $\widehat\rho_1$.
We compare these results to our default setup using the expanded hard function in \fig{decay-rate-spectrum-total-vs-expand-H}.
Here one can see that, overall, both scenarios lead to somewhat compatible results.
Nevertheless, the choice of expanding the hard function or not clearly has
a large impact at lower orders (which are expected to be more sensitive to the treatment of higher-order cross terms). Keeping the hard function unexpanded
leads to a rather unnatural reduction of scale variations in the peak region of the spectrum.
This behavior is quite dramatic at \NNLLpMatched, where the uncertainty band with
an unexpanded hard function (the gray band) barely captures the central line
from the results with expanded hard function (solid blue line), and its size
is almost as large as the scale variations at \NNNLLpMatched. This is also
visible in \fig{decay-rate-spectrum-unexpandedH}, where the blue band suspiciously
narrows down in the peak region of the spectrum and competes with the orange band.
Based on this set of observations we conclude that in our setup keeping the hard
function unexpanded is disfavored.
Hence, for our final numerical results we opt for the conventional treatment of
fully expanding the fixed-order boundary conditions of the hard, jet, and soft functions against each other, since it yields an overall more consistent picture
of perturbative uncertainties and convergence.

\subsection{Impact of short-distance schemes: \OneS vs. MSR mass schemes}\label{sec:discussion-impact-short-distance}

As already explained in \sec{short-distance-schemes} the right choice of short-distance scheme for the $b$-quark mass plays a key role in stabilizing the predictions at different orders in perturbation theory.
In this section we illustrate the numerical impact of expressing $m_b$, $\lambda_1$, and $\rho_1$ in short-distance schemes.
Moreover, we show that the \OneS mass scheme, which was successfully used in
previous works~\cite{Bernlochner:2020jlt, Ligeti:2008ac} up to NNLL$'+$NNLO,
starts to break down at \NNNLLp, while the MSR mass scheme yields convergent, stable results.

\begin{figure}[t!]
\centering
\includegraphics[width=0.5\textwidth]{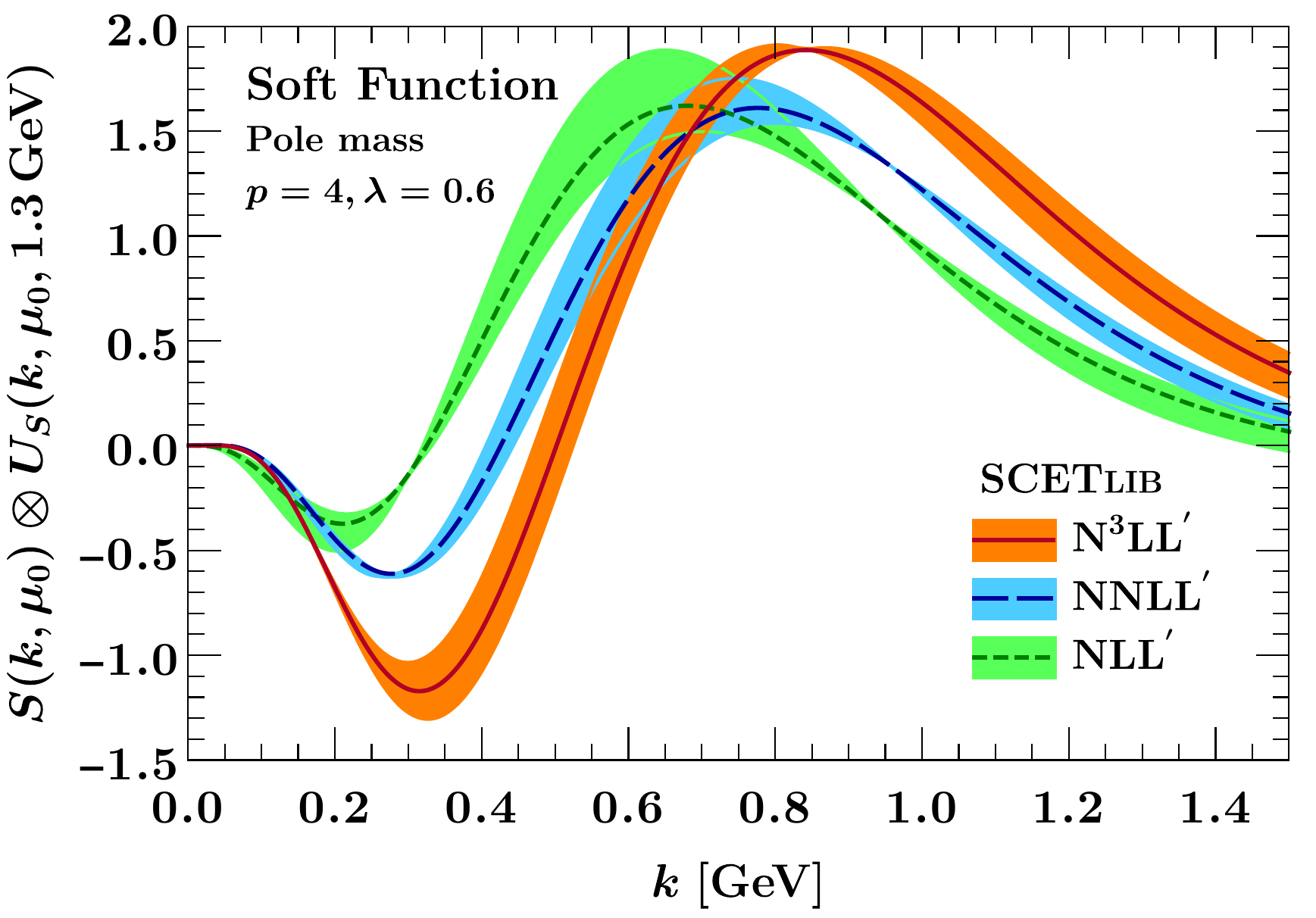}
\includegraphics[width=0.484\textwidth]{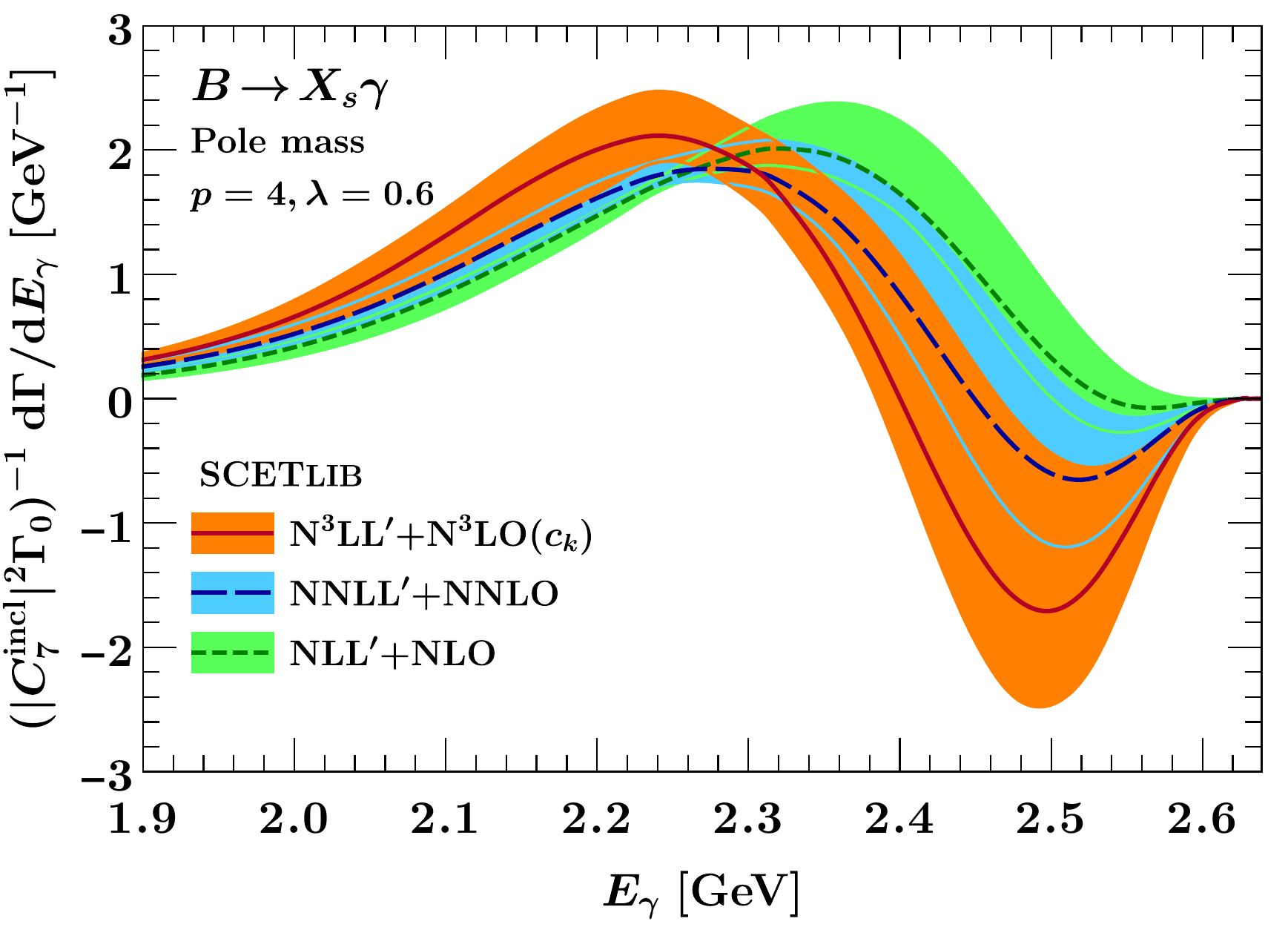}
\caption{%
  The soft function (left panel) and the \btosgamma spectrum (right panel) at different orders in the pole mass scheme.
  Small values of $k$ in the soft function correspond to large $E_\gamma$ in the spectrum.
  The pole-mass renormalon leads to the large negative dip at small $k$ and large $E_\gamma$ and an unstable position of the peak.
}
\label{fig:discussion-pole}
\end{figure}

\begin{figure}[t!]
\centering
\includegraphics[width=0.5\textwidth]{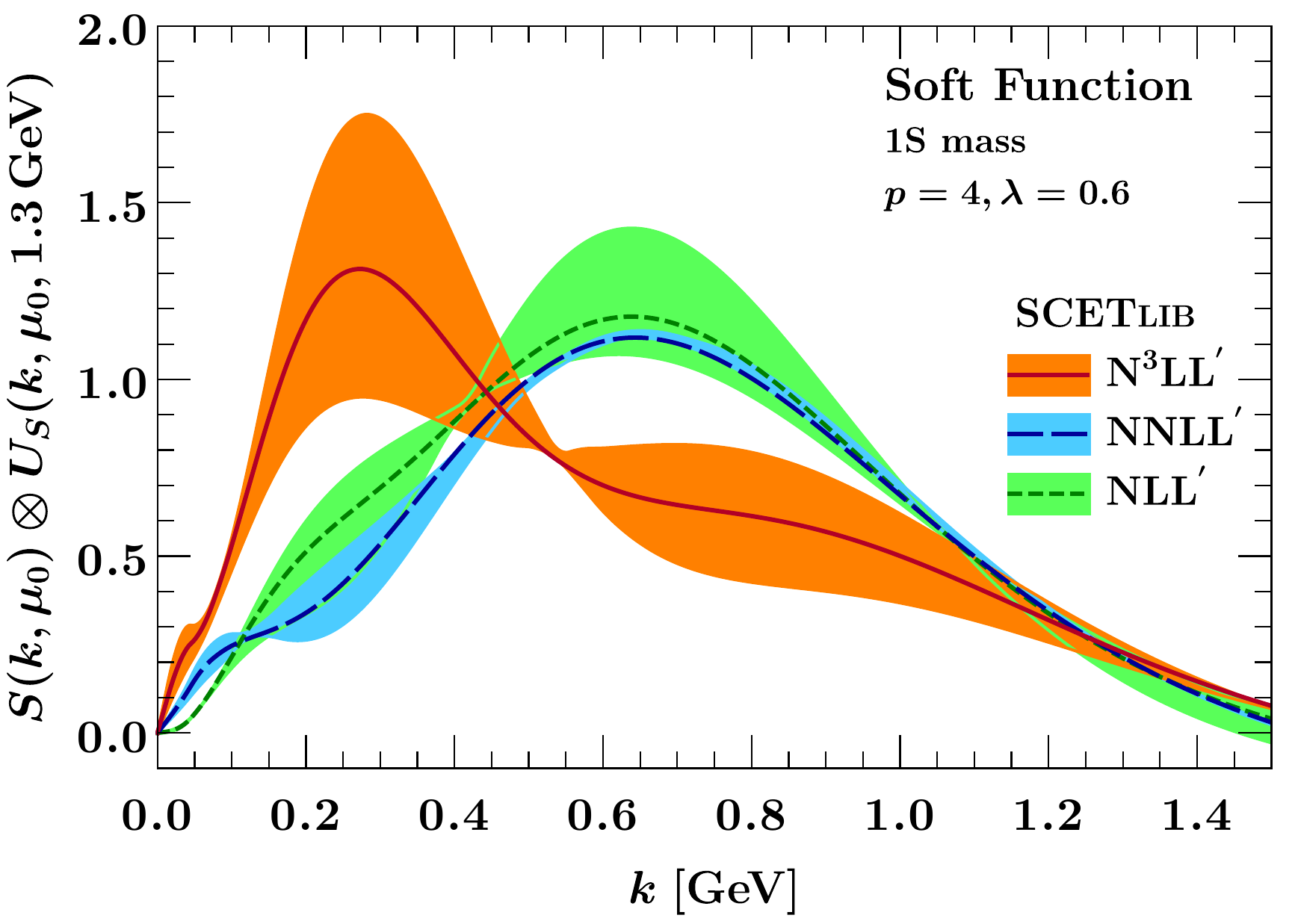}
\includegraphics[width=0.484\textwidth]{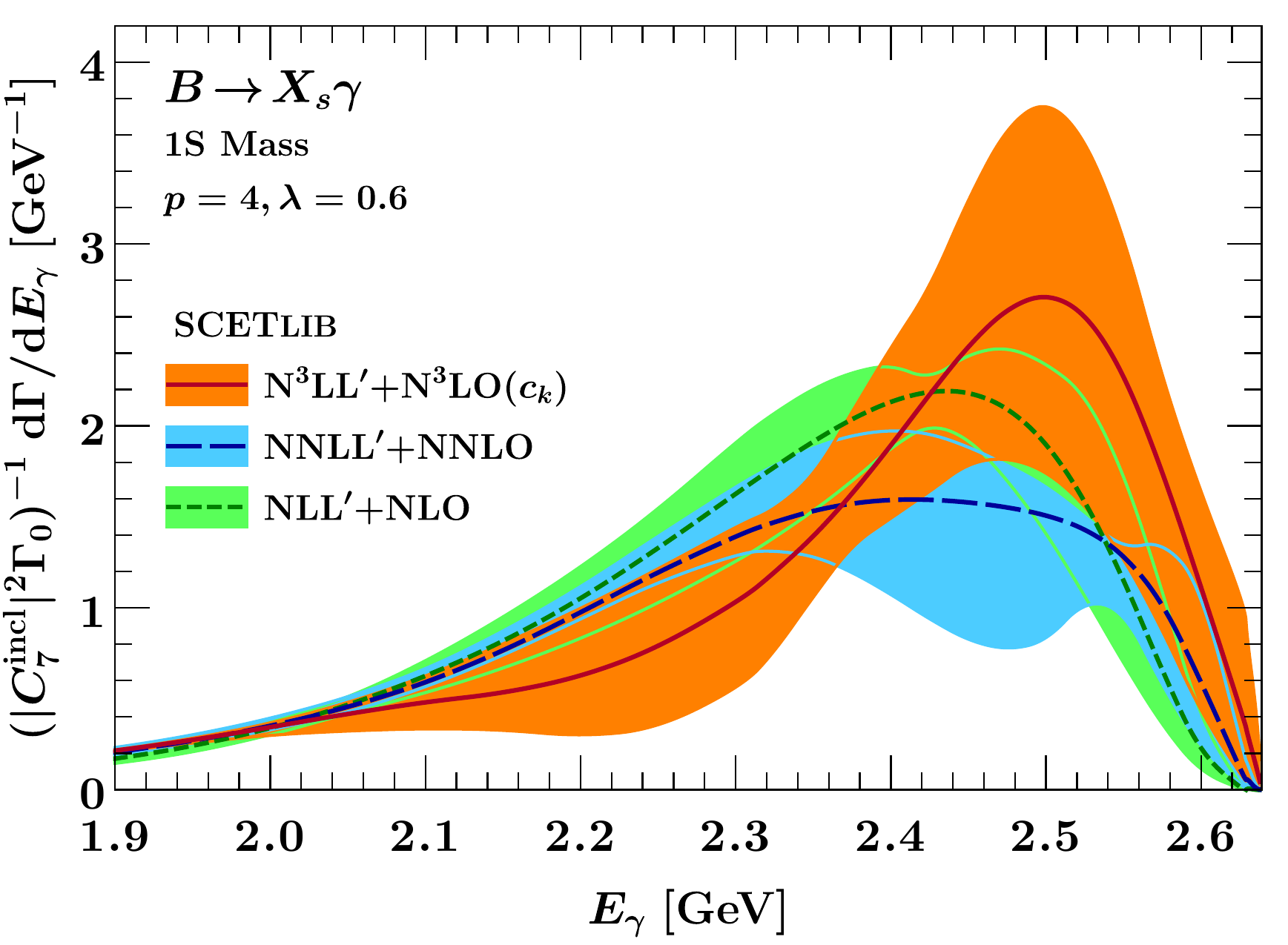}
\\
\includegraphics[width=0.5\textwidth]{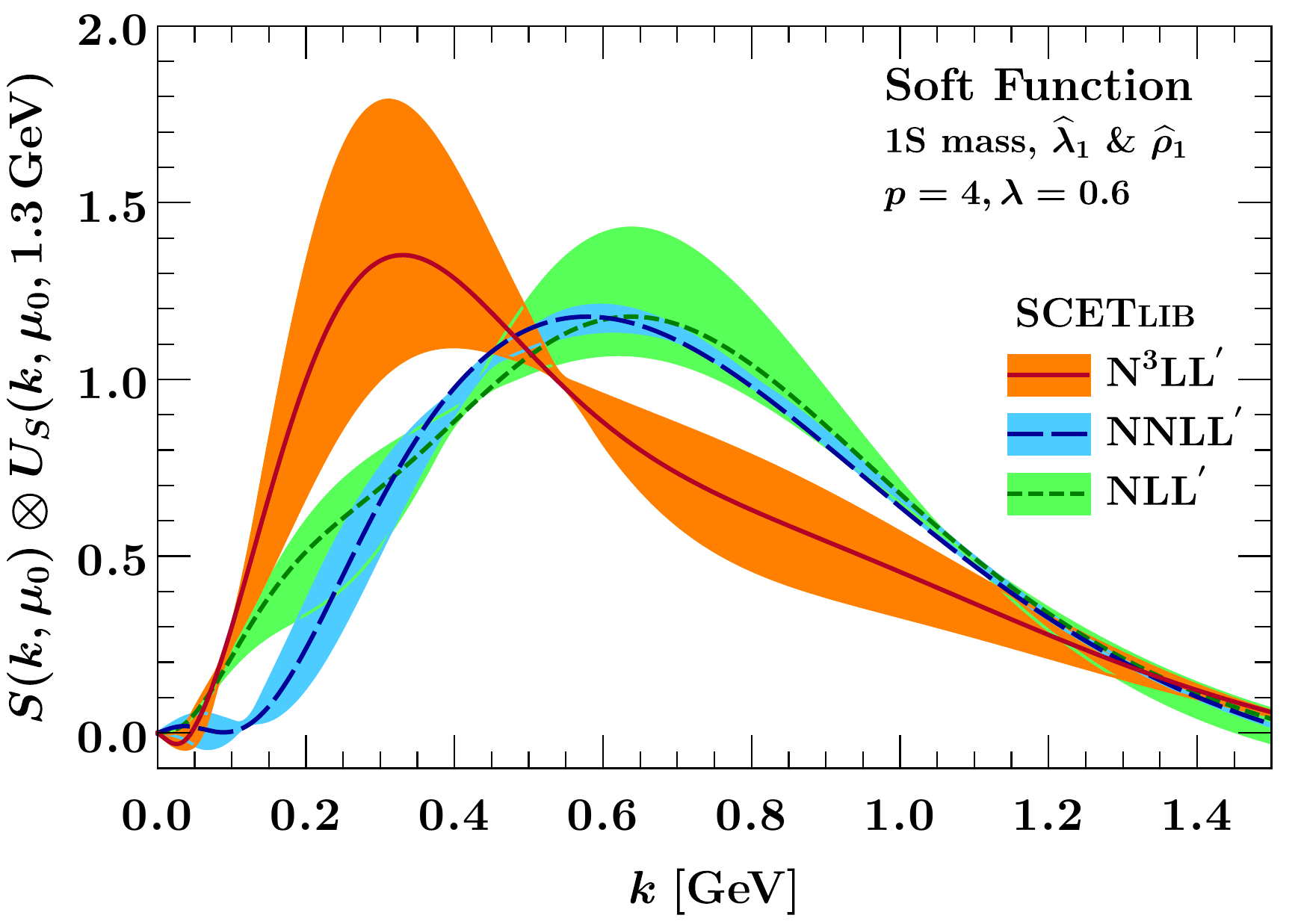}
\includegraphics[width=0.484\textwidth]{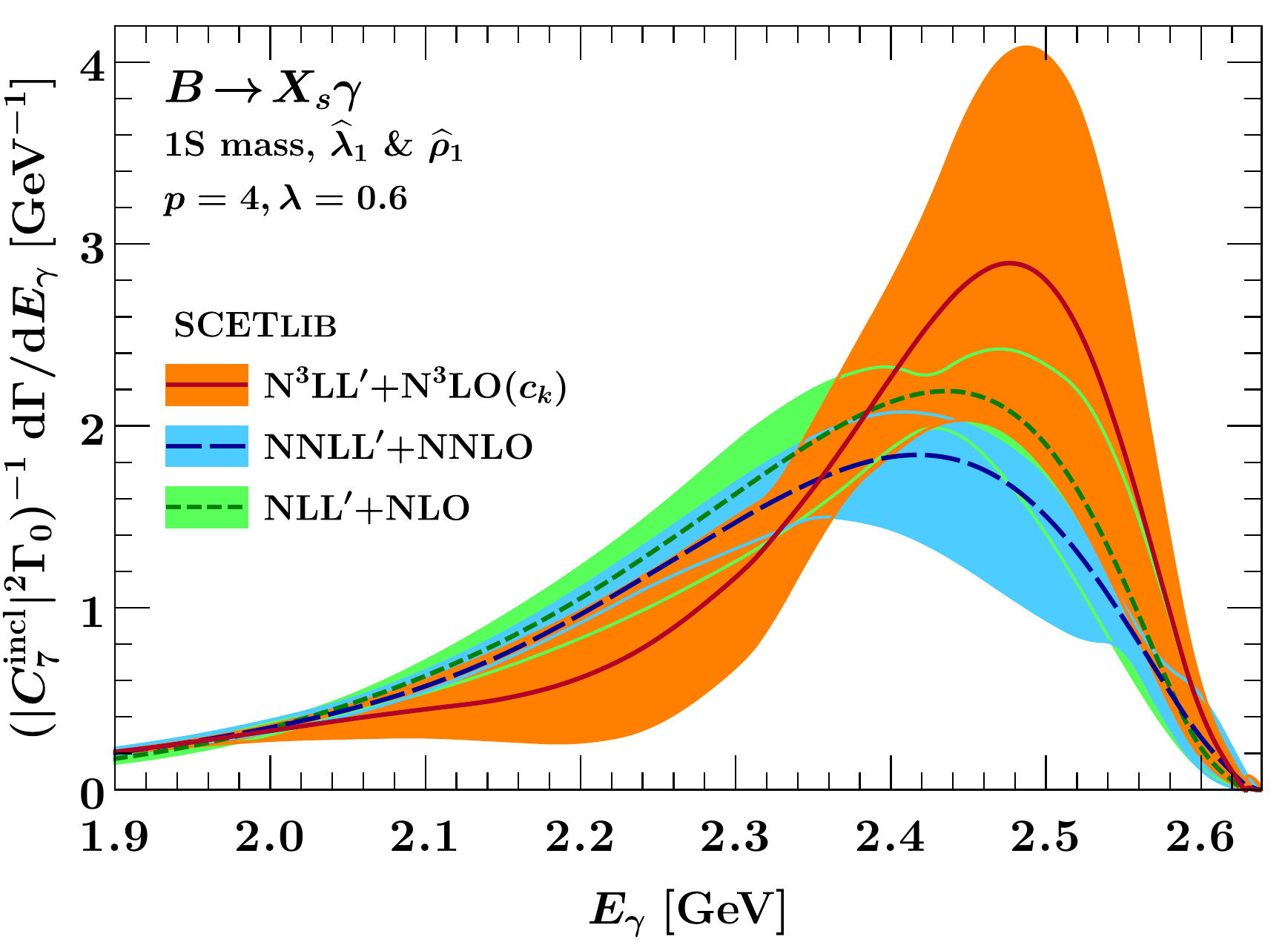}
\caption{%
  The soft function (left panel) and the \btosgamma spectrum (right panel) in the \OneS mass scheme.
  The top panels use the pole scheme for $\lambda_1$ and $\rho_1$, while the bottom panels use a short-distance scheme.
  In both cases the spectrum, and in particular the position of its peak, are stable at \NLLpMatched and \NNLLpMatched, but at \NNNLLpMatched the predictions start breaking down.
}
\label{fig:discussion-1S}
\end{figure}

\begin{figure}[t!]
\centering
\includegraphics[width=0.49\textwidth]{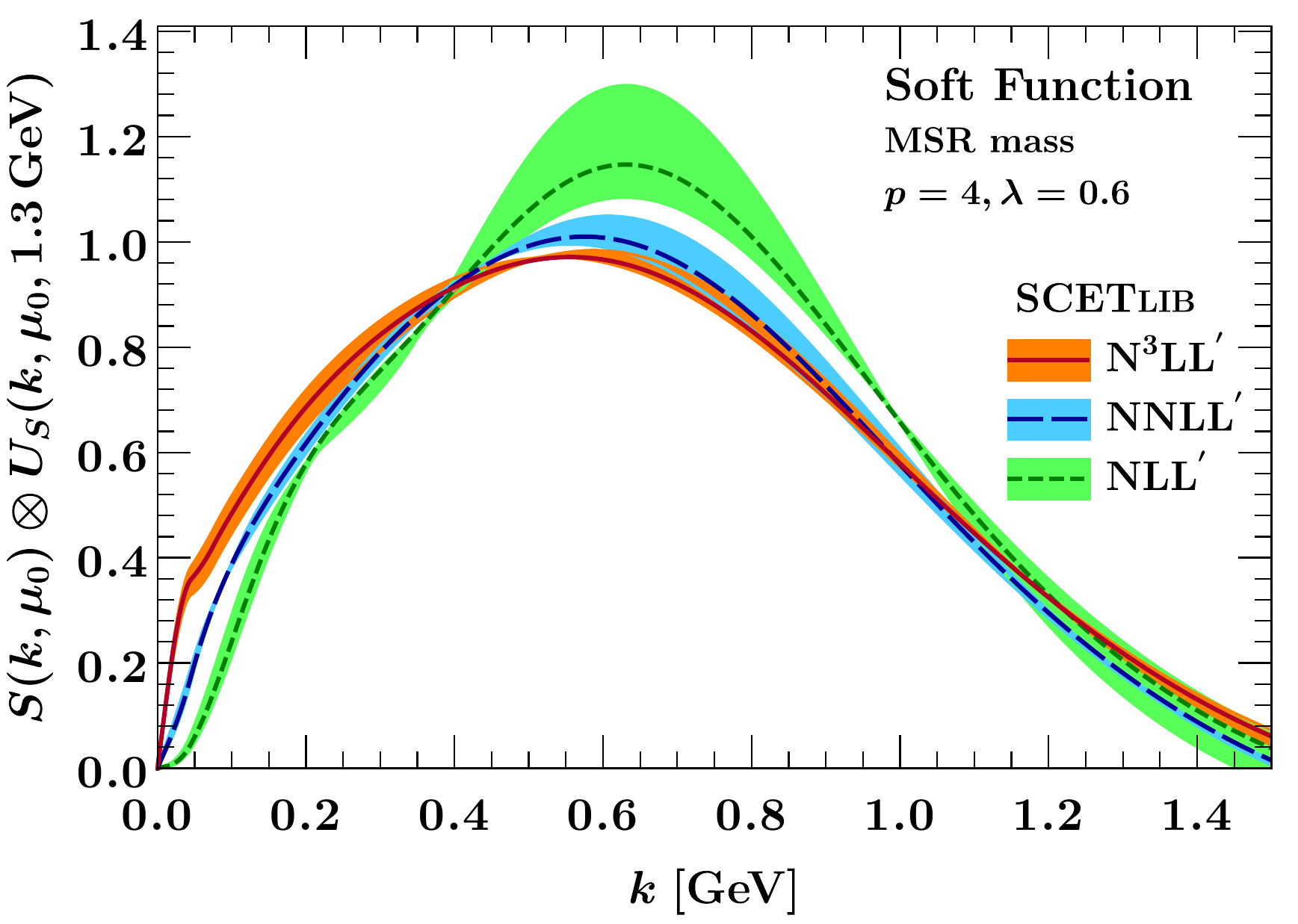}
\includegraphics[width=0.49\textwidth]{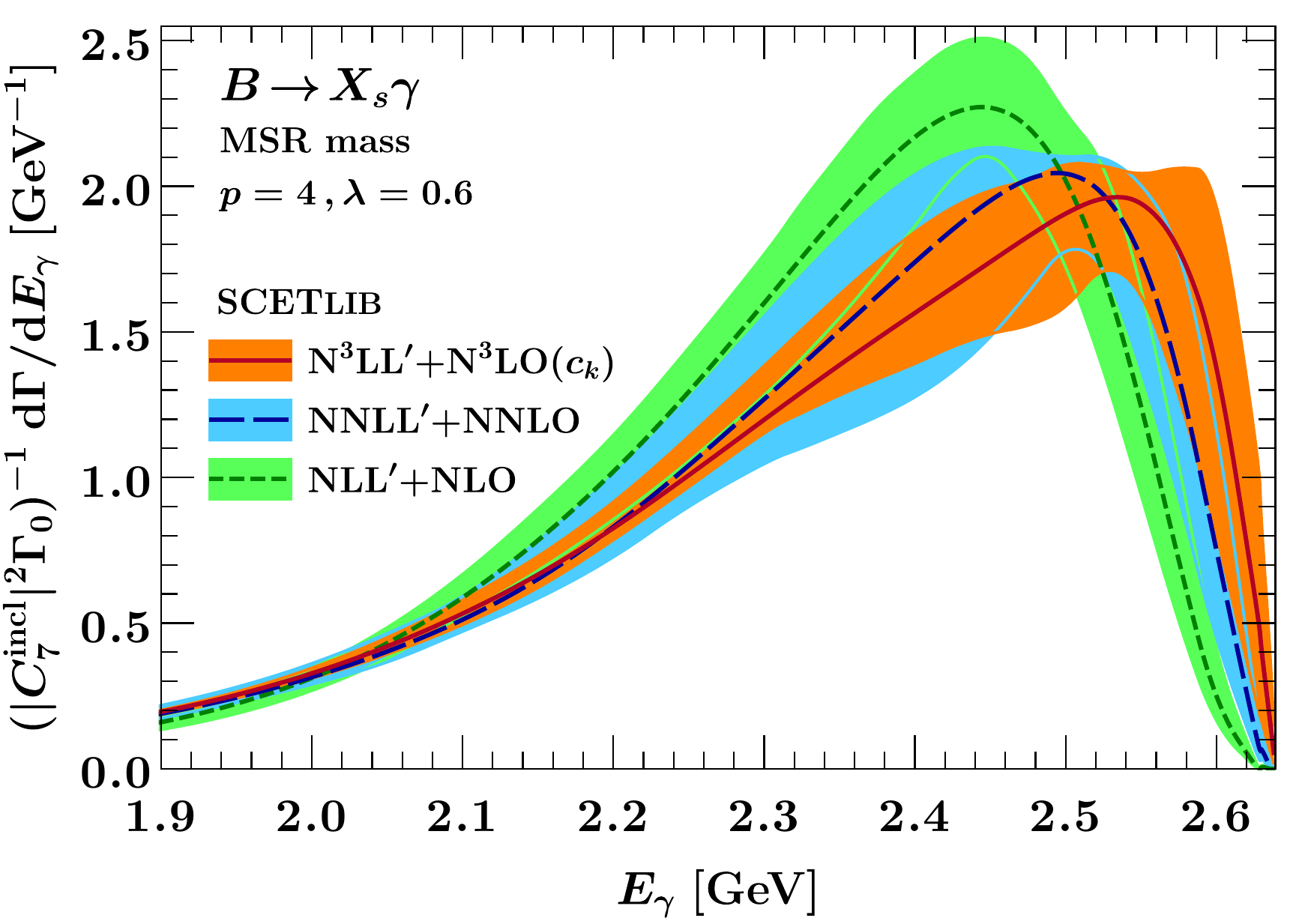}
\\
\includegraphics[width=0.49\textwidth]{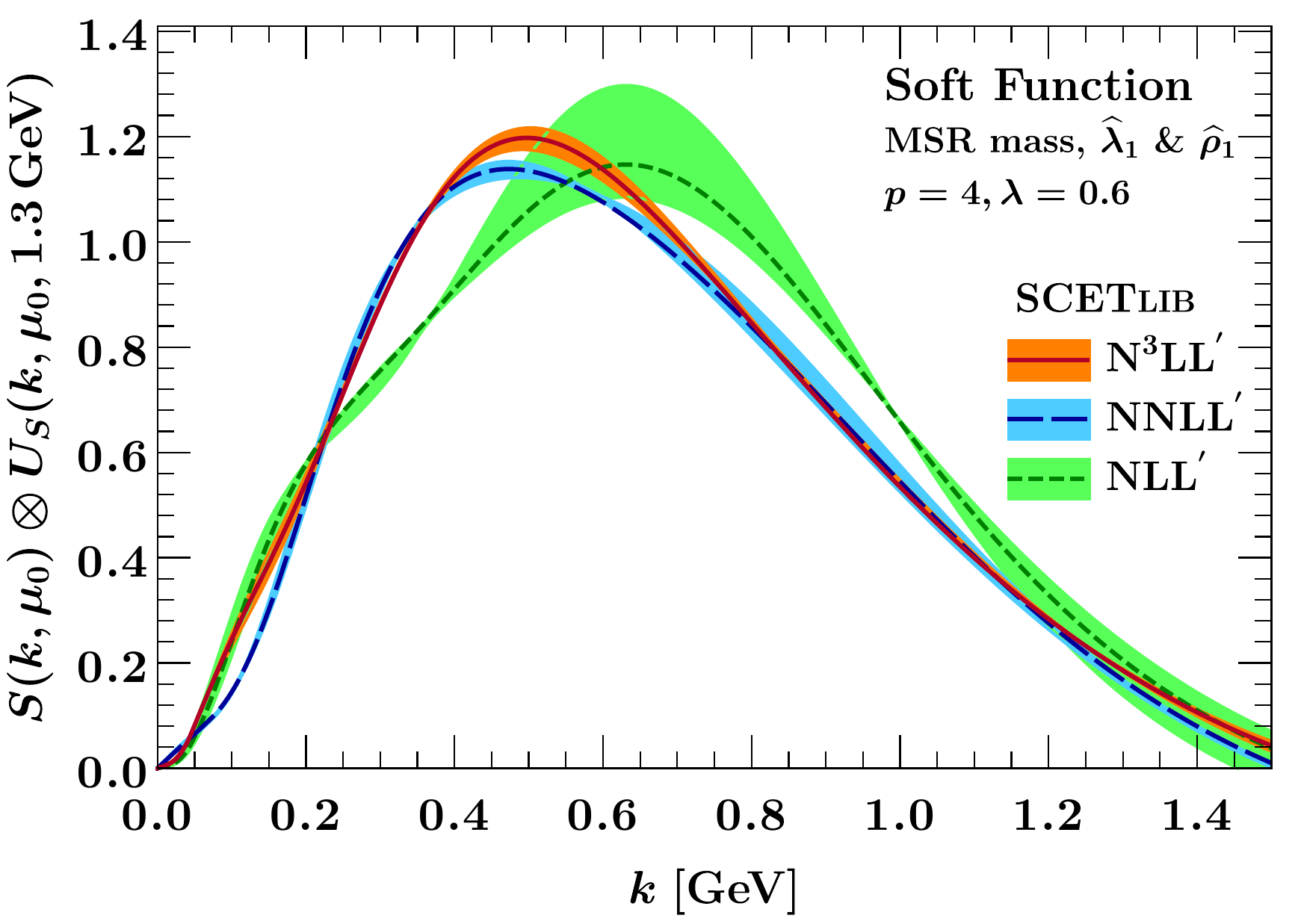}
\includegraphics[width=0.49\textwidth]{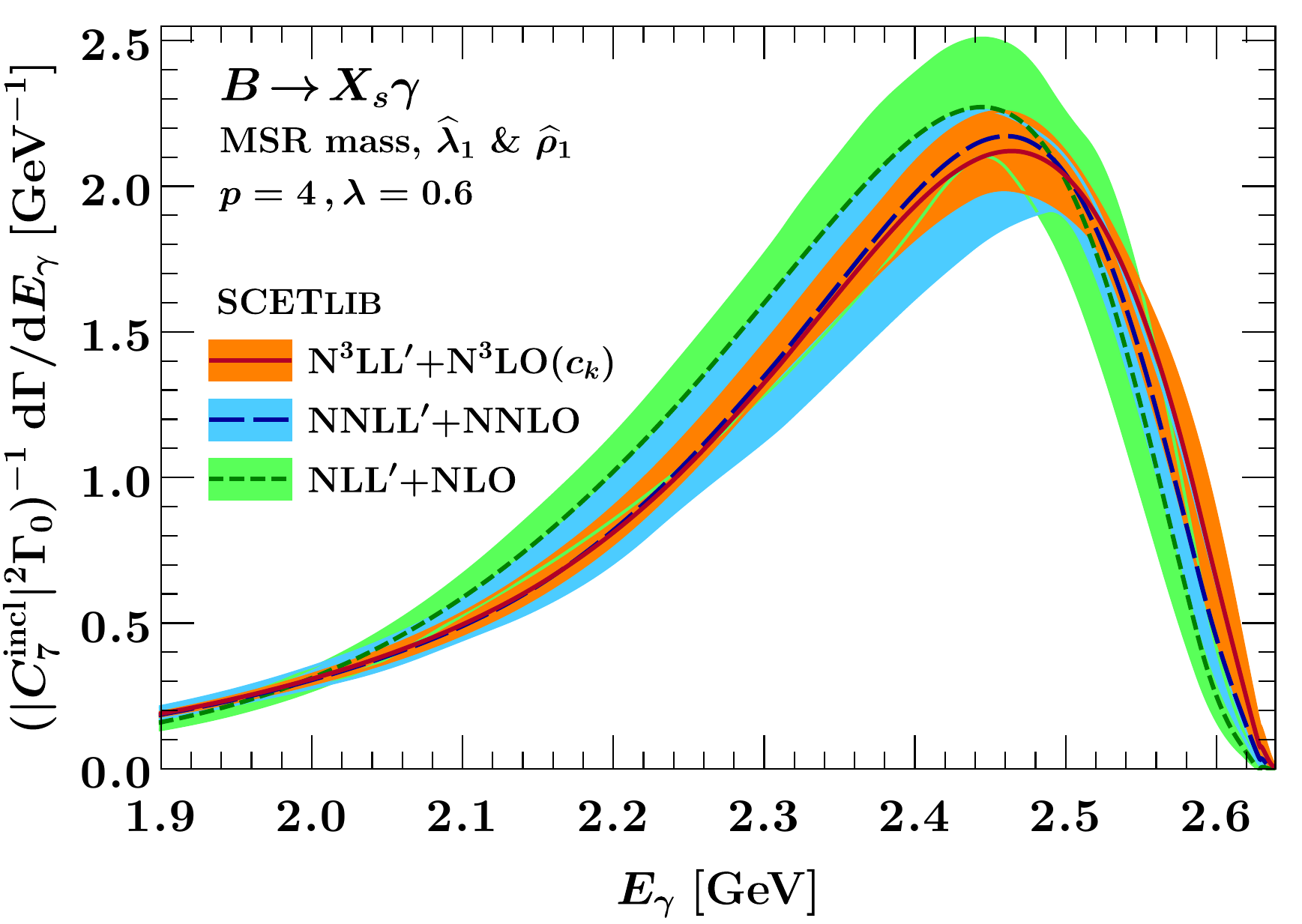}
\caption{
  The soft function (left panel) and the \btosgamma spectrum (right panel) in the MSR mass scheme.
  The top panels use the pole scheme for $\lambda_1$ and $\rho_1$, while the bottom panels use a short-distance scheme.
  The convergence is significantly better than in the other mass schemes, and is further improved by adoption of short-distance schemes for $\lambda_1$ and $\rho_1$.
}
\label{fig:discussion-MSR}
\end{figure}

\begin{figure}[t!]
\centering
\includegraphics[width=0.49\textwidth]{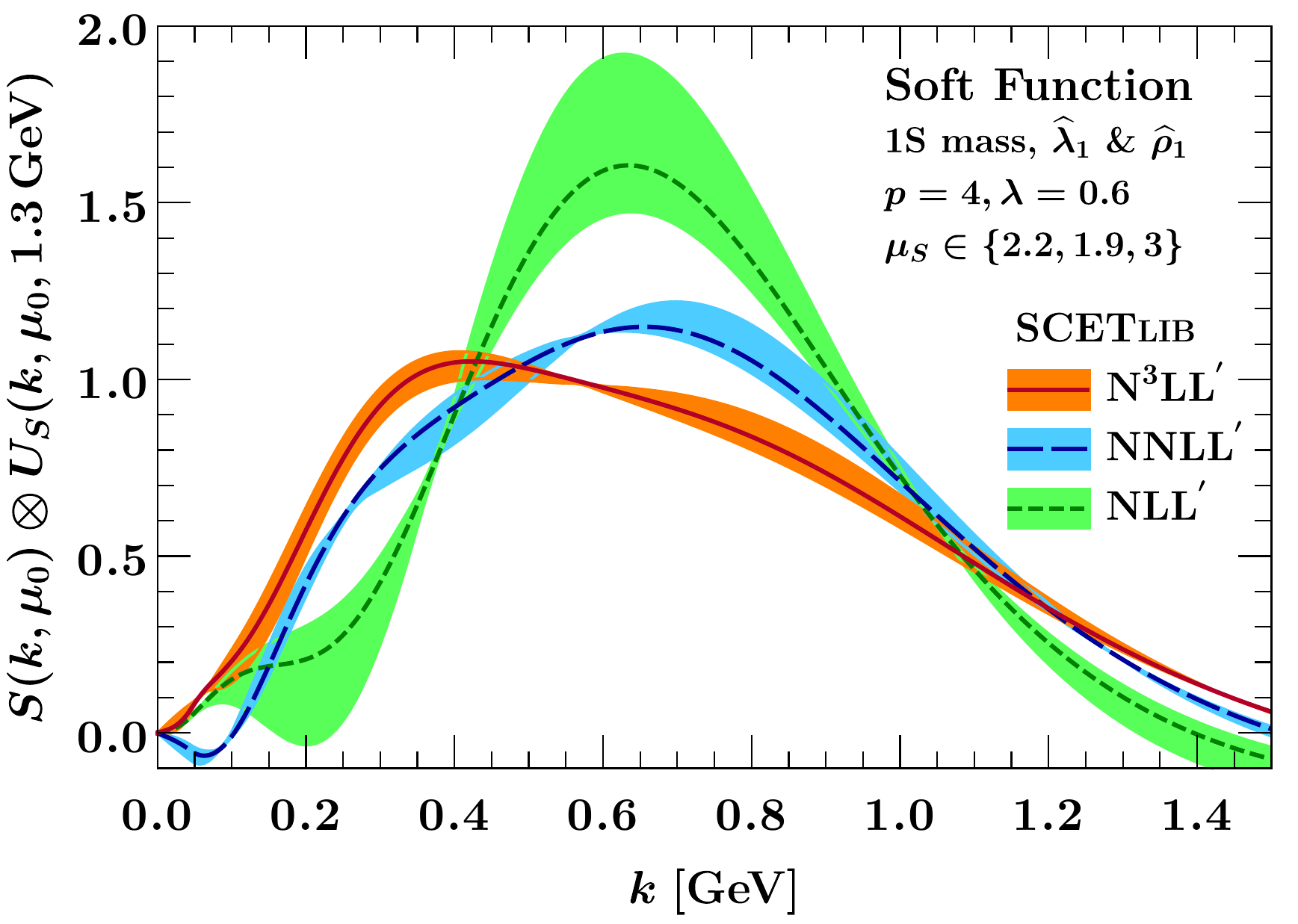}
\includegraphics[width=0.49\textwidth]{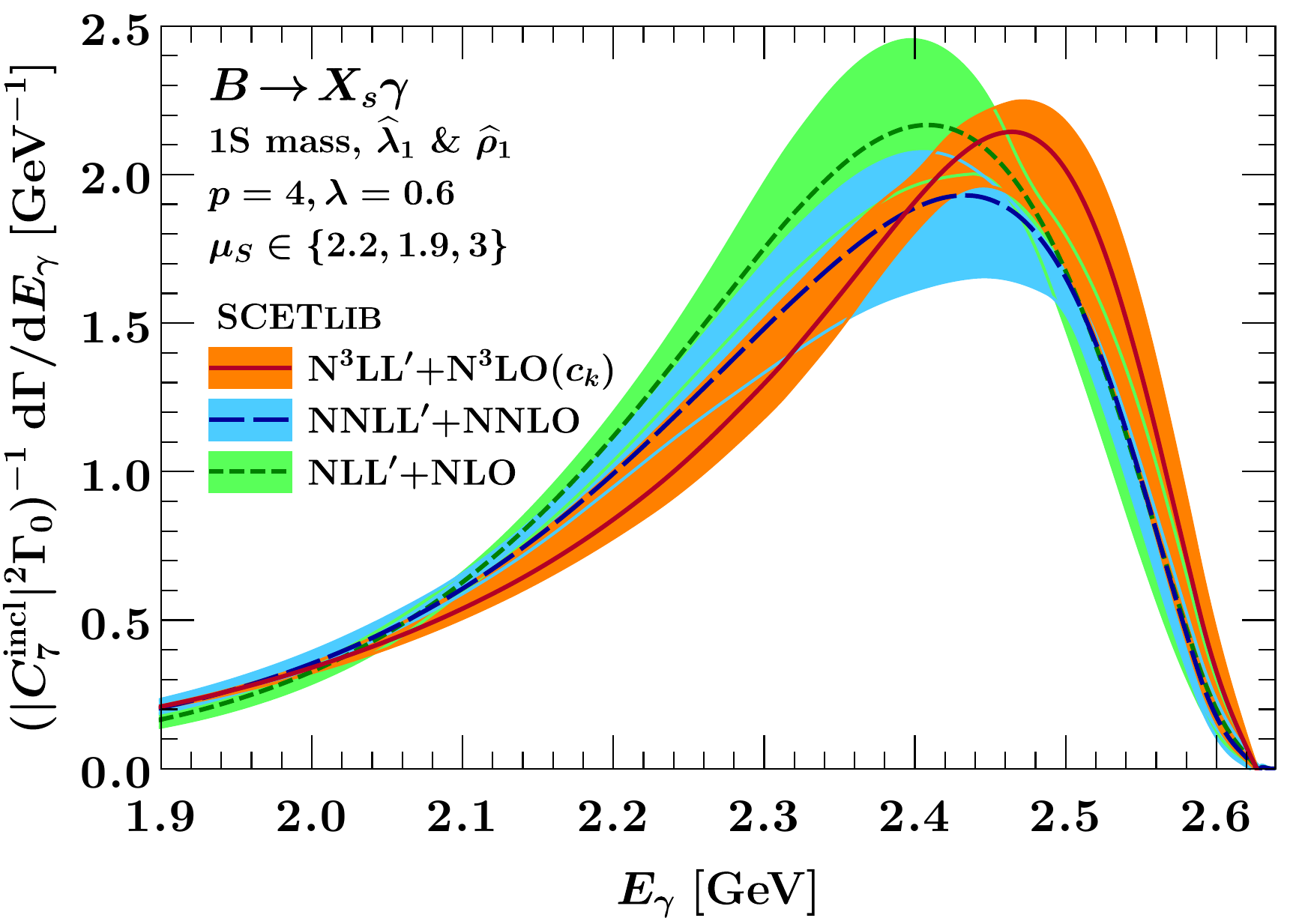}
\caption{
  The soft function (left panel) and the \btosgamma spectrum (right panel) in the \OneS mass scheme.
  A short-distance scheme is used for the hadronic parameters $\lambda_1$ and $\rho_1$.
  In contrast to the other plots, larger values are used for the soft scale: $\mu_0\in\{2.2,1.9,3\}\GeV$.
  The larger soft scale amends the breakdown of \OneS scheme at \NNNLLpMatched, but is not compatible with the SCET scale hierarchy.
}
\label{fig:discussion-1S-modified}
\end{figure}

In the following figures we show our numerical predictions for the soft function (left column) and photon energy spectrum (right column) at different orders using the pole (\fig{discussion-pole}), \OneS (\fig{discussion-1S} and \fig{discussion-1S-modified}), and MSR (\fig{discussion-MSR}) schemes. In these plots, we always use our default
setup modulo the different short-distance schemes as indicated.
For the soft function, we always show the combination $S(k, \mu_0)\otimes U_S(k, \mu_0, 1.3\GeV)$, i.e., we use the soft evolution kernel to evolve the soft function to a fixed scale $\mu = 1.3\GeV$. In this way, the $\mu_0$ dependence cancels up to higher order, so we can use our default $\mu_0$ variations to estimate the perturbative uncertainties for the soft function.

In \fig{discussion-pole} we clearly see that the soft function in the pole scheme suffers from a sizable renormalon ambiguity, which is intrinsic to the pole scheme, leading to a large negative deep before the peak.
Moreover, the peak position varies significantly from one order to another, which reflects the instabilities in the first moments of the shape function. All these features are also mirrored in the corresponding results for the spectrum.
This behaviour of the pole scheme was already observed in
\refcite{Ligeti:2008ac} up to NNLL$'$, and we see that it continues to get
worse at \NNNLLp.

The predictions in the \OneS mass scheme are presented in \fig{discussion-1S}.
Although the differential decay rate up to \NNLLp{} is somewhat stable, the prediction fails dramatically at \NNNLLp.
In particular, the uncertainty band from scale variation is completely out of control.
Adopting short-distance schemes for $\lambda_1$ and $\rho_1$ slightly improves the spectrum up to \NNLLpMatched, but the picture at \NNNLLpMatched does not change.
Neither does it seem to be possible to substantially improve the convergence by adjusting the values of the residual short-distance coefficients $\delta\lambda_1^{(2)}$, $\delta\lambda_1^{(3)}$, and $\delta\rho_1^{(3)}$. Keeping the hard function
unexpanded in the \OneS scheme somewhat improves the picture up to \NNLLpMatched
but does not help at all with the bad behaviour at \NNNLLpMatched.

The results in the MSR scheme are presented in \fig{discussion-MSR}.
In this case we observe much more stable results in comparison to the \OneS scheme.
Here in the first row, where we only switch the $b$-quark mass to the MSR scheme, we still observe that the spectrum is very sensitive to the behavior of the soft function at small $k$. This sensitivity is reflected in the uncertainty estimates from scale variation.
By subtracting the subleading renormalons present in $\lambda_1$ and $\rho_1$, we find a substantial improvement in the peak region of the spectrum, and the result exhibits excellent convergence between all orders (second row).
From these results we can conclude that the MSR scheme is indeed a much more suitable mass scheme for the \btosgamma spectrum when going beyond \NNLLp.

To understand the reason for the breakdown of the \OneS scheme at \NNNLLp we recall the relation between the pole and \OneS schemes up to the \ThreeLoop order,
\begin{equation}
  \mbPole = m_b^\OneS + R^\OneS(\mu)\sum_{n=1}^{3}\sum_{m=0}^{n-1}
  c_{nm}^\OneS\Bigl[\frac{\as(\mu)}{4\pi}\Bigr]^n\Bigl[\ln\frac{\mu}{R^\OneS(\mu)}\Bigr]^m\,.
\end{equation}
where $R^\OneS(\mu) = C_F \,m_b^\OneS\as(\mu)$ is the built-in infrared cutoff scale
of the \OneS scheme. The numerical values for the $c_{nm}$ coefficients are
\begin{align}
  c^\OneS_{10}&=2.0944\,,\nonumber\\
  c^\OneS_{20}&=135.438 -10.2393\,n_f\,,\nonumber\\
  c^\OneS_{21}&=92.1534 -5.58505\,n_f\,,\nonumber\\
  c^\OneS_{30}&=11398.2 -1372.75\,n_f + 38.9677\,n_f^2\,,\nonumber \\
  c^\OneS_{31}&=7766.02 -1077.92\,n_f + 33.5103\,n_f^2\,,\nonumber \\
  c^\OneS_{32}&=3041.06 -368.614\,n_f + 11.1701\,n_f^2\,.
\end{align}
By contrast to the MSR scheme where the $R$ scale is a parameter we can choose,
in the \OneS scheme $R^\OneS(\mu)$ depends on the renormalization scale $\mu$ via the coupling constant.
Hence its size increases when decreasing the scale $\mu$, e.g. at hard scale we find $R^\OneS(4.75\GeV) = 1.36\GeV$, whereas at the soft scale we obtain $R^\OneS(\mu_S= 1.3\GeV) = 2.40\GeV$, which is almost twice the size of the soft scale itself.
Such a large infrared scale violates the power counting of HQET that is used to describe the heavy quarks in the $B$ meson with the residual soft momenta $k\sim \Lambda_{\rm QCD}$ in the peak region.

The mismatch between the size of $R^\OneS$ and $\mu_S\sim k$ not only breaks the power counting of the EFT description of the decay rate, but also spoils the renormalon subtraction in the soft function.
To see this explicitly, we consider the conversion between the \OneS and MSR schemes $\Delta m_b(R)\defeq m_b^{\rm MSR}(R) - m_b^\OneS$, which formally does not involve a renormalon.
We obtain the following perturbative series for $\Delta m_b(R=\mu)$ at various scales,
\begin{align}\label{eq:perturbative-series-mbMSR-mb1S}
  \Delta m_b(R=\mu)\Big|_{\mu=4.2\GeV} &= - 0.35\,\epsilon - 0.12\,\epsilon^2 - 0.04\,\epsilon^3~[\!\GeV]\,,\nonumber \\
  \Delta m_b(R=\mu)\Big|_{\mu=1.93\GeV} &= - 0.15\,\epsilon - 0.06\,\epsilon^2 + 0.02\,\epsilon^3~[\!\GeV]\,,\nonumber \\
  \Delta m_b(R=\mu)\Big|_{\mu=1.3\GeV} &= - 0.06\,\epsilon - 0.06\,\epsilon^2 + 0.10\,\epsilon^3~[\!\GeV]\,,
\end{align}
where $\epsilon \defeq 1$ is an auxiliary parameter denoting the perturbative order of the corrections.
The perturbative series in the first line shows a rather good convergence for $\Delta m_b$ at the hard scale.
However, in this regime the perturbative expansion for the \OneS scheme contains large logarithms of the form $\ln(\mu/R^\OneS(\mu))|_{\mu=4.7\GeV} \sim \ln(4.7/1.36)$. These logarithms are suppressed by $R^\OneS(\mu=m_b^\OneS)/m_b^\OneS =1.36/4.7$ and are therefore harmless in the fixed-order expansion.
For example, in the context of calculating the total decay rate, it is well-known that the \OneS scheme provides a good description for the bottom mass.

We can also define a natural scale for the \OneS scheme, which we denote as $\mu_\OneS$, at which all logarithms of the form $\ln(\mu_\OneS/R^\OneS(\mu_\OneS))$ are resummed. It corresponds to the fixed point of the $R^\OneS$ scale where $R^\OneS(\mu_\OneS)=\mu_\OneS$, which yields $\mu_\OneS= 1.93\GeV$.
The perturbative series for $\Delta m_b(\mu_\OneS)$ is shown in the second line of \eq{perturbative-series-mbMSR-mb1S}.
It shows the same good convergence as the first line, but with overall smaller
corrections and a change of sign in the $\ord{\epsilon^3}$ coefficient.

Finally, the last line in \eq{perturbative-series-mbMSR-mb1S} shows the perturbative series for $\Delta m_b(\mu_S)$ at the soft scale $\mu_S=1.3\GeV$.
As one can see, the resulting series exhibits no convergence and breaks down at the \ThreeLoop order.
This behavior vividly explains the failure of the \OneS mass scheme when used
at the soft scale to remove the renormalon in the soft function.

Another interesting conclusion from the discussion above is that one can retain the use of the \OneS scheme as soon as the actual soft scale in the problem is roughly of the same order of $R^\OneS=\mu_\OneS\sim 1.93\GeV$. To examine this hypothesis, in \fig{discussion-1S-modified} we show the results for the soft function and the photon energy spectrum in the \OneS mass scheme where the soft scale is chosen to have larger values, $\mu_S\in \{2.2,1.9,3\}\GeV$.
Indeed the resulting spectrum exhibits a significant improvement at all orders
compared to \fig{discussion-1S}, and in particular the \NNNLLpMatched result is now much more well behaved. However, in practice this setup is not really ideal since
the soft scale is now much larger than $\Lambda_{\rm QCD}$ and our default
soft scale, which leads to rather large unresummed logarithms in the soft function.
Consequently, the results in \fig{discussion-1S-modified} do not reach the same level of stability as those in the MSR scheme in \fig{discussion-MSR}.

\section{Conclusions}
\label{sec:conclusion}

In this paper, we presented the photon energy spectrum in \btosgamma at \NNNLLpMatched, mediated by the electromagnetic operator in the weak Hamiltonian.
For the singular contributions we used the SCET factorization theorem to resum large logarithms, which arise in the description of the spectrum close to the kinematic endpoint.
We accounted for the complete soft and jet functions at $\rm N^3LO$ and treated the unknown nonlogarithmic constant of the \ThreeLoop hard function as a nuisance parameter.
In addition, the RG evolution is performed at complete N$^3$LL, taking advantage of the fully-known \ThreeLoop anomalous dimensions.
We matched the resummed predictions to fixed order at the formal N$^3$LO order, developing a method that allows us to consistently include the known NNLO results, while parametrizing the missing nonsingular corrections at \ThreeLoop order in terms of suitable theory nuisance parameters $c_k$.
The variation of these nuisance parameters provides an estimate of the uncertainty that arises from our ignorance of these missing terms.

We incorporated nonperturbative effects by convolving the partonic spectrum with a universal shape function.
The first moments of the shape function depend on the $b$-quark mass $m_b$ and the HQET parameters $\lambda_1$ and $\rho_1$.
It is crucial to define these parameters in a suitable short-distance scheme to avoid renormalon ambiguities that would spoil the convergence of perturbative series.
Another important aspect of our analysis was the implementation and choice of an
appropriate short-distance scheme at \NNNLLp.
By using the MSR mass scheme for $m_b$ and adopting an analogous scheme to the ``invisible'' scheme for $\lambda_1$ and $\rho_1$ we are able to obtain stable
predictions. We also find, quite unexpectedly, that the \OneS mass scheme, which has been
successfully used in the past for this process up to \NNLLp, starts to badly break down
at \NNNLLpMatched.
We demonstrated that the reason for this sudden breakdown is that the \OneS scheme is ultimately not designed for use at very low scales. This is because its built-in
infrared cutoff scales with $\as(\mu)$ and thus increases at lower scales and quickly becomes too large when the soft scale is lower than $\mu_\OneS=1.93\GeV$, such that it effectively breaks the power counting of the underlying HQET.

Our main results are presented in \sec{main-results}.
Our final predictions for the photon energy spectrum in the MSR scheme and invisible schemes for $\lambda_1$ and $\rho_1$ exhibit excellent perturbative stability, in particular considering the rather low scales involved in the problem.
In particular, we observe a substantial improvement of the perturbative theory uncertainties from \NNLLpMatched to \NNNLLpMatched. Importantly, even without the complete fixed $\ord{\as^3}$ information available, with our method
we are able to benefit from the increased perturbative precision at \NNNLLp,
allowing us to improve the precision of the theory predictions across the entire
phenomenologically important peak region.
Indeed, the uncertainties due to the missing fixed-order results at $\ord{\as^3}$
only become dominant in the fixed-order tail of the spectrum, which is phenomenologically less relevant. Nevertheless, their calculation is still
encouraged and needed to further reduce the theory uncertainties.

In the future, we look forward to confronting our improved predictions with both
existing and future experimental measurements, enabling more
precise determinations of the shape function $\ShapeFunction(k)$, the $b$-quark mass,
and the normalization of the \btosgamma rate.

\acknowledgments
This work was supported in part by the Helmholtz Association Grant W2/W3-116.

\appendix

\section{Resummation ingredients}
\label{app:resummation-ingredients}

\subsection{General definitions}

We write the perturbative series for the cusp and noncusp anomalous dimensions as
\begin{equation} \label{eq:anomdimcoeffs}
\Gamma_\cusp(\as) \equiv \Gamma_\cusp^q(\as)
= \sum_{n=0}^{\infty} \Gamma_n
\Bigl(\frac{\as}{4 \pi}\Bigr)^{n+1}
\,,\qquad
\gamma^F(\as)
= \sum_{n=0}^{\infty} \gamma_n^F \Bigl(\frac{\as}{4 \pi}\Bigr)^{n+1}
\,.\end{equation}
At \NNNLLp, we need the quark cusp anomalous coefficients up to
four loops~\cite{Korchemsky:1987wg, Moch:2004pa, Henn:2019swt, vonManteuffel:2020vjv}.
The coefficients $\beta_n$ of the QCD $\beta$ function in \MSbar{} are defined as
\begin{equation}\label{eq:beta-function}
\frac{\df\as(\mu)}{\df\ln\mu}
= \beta\bigl[\as(\mu)\bigr]
= -2\as(\mu) \sum_{n=0}^\infty
  \beta_n\,\biggl[\frac{\as(\mu)}{ 4 \,\pi} \biggr]^{n+1}
\,.\end{equation}
At \NNNLLp they are also needed up to four loops~\cite{Tarasov:1980au, Larin:1993tp, vanRitbergen:1997va, Czakon:2004bu}.

The RGE solutions are written in terms of the following standard integrals
\begin{align} \label{eq:define-eta-K}
K_{\Gamma}(\mu_0,\mu)
&= \int_{\as(\mu_0)}^{\alpha(\mu)} \frac{\df \as}{\beta(\as)}\, \Gamma_\cusp(\as)
\int_{\as(\mu_0)}^{\as}\frac{\df\as'}{\beta(\as')}
\,,\nn\\
\eta_{\Gamma}(\mu_0,\mu)
&= \int_{\as(\mu_0)}^{\as(\mu)} \frac{\df\as}{\beta(\as)}\, \Gamma_\cusp(\as)
\,,\nn\\
K_{\gamma}(\mu_0,\mu)
&= \int_{\as(\mu_0)}^{\as(\mu)} \frac{\df\as}{\beta(\as)}\, \gamma(\as)
\,.\end{align}
In principle, they can be performed analytically~\cite{Ebert:2021aoo}.
For simplicity, in our numerical results we employ the standard approximate analytic solutions~\cite{Billis:2019evv}, which are obtained by performing the integrals after expanding the numerators in $\as$. For our purposes here, the numerical accuracy of the approximate analytic solutions is sufficient~\cite{Billis:2019evv}.

Following \refcite{Ligeti:2008ac}, we define the plus distributions
\begin{align}\label{eq:plus-distribution}
\mathcal{L}_{-1}(x) &= \delta(x)
\nn\,,\\
\mathcal{L}_n(x) &= \biggl[ \frac{\theta(x) \ln^n x}{x} \biggr]_+
 = \lim_{\epsilon \to 0}\, \frac{\df}{\df x}\,
 \biggl[\theta(x-\epsilon)\, \frac{\ln^{n+1} x}{n+1} \biggr]
\qquad (n \geq 0)
\nn\,,\\
\mathcal{L}^a(x)
&= \biggl[ \frac{\theta(x)}{x^{1-a}} \biggr]_+
 = \lim_{\epsilon \to 0}\, \frac{\df}{\df x}\, \biggl[\theta(x-\epsilon)\, \frac{x^a - 1}{a} \biggr]
\,.\end{align}

\subsection{Hard function}
\label{app:hard}

We write the perturbative series for the hard function in the pole-mass scheme
as
\begin{equation}\label{eq:hard}
\HardPole(\mbPole,\mu)
= \sum_{n=0}^{\infty}\sum_{m=0}^{2\,n}\, H_{m}^{(n)}
  \biggl[\frac{\as(\mu)}{4\,\pi}\biggr]^n \ln^m\Bigl(\frac{\mu}{\mbPole}\Bigr)
\,.\end{equation}
It satisfies the following RGE,
\begin{equation}\label{eq:hard-RGE}
\frac{\df\HardPole(\mbPole,\mu)}{\df\ln\mu}
= \biggl\{\Gamma^H[\as(\mu)]\ln\frac{\mu}{\mbPole} + \gamma^H[\as(\mu)]\biggr\}\HardPole(\mbPole,\mu)
\,,\end{equation}
where $\Gamma^H(\as) \equiv -2\Gamma_\cusp(\as)$ and $\gamma^H(\as) = 2\gamma^q(\as) + 2\gamma^Q(\as)$ are
the hard cusp and noncusp anomalous dimensions.
The noncusp
anomalous coefficients $\gamma_n^H = 2\gamma_n^q + 2\gamma_n^Q$ are known to three loops~\cite{Bruser:2019yjk}.

The nonlogarithmic coefficients of the hard function in \eq{hard} are known
to two loops~\cite{Bauer:2000yr, Ligeti:2008ac},
\begin{align} \label{eq:hardconstants}
  H^{(0)}_{0} &= 1\,, \nonumber \\
  H^{(1)}_{0} &= - C_F \biggl(12 + \frac{\pi^2}{6}\biggr)\,,\nonumber \\
  H^{(2)}_{0} &=16\, C_F \,\biggl[ 3.88611\, C_F + 5.89413 \,C_A
    -\biggl(\frac{7859}{3456} + \frac{109 \pi^2}{576} +\frac{13 \zeta_3}{48}\biggr)\beta_0
    \nonumber\\&\quad
    +\frac{3563}{1296} -\frac{29 \pi^2}{108} -\frac{\zeta_3}{6}
  \biggr]
  \,,
\end{align}
The \ThreeLoop coefficient $H^{(3)}_{0}$ is currently unknown and treated as a
nuisance parameter as discussed in \sec{singular}, where $h_n \equiv H_0^{(n)}/4^n$.

The coefficients of the logarithmic terms $H^{(n)}_{m\geq 1}$ are determined by iteratively solving the RGE in \eq{hard-RGE} order by order.
Substituting \eq{hard} into \eq{hard-RGE},
we obtain a recurrence relation that expresses them in terms of the anomalous dimensions and lower-order nonlogarithmic coefficients,
\begin{equation}
  H^{(n)}_m =\frac{1}{\,m} \biggl\{
  \sum_{j=0}^{t_1}\bigl(\gamma^H_j+2\,(n-j-1)\,\beta_j\bigr)\,H^{(n-j-1)}_{m-1}
  +\theta(m\geq 2)\sum_{j=0}^{t_2}\Gamma_j^H\,H^{(n-j-1)}_{m-2}
  \biggr\}\,,
  \quad m\geq 1
\end{equation}
with summation limits $t_1=\lfloor n-(m+1)/2\rfloor$ and $t_2 = \lfloor
n-m/2\rfloor$, where the symbol $\lfloor\rfloor$ denotes the floor function. The
condition $\theta(m\geq 2)$ indicates that the second sum is present only if
$m\geq 2$. The explicit expressions up to three loops read
\begin{align}
  H^{(1)}_{1} &= \gamma_0^H\,,
  \nonumber \\
  H^{(1)}_{2} &= \frac{1}{2}\,\Gamma_0^H\,,
  \nonumber \\
  H^{(2)}_{1} &= H_0^{(1)}\,\bigl(2\,\beta_0 +\gamma^H_0\bigr)+\gamma^H_1\,,
  \nonumber \\
  H^{(2)}_{2} &= \frac{1}{2}\,\Bigl\{2\,\beta_0 \,\gamma_0^H + \bigl(\gamma_0^H\bigr)^2 + H_0^{(1)}\Gamma^H_0 +\Gamma^H_1\Bigr\}\,,
  \nonumber \\
  H^{(2)}_{3} &= \frac{1}{6}\,\Gamma_0^H\, \bigl(2\,\beta_0 +3\, \gamma^H_0\bigr)\,,
  \nonumber \\
  H^{(2)}_{4} &= \frac{1}{8}\,\bigl(\Gamma_0^H\bigr)^2\,,
  \nonumber \\
  H^{(3)}_{1} &= H_0^{(2)}\, \bigl( 4\,\beta_0 +\gamma_0^H\bigr)+H_0^{(1)} \,\bigl( 2\,\beta_1 +\gamma_1^H\bigr) + \gamma_2^H\,,
  \nonumber \\
  H^{(3)}_{2} &= \frac{1}{2}\, \Bigl\{
    2\,\beta_1\,\gamma_0^H +4\,\beta_0\,\gamma_1^H + 2 \,\gamma_0^H\, \gamma_1^H + H_0^{(2)}\Gamma_0^H
    \nonumber\\&\quad
    +H_0^{(1)} \,\Bigl(8 \,\beta_0^2+6\, \beta_0\,\gamma_0^H +\bigl(\gamma_0^H\bigr)^2 + \Gamma_1^H\Bigr) + \Gamma_2^H
  \Bigr\}\,,
  \nonumber \\
  H^{(3)}_{3} &= \frac{1}{6}\, \Bigl\{
    8\,\beta_0^2\,\gamma_0^H + \bigl(\gamma_0^H\bigr)^3 + \bigl(2\, \beta_1 + 3\,\gamma_1^H\bigr)\,\Gamma^H_0 + 3\, \gamma^H_0 \,\bigl(H_0^{(1)} \,\Gamma_0^H + \Gamma_1^H\bigr)
    \nonumber\\&\quad
    +\beta_0 \,\Bigl(6\,\bigl(\gamma_0^H\bigr)^2 + 8\, H_0^{(1)}\, \Gamma_0^H + 4\, \Gamma_1^H\Bigr)
  \Bigr\}\,,
  \nonumber \\
  H^{(3)}_{4} &= \frac{1}{24} \,\Gamma_0^H\,\bigl(8\,\beta_0^2 +20\, \beta_0\,\gamma^H_0 + 6 \,\bigl(\gamma^H_0\bigr)^2+ 3\, H_0^{(1)}\Gamma_0^H + 6\, \Gamma_1^H\bigr)\,,
  \nonumber \\
  H^{(3)}_{5} &= \frac{1}{24}\,\bigl(\Gamma_0^H\bigr)^2\,\bigl(4\,\beta_0 + 3 \,\gamma_0^H\bigr)\,,
  \nonumber \\
  H^{(3)}_{6} &= \frac{1}{48}\bigl(\Gamma_0^H\bigr)^3\,.
\end{align}

The all-order solution of the RGE in \eq{hard-RGE} is given by
\begin{align} \label{eq:hard-RGE-solution}
\HardPole(\mbPole,\mu)
&=\HardPole(\mbPole,\mu_H)\,\HardEvolutionPole(\mbPole,\mu_H,\mu)
\,,\\\nonumber
\HardEvolutionPole(m_b,\mu_H,\mu)
&= \exp\biggl[
-2K_\Gamma(\mu_H,\mu) - 2\eta_\Gamma(\mu_H,\mu)\ln\frac{\mu_H}{m_b}
+ K_{\gamma^H}(\mu_H,\mu)
\biggr]
\,.\end{align}

\subsection{Jet function}
\label{sec:jet}

The perturbative series for the renormalized jet function reads
\begin{equation}\label{eq:jet}
  \JetPole(s,\mu)= \sum_{n=0}^{\infty} \sum_{m=-1}^{2n-1}
  J^{(n)}_m \biggl[\frac{\as(\mu)}{4\pi}\biggr]^n
  \frac{1}{\mu^2}\,\mathcal{L}_m\Bigl(\frac{s}{\mu^2}\Bigr)\,,
\,.\end{equation}
The jet function is normalized such that $J^{(n)}_{-1}=1$.
Explicit expressions for the coefficients $J^{(n)}_m$ up to \ThreeLoop order in
our notation are given in \refscite{Gaunt:2015pea, Bruser:2018rad}.

For completeness, the jet function obeys the RGE
\begin{equation}\label{eq:jet-RGE}
  \frac{\df\JetPole(s,\mu)}{\df\ln\mu}
  =
  \biggl\{
    \Gamma^J[\as(\mu)]\frac{1}{\mu^2}\,\mathcal{L}_{0}\Bigl(\frac{s}{\mu^2}\Bigr)
    + \gamma^J[\as(\mu)]\,\delta(s)
  \biggr\}
  \otimes_s\JetPole(s,\mu)
\end{equation}
where $\Gamma^J(\as) = -2\Gamma_\cusp(\as)$ and
the symbol $\otimes_s$ denotes the convolution of the form
\begin{equation}
f(s)\otimes_s g(s) = \int\df s' \,f(s-s') \,g(s')
\,.\end{equation}
In our numerical implementation, we do not need to explicitly solve the
jet-function RGE, because we always evolve the hard and soft functions to the
jet scale.

\subsection{Partonic soft function}
\label{app:soft}

In the pole scheme the partonic soft function $\PartonicSoftPole(\omega, \mu)$
is given by the $b$-quark matrix element
\begin{equation}
\PartonicSoftPole(\omega,\mu)
= \langle b_v|\bar{b}_v\delta(iD_++\omega)b_v|b_v\rangle
\,.\end{equation}
Its perturbative expansion is written as
\begin{equation}\label{eq:soft}
  \PartonicSoftPole(\omega,\mu)=
    \sum_{n=0}^\infty \sum_{m=-1}^{2n-1}S^{(n)}_m
    \biggl[\frac{\as(\mu)}{4\pi}\biggr]^n\frac{1}{\mu}\,\mathcal{L}_m\Bigl(\frac{\omega}{\mu}\Bigr)
    \,.
\end{equation}
The expansion coefficients $S_m^{(n)}$ can be found in \refcite{Bruser:2019yjk}.

The RGE for the soft function reads
\begin{equation}\label{eq:soft-RGE}
  \frac{\df\PartonicSoftPole(\omega,\mu)}{\df\ln\mu}=
  \biggl\{
    \Gamma^S[\as(\mu)] \frac{1}{\mu}\,\mathcal{L}_{0}\Bigl(\frac{\omega}{\mu}\Bigr)
    + \gamma^S[\as(\mu)]\, \delta(\omega)
  \biggr\}
  \otimes_\omega\PartonicSoftPole(\omega,\mu)
  \,,
\end{equation}
where $\Gamma^S(\as) = 2\Gamma_\cusp(\as)$ and
the noncusp anomalous coefficients are known to three loops~\cite{Bruser:2019yjk}.
By solving it iteratively, we can obtain a recurrence
relation for all logarithmic coefficients $S^{(n)}_{m\geq 0}$,
\begin{equation} \label{eq:softrecurrence}
  S^{(n)}_m
  = -\frac{1}{m+\delta_{m0}}
  \biggl\{
  \sum_{j=0}^{t_1}
  \bigl[\gamma^S_j+2\,(n-j-1)\beta_j \bigr]
  \,S^{(n-j-1)}_{m-1}
  +\sum_{j=0}^{t_2}\sum_{i=t_{\rm m}-2}^{2(n-j)-3}
  \,\Gamma^S_j V^{0i}_{m-1}\,S^{(n-j-1)}_i
  \biggr\}
  \,,
\end{equation}
where the summation limits are $t_1=\lfloor n -1-m/2\rfloor \,, t_2=\lfloor n-(t_m+1)/2\rfloor$ and $t_{\rm m}=\max{(m,1)}$.
The coefficients $V^{mn}_k$ appear in the convolution algebra of plus distributions,
\begin{equation}
  \mathcal{L}_{n}(x) \otimes_x \mathcal{L}_{m}(x) =
  \sum_{\mathclap{k=-1}}^{\mathclap{n+m+1}} V_k^{nm} \, \mathcal{L}_{k}(x)
  \,,
\end{equation}
and are given in \refcite{Ligeti:2008ac}.
It is easy to check that \eq{softrecurrence} reproduces the explicit
results to three loops in \refcite{Bruser:2019yjk}.

The all-order solution of the soft RGE is~\cite{Balzereit:1998yf,Neubert:2004dd, Fleming:2007xt,Ligeti:2008ac}
\begin{align}
\PartonicSoftPole(\omega,\mu)
&=\PartonicSoftPole(\omega,\mu_S)\otimes_\omega\SoftEvolution(\omega,\mu_S,\mu)
\,,\nn\\
\SoftEvolution(\omega,\mu_S,\mu)
&= \exp\Bigl[
   - 2K_\Gamma(\mu_S, \mu)
   + K_{\gamma^S}(\mu_S,\mu)
  \Bigr]
   \mathcal{V}[2\eta_\Gamma(\mu_S,\mu),\mu_S,\omega]
\,,\nn\\
\mathcal{V}(\eta,\mu,\omega)
&= \frac{e^{-\gamma_E\eta}}{\Gamma(1+\eta)}\biggl[\frac{\eta}{\mu} \mathcal{L}^\eta\Bigl(\frac{\omega}{\mu}\Bigr) + \delta(\omega) \biggr]
\end{align}
where $K_\Gamma$, $\eta_\Gamma$, $K_\gamma$ are defined in \eq{define-eta-K},
$\gamma_E$ is the Euler-Mascheroni constant, and the $\mathcal{L}^\eta(x)$
plus distribution is defined in \eq{plus-distribution}.

Note that in the short-distance schemes we consider, the residual terms
$\delta m_b$, $\delta\lambda_1$, $\delta\rho_1$ are formally $\mu$ independent,
such that the evolution for the partonic soft function in a short-distance scheme
$\PartonicSoft(\omega, \mu)$ is the same as in the pole scheme.

\addcontentsline{toc}{section}{References}
\bibliographystyle{jhep}
\bibliography{refs}

\providecommand{\href}[2]{#2}\begingroup\raggedright\begin{thebibliography}{10}

\bibitem{Bertolini:1986th}
S.~Bertolini, F.~Borzumati and A.~Masiero, \emph{{QCD Enhancement of Radiative
  b Decays}}, \href{https://doi.org/10.1103/PhysRevLett.59.180}{\emph{Phys.
  Rev. Lett.} {\bfseries 59} (1987) 180}.

\bibitem{Grinstein:1987vj}
B.~Grinstein, R.~P. Springer and M.~B. Wise, \emph{{Effective Hamiltonian for
  Weak Radiative B Meson Decay}},
  \href{https://doi.org/10.1016/0370-2693(88)90868-4}{\emph{Phys. Lett. B}
  {\bfseries 202} (1988) 138}.

\bibitem{Misiak:2006zs}
M.~Misiak et~al., \emph{{Estimate of $\mathcal{B} (\bar B \to X_s \gamma)$ at
  $\mathcal{O}(\alpha_s^2)$}},
  \href{https://doi.org/10.1103/PhysRevLett.98.022002}{\emph{Phys. Rev. Lett.}
  {\bfseries 98} (2007) 022002}
  [\href{https://arxiv.org/abs/hep-ph/0609232}{{\ttfamily hep-ph/0609232}}].

\bibitem{Misiak:2015xwa}
M.~Misiak et~al., \emph{{Updated NNLO QCD predictions for the weak radiative
  B-meson decays}},
  \href{https://doi.org/10.1103/PhysRevLett.114.221801}{\emph{Phys. Rev. Lett.}
  {\bfseries 114} (2015) 221801}
  [\href{https://arxiv.org/abs/1503.01789}{{\ttfamily 1503.01789}}].

\bibitem{Grinstein:1987pu}
B.~Grinstein and M.~B. Wise, \emph{{Weak Radiative B Meson Decay as a Probe of
  the Higgs Sector}},
  \href{https://doi.org/10.1016/0370-2693(88)90227-4}{\emph{Phys. Lett. B}
  {\bfseries 201} (1988) 274}.

\bibitem{Hou:1987kf}
W.-S. Hou and R.~S. Willey, \emph{{Effects of Charged Higgs Bosons on the
  Processes $b \to s \gamma$, $b \to s g^*$ and $b \to s l^+l^-$}},
  \href{https://doi.org/10.1016/0370-2693(88)91870-9}{\emph{Phys. Lett. B}
  {\bfseries 202} (1988) 591}.

\bibitem{Misiak:2017bgg}
M.~Misiak and M.~Steinhauser, \emph{{Weak radiative decays of the B meson and
  bounds on $M_{H^\pm }$ in the Two-Higgs-Doublet Model}},
  \href{https://doi.org/10.1140/epjc/s10052-017-4776-y}{\emph{Eur. Phys. J. C}
  {\bfseries 77} (2017) 201}
  [\href{https://arxiv.org/abs/1702.04571}{{\ttfamily 1702.04571}}].

\bibitem{Neubert:1993um}
M.~Neubert, \emph{{Analysis of the photon spectrum in inclusive $B \to
  X_s\gamma$ decays}},
  \href{https://doi.org/10.1103/PhysRevD.49.4623}{\emph{Phys. Rev. D}
  {\bfseries 49} (1994) 4623}
  [\href{https://arxiv.org/abs/hep-ph/9312311}{{\ttfamily hep-ph/9312311}}].

\bibitem{Bigi:1993ex}
I.~I.~Y. Bigi, M.~A. Shifman, N.~G. Uraltsev and A.~I. Vainshtein, \emph{{On
  the motion of heavy quarks inside hadrons: Universal distributions and
  inclusive decays}},
  \href{https://doi.org/10.1142/S0217751X94000996}{\emph{Int. J. Mod. Phys. A}
  {\bfseries 9} (1994) 2467}
  [\href{https://arxiv.org/abs/hep-ph/9312359}{{\ttfamily hep-ph/9312359}}].

\bibitem{Neubert:1993ch}
M.~Neubert, \emph{{QCD based interpretation of the lepton spectrum in inclusive
  $\bar{B}\to X_u l\bar{\nu}$ decays}},
  \href{https://doi.org/10.1103/PhysRevD.49.3392}{\emph{Phys. Rev. D}
  {\bfseries 49} (1994) 3392}
  [\href{https://arxiv.org/abs/hep-ph/9311325}{{\ttfamily hep-ph/9311325}}].

\bibitem{BaBar:2007yhb}
{\scshape BaBar} collaboration, B.~Aubert et~al., \emph{{Measurement of the $B
  \to X_s \gamma$ branching fraction and photon energy spectrum using the
  recoil method}},
  \href{https://doi.org/10.1103/PhysRevD.77.051103}{\emph{Phys. Rev. D}
  {\bfseries 77} (2008) 051103}
  [\href{https://arxiv.org/abs/0711.4889}{{\ttfamily 0711.4889}}].

\bibitem{Belle:2009nth}
{\scshape Belle} collaboration, A.~Limosani et~al., \emph{{Measurement of
  Inclusive Radiative B-meson Decays with a Photon Energy Threshold of
  1.7-GeV}}, \href{https://doi.org/10.1103/PhysRevLett.103.241801}{\emph{Phys.
  Rev. Lett.} {\bfseries 103} (2009) 241801}
  [\href{https://arxiv.org/abs/0907.1384}{{\ttfamily 0907.1384}}].

\bibitem{BaBar:2012idb}
{\scshape BaBar} collaboration, J.~P. Lees et~al., \emph{{Measurement of
  B($B\to X_s \gamma$), the $B\to X_s \gamma$ photon energy spectrum, and the
  direct CP asymmetry in $B\to X_{s+d} \gamma$ decays}},
  \href{https://doi.org/10.1103/PhysRevD.86.112008}{\emph{Phys. Rev. D}
  {\bfseries 86} (2012) 112008}
  [\href{https://arxiv.org/abs/1207.5772}{{\ttfamily 1207.5772}}].

\bibitem{BaBar:2012eja}
{\scshape BaBar} collaboration, J.~P. Lees et~al., \emph{{Exclusive
  Measurements of $b \to s\gamma$ Transition Rate and Photon Energy Spectrum}},
  \href{https://doi.org/10.1103/PhysRevD.86.052012}{\emph{Phys. Rev. D}
  {\bfseries 86} (2012) 052012}
  [\href{https://arxiv.org/abs/1207.2520}{{\ttfamily 1207.2520}}].

\bibitem{Bernlochner:2020jlt}
{\scshape SIMBA} collaboration, F.~U. Bernlochner, H.~Lacker, Z.~Ligeti, I.~W.
  Stewart, F.~J. Tackmann and K.~Tackmann, \emph{{Precision Global
  Determination of the $B\to X_s\gamma$ Decay Rate}},
  \href{https://doi.org/10.1103/PhysRevLett.127.102001}{\emph{Phys. Rev. Lett.}
  {\bfseries 127} (2021) 102001}
  [\href{https://arxiv.org/abs/2007.04320}{{\ttfamily 2007.04320}}].

\bibitem{Bauer:2003pi}
C.~W. Bauer and A.~V. Manohar, \emph{{Shape function effects in \btosgamma{}
  and \btou{} decays}},
  \href{https://doi.org/10.1103/PhysRevD.70.034024}{\emph{Phys. Rev. D}
  {\bfseries 70} (2004) 034024}
  [\href{https://arxiv.org/abs/hep-ph/0312109}{{\ttfamily hep-ph/0312109}}].

\bibitem{Lee:2005pk}
K.~S.~M. Lee and I.~W. Stewart, \emph{{Shape-function effects and split
  matching in $B \to X_s l^+ l^-$}},
  \href{https://doi.org/10.1103/PhysRevD.74.014005}{\emph{Phys. Rev. D}
  {\bfseries 74} (2006) 014005}
  [\href{https://arxiv.org/abs/hep-ph/0511334}{{\ttfamily hep-ph/0511334}}].

\bibitem{Lee:2005pwa}
K.~S.~M. Lee, Z.~Ligeti, I.~W. Stewart and F.~J. Tackmann, \emph{{Universality
  and $m_X$ cut effects in $B \to X_s l^+ l^-$}},
  \href{https://doi.org/10.1103/PhysRevD.74.011501}{\emph{Phys. Rev. D}
  {\bfseries 74} (2006) 011501}
  [\href{https://arxiv.org/abs/hep-ph/0512191}{{\ttfamily hep-ph/0512191}}].

\bibitem{Zyla:2020zbs}
{\scshape Particle Data Group} collaboration, P.~Zyla et~al., \emph{{Review of
  Particle Physics}}, \href{https://doi.org/10.1093/ptep/ptaa104}{\emph{PTEP}
  {\bfseries 2020} (2020) 083C01}.

\bibitem{Belle-II:2022hys}
{\scshape Belle-II} collaboration, F.~Abudin\'en et~al., \emph{{Measurement of
  the photon-energy spectrum in inclusive $B\rightarrow X_{s}\gamma$ decays
  identified using hadronic decays of the recoil $B$ meson in 2019-2021 Belle
  II data}},  \href{https://arxiv.org/abs/2210.10220}{{\ttfamily 2210.10220}}.

\bibitem{Belle-II:2022cgf}
{\scshape Belle-II} collaboration, L.~Aggarwal et~al., \emph{{Snowmass White
  Paper: Belle II physics reach and plans for the next decade and beyond}},
  \href{https://arxiv.org/abs/2207.06307}{{\ttfamily 2207.06307}}.

\bibitem{Bruser:2018rad}
R.~Br\"user, Z.~L. Liu and M.~Stahlhofen, \emph{{Three-Loop Quark Jet
  Function}}, \href{https://doi.org/10.1103/PhysRevLett.121.072003}{\emph{Phys.
  Rev. Lett.} {\bfseries 121} (2018) 072003}
  [\href{https://arxiv.org/abs/1804.09722}{{\ttfamily 1804.09722}}].

\bibitem{Bruser:2019yjk}
R.~Br\"user, Z.~L. Liu and M.~Stahlhofen, \emph{{Three-loop soft function for
  heavy-to-light quark decays}},
  \href{https://doi.org/10.1007/JHEP03(2020)071}{\emph{JHEP} {\bfseries 03}
  (2020) 071} [\href{https://arxiv.org/abs/1911.04494}{{\ttfamily
  1911.04494}}].

\bibitem{TNPtalkSCET}
F.~J. Tackmann, \emph{{\it Theory Uncertainties from Nuisance Parameters}},
  {\emph{\href{https://indico.physics.lbl.gov/event/694/contributions/3344/}{Talk
  at SCET 2019 workshop}} (March 2019) }.

\bibitem{TNPs}
F.~J. Tackmann, \emph{{Beyond Scale Variations: Perturbative Theory
  Uncertainties from Nuisance Parameters}}, {\emph{DESY-19-021} }.

\bibitem{Blokland:2005uk}
I.~R. Blokland, A.~Czarnecki, M.~Misiak, M.~Slusarczyk and F.~Tkachov,
  \emph{{The Electromagnetic dipole operator effect on $\bar{B}\to X_s\gamma$
  at $\mathcal{O}(\alpha_s^2)$}},
  \href{https://doi.org/10.1103/PhysRevD.72.033014}{\emph{Phys. Rev. D}
  {\bfseries 72} (2005) 033014}
  [\href{https://arxiv.org/abs/hep-ph/0506055}{{\ttfamily hep-ph/0506055}}].

\bibitem{Melnikov:2005bx}
K.~Melnikov and A.~Mitov, \emph{{The Photon energy spectrum in \btosgamma{} in
  perturbative QCD through $\mathcal{O}(\as^2)$}},
  \href{https://doi.org/10.1016/j.physletb.2005.06.015}{\emph{Phys. Lett. B}
  {\bfseries 620} (2005) 69}
  [\href{https://arxiv.org/abs/hep-ph/0505097}{{\ttfamily hep-ph/0505097}}].

\bibitem{Asatrian:2006sm}
H.~M. Asatrian, T.~Ewerth, A.~Ferroglia, P.~Gambino and C.~Greub,
  \emph{{Magnetic dipole operator contributions to the photon energy spectrum
  in $\bar{B}\to X_s\gamma$ at $\mathcal{O}(\alpha_s^2)$}},
  \href{https://doi.org/10.1016/j.nuclphysb.2006.11.002}{\emph{Nucl. Phys. B}
  {\bfseries 762} (2007) 212}
  [\href{https://arxiv.org/abs/hep-ph/0607316}{{\ttfamily hep-ph/0607316}}].

\bibitem{Hoang:1998ng}
A.~H. Hoang, Z.~Ligeti and A.~V. Manohar, \emph{{B decay and the Upsilon
  mass}}, \href{https://doi.org/10.1103/PhysRevLett.82.277}{\emph{Phys. Rev.
  Lett.} {\bfseries 82} (1999) 277}
  [\href{https://arxiv.org/abs/hep-ph/9809423}{{\ttfamily hep-ph/9809423}}].

\bibitem{Hoang:1998hm}
A.~H. Hoang, Z.~Ligeti and A.~V. Manohar, \emph{{B decays in the upsilon
  expansion}}, \href{https://doi.org/10.1103/PhysRevD.59.074017}{\emph{Phys.
  Rev. D} {\bfseries 59} (1999) 074017}
  [\href{https://arxiv.org/abs/hep-ph/9811239}{{\ttfamily hep-ph/9811239}}].

\bibitem{Hoang:1999zc}
A.~H. Hoang and T.~Teubner, \emph{{Top quark pair production close to
  threshold: Top mass, width and momentum distribution}},
  \href{https://doi.org/10.1103/PhysRevD.60.114027}{\emph{Phys. Rev. D}
  {\bfseries 60} (1999) 114027}
  [\href{https://arxiv.org/abs/hep-ph/9904468}{{\ttfamily hep-ph/9904468}}].

\bibitem{Ligeti:2008ac}
Z.~Ligeti, I.~W. Stewart and F.~J. Tackmann, \emph{{Treating the b quark
  distribution function with reliable uncertainties}},
  \href{https://doi.org/10.1103/PhysRevD.78.114014}{\emph{Phys. Rev. D}
  {\bfseries 78} (2008) 114014}
  [\href{https://arxiv.org/abs/0807.1926}{{\ttfamily 0807.1926}}].

\bibitem{Hoang:2008yj}
A.~H. Hoang, A.~Jain, I.~Scimemi and I.~W. Stewart, \emph{{Infrared
  Renormalization Group Flow for Heavy Quark Masses}},
  \href{https://doi.org/10.1103/PhysRevLett.101.151602}{\emph{Phys. Rev. Lett.}
  {\bfseries 101} (2008) 151602}
  [\href{https://arxiv.org/abs/0803.4214}{{\ttfamily 0803.4214}}].

\bibitem{Lee:2006gs}
K.~S.~M. Lee, Z.~Ligeti, I.~W. Stewart and F.~J. Tackmann, \emph{{Extracting
  short distance information from $b \to s l^+ l^-$ effectively}},
  \href{https://doi.org/10.1103/PhysRevD.75.034016}{\emph{Phys. Rev. D}
  {\bfseries 75} (2007) 034016}
  [\href{https://arxiv.org/abs/hep-ph/0612156}{{\ttfamily hep-ph/0612156}}].

\bibitem{Korchemsky:1994jb}
G.~P. Korchemsky and G.~F. Sterman, \emph{{Infrared factorization in inclusive
  B meson decays}},
  \href{https://doi.org/10.1016/0370-2693(94)91304-8}{\emph{Phys. Lett. B}
  {\bfseries 340} (1994) 96}
  [\href{https://arxiv.org/abs/hep-ph/9407344}{{\ttfamily hep-ph/9407344}}].

\bibitem{Bauer:2001yt}
C.~W. Bauer, D.~Pirjol and I.~W. Stewart, \emph{{Soft collinear factorization
  in effective field theory}},
  \href{https://doi.org/10.1103/PhysRevD.65.054022}{\emph{Phys. Rev. D}
  {\bfseries 65} (2002) 054022}
  [\href{https://arxiv.org/abs/hep-ph/0109045}{{\ttfamily hep-ph/0109045}}].

\bibitem{Becher:2006qw}
T.~Becher and M.~Neubert, \emph{{Toward a NNLO calculation of the $\bar{B}\to
  X_s\gamma$ decay rate with a cut on photon energy. II. Two-loop result for
  the jet function}},
  \href{https://doi.org/10.1016/j.physletb.2006.04.046}{\emph{Phys. Lett. B}
  {\bfseries 637} (2006) 251}
  [\href{https://arxiv.org/abs/hep-ph/0603140}{{\ttfamily hep-ph/0603140}}].

\bibitem{Becher:2005pd}
T.~Becher and M.~Neubert, \emph{{Toward a NNLO calculation of the $\bar{B}\to
  X_s\gamma$ decay rate with a cut on photon energy: I. Two-loop result for the
  soft function}},
  \href{https://doi.org/10.1016/j.physletb.2006.01.006}{\emph{Phys. Lett. B}
  {\bfseries 633} (2006) 739}
  [\href{https://arxiv.org/abs/hep-ph/0512208}{{\ttfamily hep-ph/0512208}}].

\bibitem{Bauer:2000yr}
C.~W. Bauer, S.~Fleming, D.~Pirjol and I.~W. Stewart, \emph{{An Effective field
  theory for collinear and soft gluons: Heavy to light decays}},
  \href{https://doi.org/10.1103/PhysRevD.63.114020}{\emph{Phys. Rev. D}
  {\bfseries 63} (2001) 114020}
  [\href{https://arxiv.org/abs/hep-ph/0011336}{{\ttfamily hep-ph/0011336}}].

\bibitem{Abbate:2010xh}
R.~Abbate, M.~Fickinger, A.~H. Hoang, V.~Mateu and I.~W. Stewart, \emph{{Thrust
  at $N^{3}LL$ with Power Corrections and a Precision Global Fit for
  $\alpha_{s}(mZ)$}},
  \href{https://doi.org/10.1103/PhysRevD.83.074021}{\emph{Phys. Rev. D}
  {\bfseries 83} (2011) 074021}
  [\href{https://arxiv.org/abs/1006.3080}{{\ttfamily 1006.3080}}].

\bibitem{Martinelli:1995zw}
G.~Martinelli, M.~Neubert and C.~T. Sachrajda, \emph{{The Invisible
  renormalon}}, \href{https://doi.org/10.1016/0550-3213(95)00613-3}{\emph{Nucl.
  Phys. B} {\bfseries 461} (1996) 238}
  [\href{https://arxiv.org/abs/hep-ph/9504217}{{\ttfamily hep-ph/9504217}}].

\bibitem{Gremm:1996df}
M.~Gremm and A.~Kapustin, \emph{{Order $m_b^{-3}$ corrections to $B\to
  X_cl\bar{\nu}$ decay and their implication for the measurement of
  $\bar{\Lambda}$ and $\lambda_1$}},
  \href{https://doi.org/10.1103/PhysRevD.55.6924}{\emph{Phys. Rev. D}
  {\bfseries 55} (1997) 6924}
  [\href{https://arxiv.org/abs/hep-ph/9603448}{{\ttfamily hep-ph/9603448}}].

\bibitem{Falk:1992fm}
A.~F. Falk, M.~Neubert and M.~E. Luke, \emph{{The Residual mass term in the
  heavy quark effective theory}},
  \href{https://doi.org/10.1016/0550-3213(92)90617-K}{\emph{Nucl. Phys. B}
  {\bfseries 388} (1992) 363}
  [\href{https://arxiv.org/abs/hep-ph/9204229}{{\ttfamily hep-ph/9204229}}].

\bibitem{Hoang:2014oea}
A.~H. Hoang, \emph{{The Top Mass: Interpretation and Theoretical
  Uncertainties}},  in \emph{{7th International Workshop on Top Quark
  Physics}}, 12, 2014, \href{https://arxiv.org/abs/1412.3649}{{\ttfamily
  1412.3649}}.

\bibitem{Hoang:2017suc}
A.~H. Hoang, A.~Jain, C.~Lepenik, V.~Mateu, M.~Preisser, I.~Scimemi et~al.,
  \emph{{The MSR mass and the $
  \mathcal{O}\left({\Lambda}_{\mathrm{QCD}}\right) $ renormalon sum rule}},
  \href{https://doi.org/10.1007/JHEP04(2018)003}{\emph{JHEP} {\bfseries 04}
  (2018) 003} [\href{https://arxiv.org/abs/1704.01580}{{\ttfamily
  1704.01580}}].

\bibitem{Butenschoen:2016lpz}
M.~Butenschoen, B.~Dehnadi, A.~H. Hoang, V.~Mateu, M.~Preisser and I.~W.
  Stewart, \emph{{Top Quark Mass Calibration for Monte Carlo Event
  Generators}},
  \href{https://doi.org/10.1103/PhysRevLett.117.232001}{\emph{Phys. Rev. Lett.}
  {\bfseries 117} (2016) 232001}
  [\href{https://arxiv.org/abs/1608.01318}{{\ttfamily 1608.01318}}].

\bibitem{Czarnecki:1997wy}
A.~Czarnecki, K.~Melnikov and N.~Uraltsev, \emph{{Complete
  $\mathcal{O}(\alpha_s^2)$ corrections to zero recoil sum rules for $B\to D^*$
  transitions}}, \href{https://doi.org/10.1103/PhysRevD.57.1769}{\emph{Phys.
  Rev. D} {\bfseries 57} (1998) 1769}
  [\href{https://arxiv.org/abs/hep-ph/9706311}{{\ttfamily hep-ph/9706311}}].

\bibitem{Czarnecki:1997sz}
A.~Czarnecki, K.~Melnikov and N.~Uraltsev, \emph{{NonAbelian dipole radiation
  and the heavy quark expansion}},
  \href{https://doi.org/10.1103/PhysRevLett.80.3189}{\emph{Phys. Rev. Lett.}
  {\bfseries 80} (1998) 3189}
  [\href{https://arxiv.org/abs/hep-ph/9708372}{{\ttfamily hep-ph/9708372}}].

\bibitem{Neubert:1996zy}
M.~Neubert, \emph{{Exploring the invisible renormalon: Renormalization of the
  heavy quark kinetic energy}},
  \href{https://doi.org/10.1016/S0370-2693(96)01600-0}{\emph{Phys. Lett. B}
  {\bfseries 393} (1997) 110}
  [\href{https://arxiv.org/abs/hep-ph/9610471}{{\ttfamily hep-ph/9610471}}].

\bibitem{Bosch:2004th}
S.~W. Bosch, B.~O. Lange, M.~Neubert and G.~Paz, \emph{{Factorization and shape
  function effects in inclusive B meson decays}},
  \href{https://doi.org/10.1016/j.nuclphysb.2004.07.041}{\emph{Nucl. Phys. B}
  {\bfseries 699} (2004) 335}
  [\href{https://arxiv.org/abs/hep-ph/0402094}{{\ttfamily hep-ph/0402094}}].

\bibitem{scetlib}
M.~A. Ebert, J.~K.~L. Michel, F.~J. Tackmann et~al., \emph{{SCETlib: A C++
  Package for Numerical Calculations in QCD and Soft-Collinear Effective
  Theory}}, {\emph{DESY-17-099} (2018) }.

\bibitem{Korchemsky:1987wg}
G.~P. Korchemsky and A.~V. Radyushkin, \emph{{Renormalization of the Wilson
  Loops Beyond the Leading Order}},
  \href{https://doi.org/10.1016/0550-3213(87)90277-X}{\emph{Nucl. Phys. B}
  {\bfseries 283} (1987) 342}.

\bibitem{Moch:2004pa}
S.~Moch, J.~A.~M. Vermaseren and A.~Vogt, \emph{{The Three loop splitting
  functions in QCD: The Nonsinglet case}},
  \href{https://doi.org/10.1016/j.nuclphysb.2004.03.030}{\emph{Nucl. Phys. B}
  {\bfseries 688} (2004) 101}
  [\href{https://arxiv.org/abs/hep-ph/0403192}{{\ttfamily hep-ph/0403192}}].

\bibitem{Henn:2019swt}
J.~M. Henn, G.~P. Korchemsky and B.~Mistlberger, \emph{{The full four-loop cusp
  anomalous dimension in $\mathcal{N}=4$ super Yang-Mills and QCD}},
  \href{https://doi.org/10.1007/JHEP04(2020)018}{\emph{JHEP} {\bfseries 04}
  (2020) 018} [\href{https://arxiv.org/abs/1911.10174}{{\ttfamily
  1911.10174}}].

\bibitem{vonManteuffel:2020vjv}
A.~von Manteuffel, E.~Panzer and R.~M. Schabinger, \emph{{Cusp and collinear
  anomalous dimensions in four-loop QCD from form factors}},
  \href{https://doi.org/10.1103/PhysRevLett.124.162001}{\emph{Phys. Rev. Lett.}
  {\bfseries 124} (2020) 162001}
  [\href{https://arxiv.org/abs/2002.04617}{{\ttfamily 2002.04617}}].

\bibitem{Tarasov:1980au}
O.~V. Tarasov, A.~A. Vladimirov and A.~Y. Zharkov, \emph{{The Gell-Mann-Low
  Function of QCD in the Three Loop Approximation}},
  \href{https://doi.org/10.1016/0370-2693(80)90358-5}{\emph{Phys. Lett. B}
  {\bfseries 93} (1980) 429}.

\bibitem{Larin:1993tp}
S.~A. Larin and J.~A.~M. Vermaseren, \emph{{The Three loop QCD Beta function
  and anomalous dimensions}},
  \href{https://doi.org/10.1016/0370-2693(93)91441-O}{\emph{Phys. Lett. B}
  {\bfseries 303} (1993) 334}
  [\href{https://arxiv.org/abs/hep-ph/9302208}{{\ttfamily hep-ph/9302208}}].

\bibitem{vanRitbergen:1997va}
T.~van Ritbergen, J.~A.~M. Vermaseren and S.~A. Larin, \emph{{The Four loop
  beta function in quantum chromodynamics}},
  \href{https://doi.org/10.1016/S0370-2693(97)00370-5}{\emph{Phys. Lett. B}
  {\bfseries 400} (1997) 379}
  [\href{https://arxiv.org/abs/hep-ph/9701390}{{\ttfamily hep-ph/9701390}}].

\bibitem{Czakon:2004bu}
M.~Czakon, \emph{{The Four-loop QCD beta-function and anomalous dimensions}},
  \href{https://doi.org/10.1016/j.nuclphysb.2005.01.012}{\emph{Nucl. Phys. B}
  {\bfseries 710} (2005) 485}
  [\href{https://arxiv.org/abs/hep-ph/0411261}{{\ttfamily hep-ph/0411261}}].

\bibitem{Ebert:2021aoo}
M.~A. Ebert, \emph{{Analytic results for Sudakov form factors in QCD}},
  \href{https://doi.org/10.1007/JHEP02(2022)136}{\emph{JHEP} {\bfseries 02}
  (2022) 136} [\href{https://arxiv.org/abs/2110.11360}{{\ttfamily
  2110.11360}}].

\bibitem{Billis:2019evv}
G.~Billis, F.~J. Tackmann and J.~Talbert, \emph{{Higher-Order Sudakov
  Resummation in Coupled Gauge Theories}},
  \href{https://doi.org/10.1007/JHEP03(2020)182}{\emph{JHEP} {\bfseries 03}
  (2020) 182} [\href{https://arxiv.org/abs/1907.02971}{{\ttfamily
  1907.02971}}].

\bibitem{Gaunt:2015pea}
J.~Gaunt, M.~Stahlhofen, F.~J. Tackmann and J.~R. Walsh, \emph{{N-jettiness
  Subtractions for NNLO QCD Calculations}},
  \href{https://doi.org/10.1007/JHEP09(2015)058}{\emph{JHEP} {\bfseries 09}
  (2015) 058} [\href{https://arxiv.org/abs/1505.04794}{{\ttfamily
  1505.04794}}].

\bibitem{Balzereit:1998yf}
C.~Balzereit, T.~Mannel and W.~Kilian, \emph{{Evolution of the light cone
  distribution function for a heavy quark}},
  \href{https://doi.org/10.1103/PhysRevD.58.114029}{\emph{Phys. Rev. D}
  {\bfseries 58} (1998) 114029}
  [\href{https://arxiv.org/abs/hep-ph/9805297}{{\ttfamily hep-ph/9805297}}].

\bibitem{Neubert:2004dd}
M.~Neubert, \emph{{Renormalization-group improved calculation of the B
  ---\ensuremath{>} X(s) gamma branching ratio}},
  \href{https://doi.org/10.1140/epjc/s2005-02141-1}{\emph{Eur. Phys. J. C}
  {\bfseries 40} (2005) 165}
  [\href{https://arxiv.org/abs/hep-ph/0408179}{{\ttfamily hep-ph/0408179}}].

\bibitem{Fleming:2007xt}
S.~Fleming, A.~H. Hoang, S.~Mantry and I.~W. Stewart, \emph{{Top Jets in the
  Peak Region: Factorization Analysis with NLL Resummation}},
  \href{https://doi.org/10.1103/PhysRevD.77.114003}{\emph{Phys. Rev. D}
  {\bfseries 77} (2008) 114003}
  [\href{https://arxiv.org/abs/0711.2079}{{\ttfamily 0711.2079}}].

\end{thebibliography}\endgroup
\end{document}